\documentclass[a4paper,12pt]{article}

\usepackage{latexsym}
\usepackage{comment}
\usepackage[english]{babel}
\usepackage[font={small}]{caption}
\usepackage{epsfig,amssymb,euscript,slashed}
\usepackage[linktocpage]{hyperref}
\usepackage{amsmath}
\usepackage[noadjust]{cite}
\usepackage{mathtools}
\usepackage{mathrsfs}
\usepackage{physics}
\usepackage{extarrows}
\usepackage{xcolor}
\usepackage{float}
\usepackage{graphicx}
\usepackage{caption}
\usepackage{subcaption}
\usepackage{tcolorbox}
\usepackage{tikz}
\usetikzlibrary{positioning,shapes.multipart,arrows.meta,decorations.pathreplacing,calligraphy}
\usepackage{adjustbox}
\usepackage{nicematrix}
\usepackage{array,calc}
\usepackage[]{graphicx}
\usepackage{color}
\usepackage{bbm}
\usepackage{fancybox}
\usepackage{ytableau}
\usepackage{enumerate}
\usepackage[shortlabels]{enumitem}

\numberwithin{equation}{section}

\DeclareMathOperator{\sgn}{sgn}

\DeclarePairedDelimiter\floor{\lfloor}{\rfloor}

\def\scrA{\mathscr{A}}
\def\scrG{\mathscr{G}}
\def\Ad{\dot{A}}
\def\Bd{\dot{B}}
\def\Cd{\dot{C}}

\def\itilde{\tilde{\imath}}

\newcommand{\jt}{\tilde{\jmath}}
\def\mt{\tilde{m}}

\def\qt{\tilde{q}}
\def\xt{\tilde{x}}
\def\zt{\tilde{z}}

\def\Jt{\tilde{J}}

\def\ep{\epsilon}

\def\ald{\dot{\alpha}}

\def\alphat{\tilde{\alpha}}

\def\ty{\tilde{y}}
\def\yt{\tilde{y}}
\def\tj{\tilde{\jmath}}
\def\tq{\tilde{q}}
\def\ty{\tilde{y}}
\def\lambdat{\tilde{\lambda}}

\def\lam{\lambda}

\def\pd{\partial}

\def\lto{\longrightarrow}
\def\Lto{\Longrightarrow }

\def\CYon{\ytableausetup{centertableaux}}

\def\YTnormalsize{\ytableausetup{boxsize=1.25ex}}
\def\YTscriptsize{\ytableausetup{boxsize=0.65ex}}
\def\YTscriptscriptsize{\ytableausetup{boxsize=0.5ex}}

\definecolor{dfill1}{rgb}{0.0, 0.45, 0.73}
\definecolor{dfill2}{rgb}{0.0, 0.19, 0.33}
\definecolor{dfill3}{rgb}{0.1, 0.1, 0.5}

\def\vtikz#1{\vcenter{\hbox{\begin{tikzpicture}#1\end{tikzpicture}}}}

\newcommand{\diamcolor}{dfill3!65}
\newcommand{\diamopacity}{0.5}
\newcommand{\diam}[3][0.5]{%
\draw[fill=\diamcolor, fill opacity=\diamopacity,shift={(#2,#3)},draw opacity=0.6]
(0,#1) -- (#1,0) -- (0,-#1) -- (-#1,0) -- cycle;
}

\RenewDocumentCommand\ket{s m}{%
   \IfBooleanTF{#1}
     {\left\lvert\smash{#2}\right\rangle}%
     {\left\lvert{#2}\right\rangle}%
}

\definecolor{garrow1}{rgb}{0.90,0.75,0.5}

\def\p{\partial}
\def\Kt{\tilde{K}}
\def\St{\tilde{S}}
\def\alphad{{\dot{\alpha}}}
\def\betad{{\dot{\beta}}}
\def\gammad{{\dot{\gamma}}}
\def\psit{{\tilde{\psi}}}
\def\Vt{{\tilde{V}}}
\def\half{\frac12}
\def\thalf{\tfrac12}
\def\etat{\tilde{\eta}}
\def\etatb{\tilde{\boldsymbol{\eta}}}
\def\ytb{\tilde{\boldsymbol{y}}}
\def\te{\tilde{\eta}}
\def\cGt{\tilde{\mathcal{G}}}

\makeatletter
\g@addto@macro\bfseries{\boldmath}
\makeatother

\def\cA{{\cal A}}

\def\cD{{\cal D}}
\def\cE{{\cal E}}

\def\cG{{\cal G}}
\def\cH{{\cal H}}

\def\cM{{\cal M}}
\def\cN{{\cal N}}
\def\cO{{\cal O}}
\def\cP{{\cal P}}

\def\cR{{\cal R}}

\def\cT{{\cal T}}

\def\cZ{{\cal Z}}

\def\bbC{{\mathbb{C}}}

\def\bbZ{{\mathbb{Z}}}

\usepackage{pict2e}

\makeatletter
\DeclareRobustCommand{\loplus}{\mathbin{\mathpalette\dog@lsemi{+}}}
\DeclareRobustCommand{\lotimes}{\mathbin{\mathpalette\dog@lsemi{\times}}}
\DeclareRobustCommand{\roplus}{\mathbin{\mathpalette\dog@rsemi{+}}}
\DeclareRobustCommand{\rotimes}{\mathbin{\mathpalette\dog@rsemi{\times}}}

\newcommand{\dog@rsemi}[2]{\dog@semi{#1}{#2}{-90,90}}
\newcommand{\dog@lsemi}[2]{\dog@semi{#1}{#2}{270,90}}
\newcommand{\dog@semi}[3]{%
  \begingroup
  \sbox\z@{$\m@th#1#2$}%
  \setlength{\unitlength}{\dimexpr\ht\z@+\dp\z@\relax}%
  \makebox[\wd\z@]{\raisebox{-\dp\z@}{%
    \begin{picture}(1,1)
    \linethickness{\variable@rule{#1}}
    \roundcap
    \put(0.5,0.5){\makebox(0,0){\raisebox{\dp\z@}{$\m@th#1#2$}}}
    \put(0.5,0.5){\arc[#3]{0.5}}
    \end{picture}%
  }}%
  \endgroup
}
\newcommand{\variable@rule}[1]{%
  \fontdimen8  
  \ifx#1\displaystyle\textfont3\else
    \ifx#1\textstyle\textfont3\else
      \ifx#1\scriptstyle\scriptfont3\else
        \scriptscriptfont3\relax
  \fi\fi\fi
}
\makeatother
\def\Res{\mathop{\mathrm{Res}}\limits}

\begin{document}
\font\cmss=cmss10 \font\cmsss=cmss10 at 7pt

\begin{flushright}{  
\scriptsize YITP-26-27}
\end{flushright}
\hfill
\vspace{18pt}
\begin{center}
{\Large 
\textbf{The Resolved Elliptic Genus and the D1-D5 CFT}}

\end{center}

\vspace{8pt}
\begin{center}
{\textsl{Marcel R. R. Hughes$^{\,a}$ and Masaki Shigemori$^{\,a,b}$}}

\vspace{1cm}

\textit{\small ${}^a$ Department of Physics, Nagoya University\\
Furo-cho, Chikusa-ku, Nagoya 464-8602, Japan} \\ \vspace{6pt}

\textit{\small ${}^b$ 
Center for Gravitational Physics,\\
Yukawa Institute for Theoretical Physics, Kyoto University\\
Kitashirakawa Oiwakecho, Sakyo-ku, Kyoto 606-8502, Japan
}\\
\vspace{6pt}

\end{center}

\vspace{12pt}

\begin{center}
\textbf{Abstract}
\end{center}

\vspace{4pt} {\small
\noindent
This paper is a follow-up to the short paper~\cite{Hughes:2025oxu}, 
greatly expanding the discussion with examples and providing derivations and justifications of the results presented there.

We introduce a new supersymmetry index for the D1-D5 CFT on $T^4$, which we call the resolved elliptic genus (REG)\@.  It is a one-parameter generalisation of the standard supersymmetry index, the modified elliptic genus (MEG), and arises naturally in the free symmetric orbifold description of the theory within a new formalism, based on Schur-Weyl duality, that we develop.
In this formalism, the Hilbert space of the symmetric orbifold CFT is decomposed into symmetry sectors in which the structure of the states contributing to the MEG is transparent.  By examining the action of the supercharge deformed by an exactly marginal operator on the relevant symmetry algebra, we propose a superselection rule governing the lifting process of BPS states, and use it to construct the REG by summing only over those symmetry sectors that can mix according to this rule.
The REG exhibits detailed agreement between the CFT and supergravity below the black-hole threshold, a regime in which the MEG is essentially trivial. Above the threshold, the REG is dominated by black-hole microstates, which are now distributed amongst distinct sectors that are invisible to the MEG\@.
We expect both the new formalism and the REG to provide useful new tools for studying the structure of black-hole microstates. In particular, we comment on their possible relevance to the fortuity program for understanding black-hole microstates within CFT.


}

\vspace{1cm}

\thispagestyle{empty}

\vfill
\vskip 5.mm
\hrule width 5.cm
\vskip 2.mm
{
\noindent  {\scriptsize e-mails:  {\texttt{hughes.mrr@eken.phys.nagoya-u.ac.jp, masaki.shigemori@nagoya-u.jp}} }
}

\setcounter{footnote}{0}
\setcounter{page}{0}

\newpage

\tableofcontents





\section{Introduction}
\label{sec:intro}


The D1-D5 CFT has played a central role in the development of holography.  Arising in the decoupling limit of the D1-D5 system, it is one of the most studied realizations of the AdS/CFT correspondence \cite{Maldacena:1997re} and has been particularly important in the search for an understanding of black-hole physics within string theory.  The CFT is a two-dimensional $\cN=(4,4)$ SCFT and is dual to type IIB superstring theory in AdS$_3\times S^3\times \cM$ with $\cM=T^4$ or $K3$.
On a certain locus in its moduli space, the CFT's target space becomes a symmetric orbifold \cite{Dixon:1986jc,Dixon:1986qv}, a structure that universally exhibits many large $N$ features key to the holographic nature of the theory~\cite{Hartman:2014oaa,Belin:2025nqd}.

As one of the earliest examples of the AdS/CFT correspondence, the D1-D5 system has been the setting of extensive comparisons between bulk and boundary states.  It was in this system where the first match between thermodynamic entropy in the CFT and the Bekenstein-Hawking entropy of AdS black holes was obtained above the black-hole threshold~\cite{Strominger:1996sh, Breckenridge:1996is}.  Below this threshold, precise agreement between the CFT and bulk excitation spectra has also been established~\cite{Maldacena:1998bw, deBoer:1998ip, Deger:1998nm, Larsen:1998xm}.
Supersymmetry indices \cite{Witten:1982df, deBoer:1998us, Maldacena:1999bp} have been integral to such matching, being independent of the coupling that relates the symmetric orbifold and weakly-coupled gravity regimes.
In particular, for $\cM=K3$, a precise agreement between the elliptic genus computed in CFT and in bulk supergravity was found in \cite{deBoer:1998us} below the black-hole threshold when a stringy exclusion principle \cite{Maldacena:1998bw} is imposed. Analogously for $\cM=T^4$, similar agreement was found in~\cite{Maldacena:1999bp}. However, this agreement is rather vacuous in the sense that, except for the vacuum contribution, the index identically vanishes below the threshold.  This is because the relevant index for the $T^4$ theory, the modified elliptic genus (MEG), is not especially sensitive to the detailed spectrum.

Recent developments are beginning to uncover finer structure of the matched spectra: fortuity \cite{Chang:2024zqi, Chang:2025rqy} sharpens the notion of typicality for black-hole microstates, while CFT techniques \cite{Gava:2002xb, Guo:2019ady, Guo:2020gxm, Hampton:2018ygz, Guo:2019pzk, Benjamin:2021zkn, Guo:2022ifr, Gaberdiel:2024nge} make it possible to determine whether particular microstates remain supersymmetric when interactions are turned on. Again, existing supersymmetry indices are of limited use for these purposes: defined purely from the supersymmetry algebra, they are largely insensitive to the details of interactions.

In this paper, we introduce a new formulation for symmetric orbifold CFTs based on Schur-Weyl duality, and use it to rewrite the modified elliptic genus (MEG) of the D1-D5 CFT on $T^4$ in a form that makes the structure of contributing states more transparent.
Using this framework, we are then motivated to propose a new supersymmetry index -- the \emph{resolved elliptic genus (REG)} -- that incorporates superselection rules based on the nature of the interaction in the symmetric orbifold CFT\@. For the D1-D5 CFT on $T^4$, we demonstrate that the REG provides much more information about the structure of microstates than the MEG\@.

\bigskip

In the remainder of this section, we summarize the main content of this paper in more detail.

First, we introduce a new formulation of symmetric orbifold CFTs.  The Hilbert space of a symmetric orbifold CFT, ${\rm Sym}^N(\cM)$, decomposes into so-called twist sectors, each consisting of many ``strands'' of various lengths.  Intuitively, a strand can be viewed as a string propagating in $\cM$, whose length is given by the strand length, so that the Hilbert space describes a system of multiple such strings.
In the case of the D1-D5 CFT on ${\rm Sym}^N (T^4)$, this Hilbert space factorizes into the left- and right-moving sectors.  This is equally true in the $\frac14$-BPS sector that we focus on, in which the left-moving sector is in an arbitrary state while the right-moving sector is in a Ramond ground state (per strand, there are two bosonic and two fermionic right-moving ground states).
In the case of there being $n$ strands, the total state on these strands must be invariant under $S_n$ permuting them. This is analogous to the wavefunction of $n$ spin-$\frac12$ particles, which must be antisymmetric under exchange; in that case, the wavefunction can be decomposed into orbital and spin parts, each transforming in an $S_n$-representation in such a way that the total wavefunction is antisymmetric.
Analogously, the state of $n$ strands can be decomposed into left- and right-moving parts, each transforming nontrivially under $S_n$ such that the total state is $S_n$-invariant.  In this case, the left- and right-moving states must be in the same $S_n$ representation, labeled by a Young diagram $\lambda$ with $n$ boxes.  Accordingly, the $n$-strand Hilbert space $\cH_n$ decomposes as
\begin{align}
    \cH_n = \bigoplus_\lambda V_\lambda\otimes \Vt_\lambda \ ,
    \label{SW-decomp_intro}
\end{align}
where $V_\lambda$ is the subspace of the left-moving $n$-strand Hilbert space transforming in the $S_n$-representation labeled by $\lambda$, and $\Vt_\lambda$ is the corresponding right-moving subspace.
Furthermore, the Schur-Weyl duality implies that $\Vt_\lambda$ is an irreducible representation space of $GL(2|2)$ labeled by the same Young diagram $\lambda$, where $GL(2|2)$ rotates $2+2$ right-moving ground states on each strand.  Thus, each term in the sum corresponds to a $GL(2|2)$ multiplet.
The decomposition \eqref{SW-decomp_intro} implies that the partition function admits a corresponding decomposition into $\lambda$-sectors, along with the index of the $T^4$ theory -- the modified elliptic genus:
\begin{align}
    \cE=\sum_\lambda S_\lambda\, \cD\St_\lambda \ ,
    \label{MEG-decomp_intro}
\end{align}
where $S_\lambda$ is the contribution from the left-moving sector in the $\lambda$-sector, while $\cD\St_\lambda$ is the right-moving contribution, to be fully explained later.

In the decomposition \eqref{MEG-decomp_intro}, each $\Vt_\lambda$ corresponds to a $GL(2|2)$ multiplet.  However, in fact, $gl(2|2)$ is not a symmetry algebra of the D1-D5 CFT; once interaction is turned on, the relevant symmetry algebra is instead a certain subalgebra, $\scrA\subset gl(2|2)$.   Its irreducible representations, denoted by $\cR^{\scrA}_{\jt,\jt_2}$, are labeled by spins $(\jt,\jt_2)$ of free-theory $su(2)$ symmetries and may be graphically represented by ``diamond diagrams'', examples of which are shown in Figure~\ref{fig:A-alg_reps_ex} (page \pageref{fig:A-alg_reps_ex}), where each diamond corresponds to a quartet generated by fermionic zero modes.
Thus, each $\lambda$-sector in \eqref{MEG-decomp_intro} is decomposed into multiple $\scrA$-representations, or equivalently, diamond diagrams.

When interactions are turned on, BPS states of the free orbifold theory become non-BPS or remain BPS at the level of $\scrA$-representations: each $\scrA$-representation either lifts as a whole or remains BPS as a whole.
By analyzing the structure of the first-order deformed supercharge, we derive a superselection rule that forbids mixing between $\scrA$-representations with different values of $\jt_2$.  This naturally leads to a new protected quantity by restricting the sum \eqref{MEG-decomp_intro} as follows: first decompose the $gl(2|2)$-representation labeled by $\lambda$ into $\scrA$-representations labeled by $(\jt,\jt_2)$, and then sum only over terms with a fixed value of $\jt_2$.  This defines the resolved elliptic genus (REG), denoted by $\cE_{N,\jt_2}$, where $N$ means restriction to states with total strand length $N$, so that it defines an index for the ${\rm Sym}^N(T^4)$ theory.  For an explicit expression, see \eqref{eq.REG1}.  We also derive a closed-form expression for the generating function of the REG (Eq.~\eqref{genFuncREG}).

We then apply the REG to compare the CFT and supergravity spectra below the black-hole threshold.   We explicitly see that, although the matching is rather trivial for the MEG, the REG resolves it into $\jt_2$-sectors that individually exhibit nontrivial agreement.  Also, above the black-hole threshold, we explicitly see how black-hole states counted by the MEG decompose into multiple $\jt_2$-sectors counted by the REG\@.

\bigskip
This paper is a follow-up to \cite{Hughes:2025oxu}, in which some of the results presented in this paper were briefly announced.  The current paper considerably expands the discussion, with full details and more examples.

The plan of the paper is as follows.  
In Section~\ref{sec:background}, we explain the structure of the Hilbert space of symmetric orbifold CFTs in general, and then turn to the specific case relevant to the D1-D5 CFT on $T^4$, briefly reviewing its fields, states, and symmetry algebras.
In Section~\ref{sec:S-W}, we derive the Schur-Weyl decomposition of the Hilbert space of $\mathrm{Sym}^N(T^4)$ (as in \eqref{SW-decomp_intro}).  We also discuss the Schur-Weyl decomposition of relevant partition functions and indices (as in \eqref{MEG-decomp_intro}), focusing on the MEG in particular.
In Section~\ref{ssec:Lifting}, we study the structure and representation theory of the symmetry algebra relevant for lifting, and derive a superselection rule.
In Section~\ref{sec:REG}, we apply the superselection rule to the Schur-Weyl decomposition of the MEG and derive the REG as a partial sum over $\jt_2$-sectors.  We then demonstrate that the REG allows for a finer comparison between CFT and supergravity than the MEG\@.
Section~\ref{sec:summary} is devoted to discussion and outlook.
In the appendices we present details and derivations not discussed in the main text.

\section{Background}
\label{sec:background}

Here we briefly present some basic aspects of the symmetric orbifold CFT for $T^4$ and its relation to the D1-D5 CFT\@. For a more detailed introduction we refer the reader, for example, to \cite{David:2002wn,Avery:2010qw,Fabri:2025rok}. We will, for the most part, use the same notation and conventions as \cite{Avery:2010qw,Hughes:2025tdy,Hughes:2025oxu}.

The symmetric orbifold CFT on $\mathrm{Sym}^N(\cM)$ is defined from a ``seed'' CFT$_2$  on $\cM$ defined on a spatial $S^1$ and with central charge $c_0$, by taking the $N$-fold tensor product $\cM^{\otimes N}$ and then orbifolding by the symmetric group $S_N$.
The action of $S_N$ on $\cM^{\otimes N}$ is to permute the tensor factors -- the $N$ ``copies'' of $\cM$. The resulting orbifold theory has a central charge $c=Nc_0$ and its Hilbert space $H(\mathrm{Sym}^N(\cM))$ contains only states which are invariant under $S_N$. This orbifold Hilbert space decomposes into so-called twist sectors \cite{Dixon:1986jc,Dixon:1986qv} labelled by the conjugacy classes $[g]$ of $S_N$ as
\begin{equation} \label{eq.symm_orb_H}
    H\big(\mathrm{Sym}^N(\cM)\big) \cong \bigoplus_{[g]} H_{[g]} \ .
\end{equation}  
For the symmetric group $S_N$, there is a one-to-one correspondence between its conjugacy classes $[g]$ and  integer partitions of $N$\@. Therefore, we can label a twist sector by $\{n_k\}\vdash N$, where $\vdash$ denotes ``is a partition of'', namely, $\sum_{k=1}^N k n_k=N$.   The corresponding Hilbert space takes the form
\begin{equation} \label{eq.HtwistSector}
    H_{\{n_k\}} = \bigotimes_{k=1}^{N} \bigg(\underbrace{H_{(k)}\otimes\cdots\otimes H_{(k)}}_{n_k}\bigg)_{\text{$S_{n_k}$-inv}} \ ,
\end{equation}
where $(H\otimes\cdots\otimes H)_{\text{$S_{n}$-inv}}$ represents the $S_n$-invariant subspace of the $n$-fold tensor product space $H\otimes\cdots\otimes H$. 
In \eqref{eq.HtwistSector}, $H_{(k)}$ is the Hilbert space of the theory defined on a $k$-times longer spatial $S^1$ -- a ``strand'' of length $k$.  It is natural to then introduce a ``covering'' Hilbert space, or Fock space, $\cH$, in which all values of $N$ are included by summing over $n_k\in\mathbb{Z}_{\ge 0}$
\begin{equation} \label{eq.symm_orb_H_cover}
    \cH \equiv \bigoplus_{N=1}^{\infty} H\big(\mathrm{Sym}^N(\cM)\big) \cong \bigoplus_{n_k}H_{\{n_k\}} \ .
\end{equation}
The Hilbert space \eqref{eq.symm_orb_H} at fixed $N$ is then obtained by projection onto the subspace with total strand number $N$.\footnote{For a recent discussion of the structure of symmetric orbifold theories in terms of topological defect lines, see \cite{Benjamin:2025knd}.}

Besides the operator contents of the seed theory, symmetric orbifold CFTs also contain the twist operators $\sigma_k$, labeled by an order $k$ cyclic permutation, which generate the ground states of the single-cycle\footnote{Twist operators that generate ground states of multi-cycle twist sectors can be decomposed into a product of single-cycle twist operators, with this product being normal ordered such that the cycles act on distinct subsets of copies.} Hilbert space $H_{(k)}$. In particular, $\sigma_2$, which splits a strand into two shorter strands or glues together two strands into a longer one, is used to construct a marginal deformation operator and plays an essential role in linking symmetric orbifold CFTs to a holographic setup.

In the case of the D1-D5 CFT for $T^4$, which is the focus of this paper, the seed theory is an $\cN=(4,4)$ supersymmetric sigma model on $\cM=T^4$ with central charge $c_0=6$. This seed theory contains four bosons $X^{\Ad A}$ from the coordinates on the target space $T^4$, along with four left-moving $\psi^{\alpha \Ad}$ and four right-moving $\tilde{\psi}^{\ald \Ad}$ fermionic superpartners\footnote{In this paper, right-moving quantities will always be denoted with a tilde.}. The expansion modes of the left-moving $\pd X^{\Ad A}$, $\psi^{\alpha \Ad}$ and the right-moving $\tilde{\pd} X^{\Ad A}$, $\tilde{\psi}^{\ald \Ad}$ are denoted respectively by
\begin{align} \label{eq.Xpsi_modes}
    \alpha^{\Ad A}_{r} \ ,\ \ \psi^{\alpha \Ad}_{r} \ ,\ \ \tilde{\alpha}^{\Ad A}_{r}\ ,\ \ \tilde{\psi}^{\ald \Ad}_r \ ,
\end{align}
where the mode number $r$ is integer in the case of the Ramond sector which we work in. Spectral flow can be used to map to the Neveu-Schwarz (NS) sector \cite{Schwimmer:1986mf}, in which the fermion modes are half-integer moded, as defined in \eqref{eq.symmModeDeform}. The $SU(2)$ doublet indices used above take the values
\begin{equation} \label{eq.doubletdefs}
    \alpha\in\{+,-\}\ ,\ \ald\in\{\dot{+},\dot{-}\}\ ,\ A\in\{1,2\}\ ,\ \Ad\in\{\dot{1},\dot{2}\} \ ,
\end{equation}
where $\alpha$ and $\ald$ are doublet indices for the left and right $SU(2)$ R-symmetry groups from the ``external"
\begin{equation} \label{eq:SO(4)_E}
    SO(4)_E\cong SU(2)_L\times SU(2)_R \ ,
\end{equation}
coming from the isometry group of $S^3$ transverse to the D1-D5 brane worldvolume. On top of this, the free bosons transform in the vector representation of an ``internal" $SO(4)_I$ originating from the isometry group of $T^4$ longitudinal to the D5 worldvolume. This is generically a broken symmetry of the theory due to the compactification, but it will play an important role in this paper for organising the spectrum and in particular we will make use of the isomorphism
\begin{equation} \label{eq:SO(4)_I}
    SO(4)_I\cong SU(2)_1\times SU(2)_2 \ , 
\end{equation}
with the doublet representation indices $A$ and $\Ad$ in \eqref{eq.doubletdefs} being for $SU(2)_1$ and $SU(2)_2$ respectively. Although the $SU(2)_1\times SU(2)_2$ symmetry is broken by states carrying non-trivial winding and momentum on the torus, we will be focusing on the BPS sector of the CFT which is contained in the sector with zero momentum and winding along the internal $T^4$, and hence these $SU(2)$ will be used to organize states.

The symmetry algebra of this theory contains as a subalgebra the 2d small $\cN=(4,4)$ superconformal algebra generated by the left-moving stress tensor $T$, $SU(2)_L$ R-symmetry currents $J^{\pm,3}$ and supersymmetry generators $G^{\alpha A}$, plus the respective right-moving currents. In addition to these generators, the full symmetry algebra -- the 2d contracted large $\cN=(4,4)$ superconformal algebra -- also contains the free bosons and fermions as generators. We denote the modes of the small $\cN=(4,4)$ algebra as
\begin{equation} \label{eq.currentModes}
    \Big\{ L_{r}\ ,\ J^{\pm,3}_{r}\ ,\ G^{\alpha A}_{r}\Big\}\ \ ,\ \ \Big\{ \tilde{L}_{r}\ ,\ \tilde{J}^{\dot{\pm},3}_{r}\ ,\ \tilde{G}^{\ald A}_{r}\Big\} \ ,
\end{equation}
with $r\in \mathbb{Z}$ again in the Ramond sector. The algebra satisfied by these modes is given in Appendix~\ref{app:conventions}.

In the symmetric orbifold $\mathrm{Sym}^N(T^4)$, on an individual strand of length $k$, we have the same free boson and fermion modes \eqref{eq.Xpsi_modes}. However, due to the periodicity of the fields on the $k$-times longer $S^1$ of the theory on that strand, the mode numbers are fractional in $k$, defined in terms of the fields in \eqref{eq.strand_mode_def}. For the $I$th strand whose length is $k_I$, these modes are
\begin{align} \label{eq.Xpsi_modes_k}
    \alpha^{\Ad A[I]}_{s} \ ,\ \ \psi^{\alpha \Ad[I]}_{s} \ ,\ \ \tilde{\alpha}^{\Ad A[I]}_{s}\ ,\ \ \tilde{\psi}^{\ald \Ad[I]}_s \ ,
\end{align}
with $s=\frac{r}{k_I}$ and $r\in\mathbb{Z}$, where $\sum_{I=1}^n k_I = N$ for a given twist sector with $n$ strands. The surviving symmetry algebra from the $N$ copies of the seed theory is the diagonal contracted large $\cN=(4,4)$ superconformal algebra, with central charge $c=6N$. 
The generators of this diagonal algebra will be referred to as ``total modes''.  They are defined by summing the corresponding generator over the $N$ copies; for example, in the case of the Virasoro generators,
\begin{equation} \label{eq.total_mode_def_copy}
    L_m^{({\rm T})} \equiv \sum_{i=1}^{N} L_m^{(i)} \ ,
\end{equation}
where $L_m^{(i)}$ are the generators on the $i$th copy. 
The other generators of the total mode algebra are defined analogously. We will choose to drop the $({\rm T})$ label on total modes, clarifying explicitly if unclear from the context. 
In this paper, we will prioritise the strand picture of the symmetric orbifold theory, rather than the copy picture.  Accordingly, it is more convenient to write total mode generators in terms of generators on individual strands, \textit{e.g.}\ $L_m^{[I]}$, defined by summing $L^{(i)}_m$ over the copies belonging to a given strand (see  \eqref{eq.strand_mode_def}). 
In a particular twist sector containing $n$ strands, the total modes are given by \textit{e.g.}
\begin{equation} \label{eq.total_mode_def_strand}
    L_m^{({\rm T})} = \sum_{I=1}^{n} L_m^{[I]} \ .
\end{equation}

The Ramond sector of this theory has a highly degenerate ground state with conformal dimensions $h=\tilde{h}=\frac{N}{4}$; on an individual strand of length $k$, the RR (Ramond-Ramond) ground states are tensor products of the left-moving ground states $\ket*{\alpha}_k,\ket*{\Ad}_k$ and the right-moving ground states $\ket*{\dot\alpha}_k, \ket*{\Ad}_k$, generated respectively by the left- and right-moving fermion zero modes on that strand. The explicit form of the right-moving ground states is
\begin{equation}
\begin{aligned} \label{eq.R_GS_rels}
    &\ket*{\dot{-}}_{k_I} & \tilde{h}&=\frac{k_I}{4}\,,\ \mt = -\frac12 \ ,\\
    &\ket*{\dot{A}}_{k_I} = \frac{1}{\sqrt{k_I}}\tilde{\psi}^{\dot{+}\dot{A}[I]}_0 \ket*{\dot{-}}_{k_I} & \tilde{h}&=\frac{k_I}{4}\,,\ \mt = 0\ ,\\
    &\ket*{\dot{+}}_{k_I} = \frac{1}{k_I}\tilde{\psi}^{\dot{+}\dot{1}[I]}_0 \tilde{\psi}^{\dot{+}\dot{2}[I]}_0 \ket*{\dot{-}}_{k_I} \qquad
    &\tilde{h}&=\frac{k_I}{4}\,,\ \mt = +\frac12 \ ,
\end{aligned}
\end{equation}
where $\mt$ is the right-moving R-charge, and the left-moving ground states follow analogously. The states $\ket*{\alpha}_k,\ket*{\dot\alpha}_k$ are bosonic while the $\ket*{\Ad}_k$ are fermionic in our conventions.

While the above is the standard description of the Ramond ground states of this theory and the representations in which they transform, for our purposes it will prove important to distinguish the action of $SU(2)_2$ on the left-moving and right-moving parts of states separately. We define the group $\widetilde{SU}(2)_2$ as the right-moving action of $SU(2)_2$, under which the right-moving $\ket*{\Ad}$ Ramond ground states transform as a doublet and the analogous left-moving ground states as singlets.

In this $\mathrm{Sym}^N(T^4)$ theory, 1/4-BPS states have their right-moving part in Ramond ground states on each strand, while the left-moving part can be in any states. It is this restriction of the full Hilbert space \eqref{eq.symm_orb_H}, or rather the covering Hilbert space \eqref{eq.symm_orb_H_cover}, that we will be concerned with in this paper.

The full D1-D5 CFT is described by the moduli space\footnote{The D1-D5 CFT for $T^4$ has 20 exactly marginal operators which preserve the superconformal algebra, 16 of which are in the untwisted sector of the symmetric orbifold and are related to the moduli of the target space $T^4$, and 4 are in the twist-2 sector.} of the symmetric orbifold of $T^4$, over which the 1/4-BPS spectrum is not generally protected. In particular, by deforming the symmetric orbifold by the twisted-sector exactly marginal operator\footnote{Holographically, this particular deformation is equivalent to turning on Wilson lines for the RR potentials in the bulk description.}
\begin{equation} \label{eq:deformation_op_def}
    \Phi = \frac14\ep_{\alpha\beta}\ep_{\alphad\betad} \ep_{AB} G^{\alpha A}_{-\frac12} \tilde{G}^{\alphad B}_{-\frac12} \sigma_2^{\beta\betad} \ ,
\end{equation}
the 1/4-BPS spectrum of $\mathrm{Sym}^N(T^4)$ is partially lifted due to multiplet recombination: short multiplets of the free theory combining into long multiplets in the deformed theory. This will be the topic of Section~\ref{ssec:Lifting}.

\section{A Schur-Weyl formalism}
\label{sec:S-W}

In the previous section, we reviewed the standard way to construct the covering Hilbert space $\cH$ of a symmetric orbifold theory.  Here we present an alternative construction\footnote{For the mathematical background for this section, such as the Schur-Weyl duality, Schur functions, and Cauchy identities, see \cite{Fulton:1991} for the purely bosonic case and \cite{469926,BERELE1987118,BERELE1985225,BERELE3} for supersymmetric extensions.} which is based on factoring the theory into left- and right-moving sectors and provides a useful framework for the new index with which the current paper is concerned.

In a symmetric orbifold theory, physical states are $S_n$ invariant.  However, when states factor into the left- and right-moving parts, these parts can be individually in non-trivial representations of $S_n$ so long as the total state is invariant.  This is analogous to the situation for a system of spin-1/2 particles in quantum mechanics, where the total wavefunction is required to be antisymmetric under the exchange of any pair of particles. Namely, the total wavefunction must be in the totally antisymmetric representation (the alternating representation) of $S_n$ where $n$ is the number of particles. The standard way to construct the wavefunction, found in any textbook~\cite{LandauLifshitzQM}, is to allow the orbital wavefunction and the spin wavefunction to be in separate representations of $S_n$ such that the total wavefunction is totally antisymmetric.

This is exactly the spirit of what we do below for symmetric orbifold CFTs. For other approaches to symmetric orbifold CFTs based on left-right factorization and the Schur-Weyl duality, see {\it e.g.}~\cite{Jevicki:2015irq}, as well as applications in other settings in~\cite{Bantay:1999us,Bjornberg:2023jsi,Leiner:2018odg}. To our knowledge the approach that we present in the present paper, however, differs substantially from these previous formulations in symmetric orbifold theories and critically will lead to new insights into the supersymmetric indices of these theories.

\subsection{A bosonic example} \label{ssec:bosonEx}

\subsubsection*{Single strand}

Let us consider the Hilbert space of states on a single strand of length $k$ $(k\ge 1)$.
Throughout this paper, we assume that this Hilbert is the tensor product of 
left- and right-moving Hilbert spaces.

To illustrate the idea, we consider a concrete situation where the left-moving Hilbert space $V_k$ on a strand of length $k$ is obtained by the action of Virasoro generators $L_{-r}$ ($r\ge 1$) on a Virasoro primary state $\ket{\psi}_k$ with conformal weight $h_k$.
The right-moving Hilbert space $\Vt$, on the other hand, is assumed to be independent of $k$ and contains two states, $\ket*{\dot{+}}$ and $\ket*{\dot{-}}$, which form an $SU(2)$ doublet with the eigenvalues of $\Jt^3_0$ being $\mt=\pm \frac12$. 
This might seem too simple, but the physical example discussed in Section~\ref{ssec:SW_T4} will be of this type.
All states and operators in this toy example are assumed to be bosonic.  
Explicitly, the left- and right-moving Hilbert spaces on a single strand of length $k$ are
\begin{align}
 V_k={\rm span}\left\{\cdots L_{-3}^{\nu_3} L_{-2}^{\nu_2} L_{-1}^{\nu_1}
 \ket{\psi}_k\right\}_{\{\nu_r\}}^{}\ ,\qquad
 \Vt={\rm span}\left\{\ket*{\dot{+}},\ket*{\dot{-}}\right\}\ ,
 \label{eq:V_Vt_bosonic}
\end{align}
where $\nu_r\in \bbZ_{\ge 0}$.
The corresponding left- and right-moving partition functions are
\begin{align}
z_k(q)&=\tr_{V_k}^{}\!\big[q^{L_0}\big]={q^{h_k}\over \prod_{r=1}^\infty (1-q^r)}\ ,\label{z_k_bosonic}
\\
\zt(\yt)&=\tr_{\Vt}^{}\!\big[\yt^{2\Jt_0^3}\big]=\yt+\yt^{-1} \ ,
\label{zt_bosonic}
\end{align}
where $q,\yt$ are fugacities that count $L_0,\Jt^3_0$.

It is useful to then introduce a \emph{single-strand} Hilbert space that sums over all strand lengths:
\begin{align}
 V=\bigoplus_{k\ge 1} V_k\ .\label{V_bosonic}
\end{align}
More precisely, this is the left-moving single-strand Hilbert space; the total single-strand Hilbert space is the left-right product
\begin{align}
    \cH_1=V\otimes \Vt=\bigoplus_{k\ge 1} H_{(k)}\ ,\qquad H_{(k)}\equiv V_k\otimes \Vt\ .
    \label{eq:H1}
\end{align}
Note that we do not need a summation over $k$ for the right-moving sector because of the assumption that it is given by the same space $\Vt$, independent of the strand length $k$.
In the ``covering'' Hilbert space \eqref{V_bosonic} (and also in \eqref{eq:H1}), we can say that even the strand length $k$ is regarded as a quantum number to specify the state on the strand, just like $L_0$ and $\Jt^3_0$.
This covering Hilbert space is a natural object to consider due to the one-to-one correspondence between this single-strand Hilbert space and the Hilbert space of a string \cite{Dijkgraaf:1996xw}.   Correspondingly, the left-moving, single-stand partition function is
\begin{align}
z(p,q)=\tr_{V}^{}\!\big[p^{\hat{k}}q^{L_0}\big]=\sum_{k\ge 1}p^k z_k(q)\ ,
\label{z_bosonic}
\end{align}
where we introduced a fugacity $p$ that keeps track of the strand length, and $\hat{k}$ is the operator defined in $V$ that gives $k$ on a state in $V_k$.

If we regard the left-moving Hilbert space $V$ as the space of vectors $T^i$ with $i=1,2,\dots$, $\dim V$ (where really $\dim V=\infty$ here) representing a wavefunction, an operator $g$ on $V$ acts on the vector $T^i$ as a $GL(\infty)$ matrix $g^i{}_j$ as $T^i\to g^i{}_j\, T^j$.  Namely, $V$ can be regarded as a representation space of the fundamental representation of $GL(\infty)$ denoted by the Young diagram $\YTnormalsize\ydiagram{1}\,$. 
Likewise,
the right-moving Hilbert space $\Vt$, having a dimension of 2, can be regarded as the space of vectors $\tilde{T}^{\itilde}$ with $\itilde=1,2$, or a representation space of the fundamental representation $\ydiagram{1}$ of $GL(2)$.

\subsubsection*{Multiple strands}

As a first step towards the general multi-strand case, consider the case with two strands, where the left- and right-moving Hilbert spaces are the tensor product spaces $V\otimes V$ and $\Vt\otimes \Vt$, respectively (before imposing $S_2$ invariance under the exchange of strands, to be discussed below). We can regard $V\otimes V$ as the space of 2-tensors $T^{i_1 i_2}$ with $i_1,i_2=1,2,\dots,\dim V$. As is well-known, this space can be decomposed into $\YTscriptsize V_{\ydiagram{2}}$, the space of symmetric tensors $T^{(i_1 i_2)}={1\over 2}(T^{i_1 i_2}+T^{i_2 i_1})$, and $V_{\ydiagram{1,1}}$, the space of antisymmetric tensors $T^{[i_1 i_2]}={1\over 2}(T^{i_1 i_2}-T^{i_2 i_1})$.
Note that the tensors $T^{(i_1 i_2)}$ and $T^{[i_1 i_2]}$ each transform in an irreducible representation of $GL(\infty)$ that acts on the tensor indices in the usual way by matrix multiplication with $g^i{}_j$ on each index.  Furthermore, under exchanging the indices $i_1\leftrightarrow i_2$, \textit{i.e.}\ under $S_2$, the tensors $T^{(i_1 i_2)}$ and $T^{[i_1 i_2]}$ transform as trivial and alternating representations, respectively.  Therefore, the Young diagrams $\YTnormalsize\CYon\lambda=\ydiagram{2},\ydiagram{1,1}$ serve the dual purpose of labelling both $GL(\infty)$ and $S_2$ representations. We can similarly decompose the right-moving Hilbert space as $\YTscriptsize\Vt\otimes \Vt\cong \Vt_{\ydiagram{2}}\oplus \Vt_{\ydiagram{1,1}}$, and the relevant 2-tensor $\tilde{T}^{\itilde_1 \itilde_2}$ into 
$\tilde{T}^{(\itilde_1 \itilde_2)}$ 
and $\tilde{T}^{[\itilde_1 \itilde_2]}$ ($\itilde_1,\itilde_2=1,2$).

The physical two-strand Hilbert space is obtained by projection of $\cH_1\otimes \cH_1$ onto the $S_2$-invariant subspace as\footnote{This $S_2$ projection can be understood as follows.  Recall  \cite{Dijkgraaf:1996xw}  that, if the strand lengths $k$ and $k'$ are different, the two-strand Hilbert space of the symmetric orbifold CFT is $H_{(k)}\otimes H_{(k')}$, while for $k=k'$ it is $(H_{(k)}^{\otimes 2})_{\text{$S_2$-inv}}$, where $H_{(k)}$ is the full (left times right) Hilbert space of a single strand of length $k$, and $S_2$ interchanges the two strands. 
On the other hand, using \eqref{eq:H1},
\begin{align}
(\cH_1^{\otimes 2})_{\text{$S_2$-inv}}=
\Bigl[\,
\bigoplus_{k<k'}
\left(
(H_{(k)}\otimes H_{(k')})\oplus (H_{(k')}\otimes H_{(k)})
\right)_{\text{$S_2$-inv}}
\Bigr]
\oplus
\Bigl[\bigoplus_k (H_{(k)}^{\otimes 2})_{\text{$S_2$-inv}}\Bigr]\ .
\label{eq:H2_ftnt}
\end{align}
Because $((H_{(k)}\otimes H_{(k')})\oplus (H_{(k')}\otimes H_{(k)}))_{\text{$S_2$-inv}}$ with $k\neq k'$ is isomorphic to $H_{(k)}\otimes H_{(k')}$, \eqref{eq:H2_ftnt} is isomorphic to the two-strand Hilbert space $\cH_2$.  More generally, the $n$-strand Hilbert space is isomorphic to $(\cH_1^{\otimes n})_{\text{$S_n$-inv}}$.
\label{ftnt:H1^n/Sn}
}
\begin{align}
    \cH_2&=
    [(\cH_1)^{\otimes 2}]_{\text{$S_2$-inv}}
    =
    \big[(V\otimes \Vt)^{\otimes 2}\big]_{\text{$S_2$-inv}}
    =
    \big[(V_{\ydiagram{2}}\oplus V_{\ydiagram{1,1}})
        \otimes (\Vt_{\ydiagram{2}}\oplus \Vt_{\ydiagram{1,1}})
    \big]_{\text{$S_2$-inv}}
    \notag\\
    &=
    (V_{\ydiagram{2}}\otimes \Vt_{\ydiagram{2}})\oplus 
    (V_{\ydiagram{1,1}}\otimes \Vt_{\ydiagram{1,1}}) \ .
    \label{eq:H2}
\end{align}
In terms of the tensors, this means that the physical wavefunction is given by
\begin{align}
    T^{(i_1 i_2)}\tilde{T}^{(\itilde_1 \itilde_2)}
+    T^{[i_1 i_2]}\tilde{T}^{[\itilde_1 \itilde_2]}\ ,
\end{align}
which is invariant under $S_2$.
This is exactly analogous to the construction of the totally antisymmetric wavefunction of the system of two spin-1/2 particles in quantum mechanics that we mentioned before.

One might wonder about the meaning of the spaces such as $V_{\ydiagram{2}}$ or  $V_{\ydiagram{1,1}}$.  To clarify this, consider the situation where the left-moving part of two strands are in two distinct states 1 and 2, which can even have different strand lengths, while the right-moving part are in two distinct states $\tilde{1}$ and $\tilde{2}$.  If we combine the left- and right-moving parts, the possible 2-strand states are
\begin{align}
\ket*{1\tilde{1}}\ket*{2\tilde{2}},\qquad\ket*{1\tilde{2}}\ket*{2\tilde{1}}.
\label{eq:2-strand_states(1)}
\end{align}
On the other hand, in the above formulation \eqref{eq:H2}, we consider symmetric and antisymmetric combinations of left and right states separately:
\begin{align}
    \Bigl(\ket{1}^{[1]}\ket{2}^{[2]}\pm \ket{2}^{[1]}\ket{1}^{[2]}\Bigr)
    \,
    \Bigl(\ket*{\tilde{1}}^{[1]}\ket*{\tilde{2}}^{[2]}\pm \ket*{\tilde{2}}^{[1]}\ket*{\tilde{1}}^{[2]}\Bigr),
\label{eq:2-strand_states(2)}
\end{align}
where ${[I]}$ denotes the strand number.
If we expand this and combine the left and right parts, we get
\begin{align}
    \Bigl(\ket*{1\tilde{1}}^{[1]}\ket*{2\tilde{2}}^{[2]}
    +\ket*{2\tilde{2}}^{[1]}\ket*{1\tilde{1}}^{[2]}\Bigr)
    \pm\Bigl(\ket*{1\tilde{2}}^{[1]}\ket*{2\tilde{1}}^{[2]}
    +\ket*{2\tilde{1}}^{[1]}\ket*{1\tilde{2}}^{[2]}\Bigr),
\label{eq:2-strand_states(3)}
\end{align}
which has the same content as \eqref{eq:2-strand_states(1)}, with the two terms here corresponding to the two states there (see footnote \ref{ftnt:H1^n/Sn}).
The states in the left- and right-moving sectors that appear in \eqref{eq:2-strand_states(2)} are not separately physical but are formal, intermediate expressions used for constructing the full strand state.  The (anti)symmetry of the one-sided wavefunction in \eqref{eq:2-strand_states(2)} manifests itself in the total wavefunction in \eqref{eq:2-strand_states(3)} as a choice of the linear combination.  For example, if we take $\tilde{1}=\dot{+}$ and $\tilde{2}=\dot{-}$ for the right-moving sector states (see \eqref{eq:V_Vt_bosonic}), the full wavefunction \eqref{eq:2-strand_states(3)} in the form of \eqref{eq:2-strand_states(1)} is
\begin{align}
    \ket*{1\dot{+}}\ket*{2\dot{-}}\pm \ket*{1\dot{-}}\ket*{2\dot{+}} \ ,
\end{align}
which is an $SU(2)$ triplet/singlet state depending on the $\pm$ sign.  So, the symmetry sector dictates  the $SU(2)$ multiplet structure on the right.

\bigskip
We can generalize this to the case with $n$ strands. Here we will state only the essential features necessary for our purposes, relegating the details to Appendix \ref{app:S-W++}\@.
The left- and right-moving Hilbert spaces on $n$ strands, before projection onto $S_n$-invariant states, are the tensor product spaces $V^{\otimes n}$ and $\Vt^{\otimes n}$. The Schur-Weyl duality \cite{Fulton:1991} states that these spaces decompose as
\begin{align}
 V^{\otimes n}\cong\bigoplus_{\lambda\,\vdash n} V_\lambda\otimes M_\lambda\ ,\qquad
 \Vt^{\otimes n}\cong\bigoplus_{\lambdat\,\vdash n} \Vt_{\lambdat}\otimes M_{\lambdat}\ ,
 \label{SchurWeyl_bosonic}
\end{align}
where the sums are over Young diagrams $\lambda,\tilde{\lambda}$, each with $n$ boxes (as denoted by $\lambda,\tilde{\lambda}\vdash n$). 
Here, $V_\lambda\subset V^{\otimes n}$ is the subspace of tensors $T^{i_1\dots i_n}$ with symmetry under exchanging indices dictated by $\lambda$, and spans an irreducible representation of $GL(\infty)$.  More precisely, the Young diagram $\lambda$ alone does not fix the exchange symmetry: we must choose a Young tableau, \textit{i.e.}\ a filling of $\lambda$ with the integers $1,\dots,n$.  For a fixed $\lambda$, there are generally multiple such tableaux that give different exchange symmetries of the indices, and those tableaux can be permuted into each other by the action of $S_n$. 
In this sense, the space of tensors $T^{i_1\dots i_n}$ with symmetry type $\lambda$ spans a representation of $S_n$, which is denoted by $M_\lambda$ in \eqref{SchurWeyl_bosonic}.  
In the $n=2$ case where $\lambda=\YTnormalsize\ydiagram{2}$ and $\ydiagram{1,1}\,$, as we discussed before, $\YTscriptsize M_{\ydiagram{2}}$ is the trivial representation and $M_{\ydiagram{1,1}}$ is the alternating representation, both of which are one-dimensional.  However, for $n\ge 3$, $M_\lambda$ is generically higher dimensional.
To keep the presentation simple,
we will not spell out the explicit constructions of $V_\lambda$ and $M_\lambda$ for general $n$ here.  These details are not essential for the discussion in the main text; we refer the reader to Appendix~\ref{app:S-W++} for a more detailed account and examples.
The right-moving 
$\Vt_{\lambdat}\subset \Vt^{\otimes n}$ and $M_{\lambdat}$ are defined similarly, and span representations of $GL(2)$  and $S_n$, respectively.
Because we cannot antisymmetrize more than two $GL(2)$ indices of $\Vt$, the space $\Vt_{\lambdat}$ vanishes if the Young diagram $\lambdat$ has more than two rows.

The physical $n$-strand Hilbert space $\cH_n$ is the $S_n$-invariant subspace of $V^{\otimes n}\otimes \tilde{V}^{\otimes n}$ (see footnote \ref{ftnt:H1^n/Sn}).  Because the $S_n$ product representation $M_{\lambda}\otimes M_{\tilde{\lambda}}$ contains the one-dimensional trivial representation $M_{\{n\}}$ if and only if $\lambda=\tilde{\lambda}$, we find\footnote{This simple form is because $\tilde{V}$ does not involve summation over the strand number. If it did, we would also have to project \eqref{eq:Hn_LxR} onto the subspace in which the left and right strand numbers are the same.}
\begin{equation}
    \cH_n=\bigoplus_{\lambda\,\vdash n} V_\lambda \otimes \tilde{V}_{\lambda}\ ,
\label{eq:Hn_LxR}
\end{equation}
where for the $S_n$ part it is understood that we take only the $S_n$ invariant part, $(M_\lambda\otimes M_{\lambda})_{\text{$S_n$-inv}}$ (see Appendix \ref{app:S-W++} for detail).
Then the covering (or multi-strand) Hilbert space, summed over all strand numbers, is 
\begin{equation}
    \cH = \bigoplus_{\lambda} V_\lambda \otimes \tilde{V}_{\lambda}\ ,
\label{eq:H_LxR}
\end{equation}
where now the direct sum is over $\lambda$ with arbitrary numbers of boxes.  The Hilbert space for fixed $N$ is obtained by projection onto the subspace with total strand number~$N$.  Eq.~\eqref{eq:H_LxR} is the same multi-strand Hilbert space that we discussed in \eqref{eq.symm_orb_H_cover}. The alternative, ``Schur-Weyl'' form \eqref{eq:H_LxR} of the covering Hilbert space turns out to be useful for constructing the new supersymmetry index in Section~\ref{ssec:REG}.


\subsubsection*{Partition function}

The Schur-Weyl decomposition \eqref{eq:H_LxR} immediately allows us to write down the multi-strand partition function as a sum over different symmetry sectors labelled by Young diagrams as
\begin{align}
\cZ(p,q,\yt)= \sum_{\lambda} S_{\lambda}(p,q)\,  \St_{\lambda}(\yt)\  ,
\label{Z_sum_lambda-1}
\end{align}
where $S_\lambda(p,q)=\tr_{V_\lambda}[p^{\hat{k}}q^{L_0}]$ is the trace of the operator that appeared in \eqref{z_bosonic} now on the space $V_{\lambda}$. Similarly $\St_{\lambda}(\yt)=\tr_{\Vt_\lambda}[y^{2\Jt^3_0}]$ is the $\Vt_\lambda$ version of $\zt(\yt)$ in \eqref{zt_bosonic}.
These are nothing but the Schur functions (see \textit{e.g.}~\cite[\S6]{Fulton:1991}) which are related to 
$z(p,q),\zt(\yt)$ via the use of so-called power sum polynomials, as reviewed in Appendix~\ref{app:symmPoly}\@.
For example,
\begin{align}
\YTscriptsize
    S_{\ydiagram{2}}(p,q)=\frac12\big(z(p,q)^2+z(p^2,q^2)\big)\ ,\quad
    S_{\ydiagram{1,1}}(p,q)=\frac12\big(z(p,q)^2-z(p^2,q^2)\big)\ .
\end{align}

It is useful here to be slightly abstract and generalize the above to the case where the single-strand left-moving partition function \eqref{z_bosonic} is the trace of an arbitrary operator $g\in GL(\infty)$:
\begin{align}
    z(x)=\tr_V[g]=\sum_i x_i\ ,
    \label{z(x)_gen}
\end{align}
where $x_i$ ($i=1,\dots,\dim V$) are the eigenvalues of $g$.  
Eq.~\eqref{z_bosonic} corresponds to the specific choice of $g=p^{\hat{k}}q^{L_0}$.
Mathematically, $z(x)$ is the $GL(\infty)$-character of $V$ on $g$. Likewise, for the right-moving sector, we generalize $\zt(\yt)$ in \eqref{zt_bosonic} to
\begin{align}
    \zt(\xt)=\tr_{\Vt}[\tilde{g}]=\sum_{\itilde} \xt_{\itilde}\ ,
    \label{zt(xt)_gen}
\end{align}
where $\tilde{g}$ is an arbitrary operator and $\xt_{\itilde}$ ($\itilde=1,\dots,\dim \Vt$) are its eigenvalues.\footnote{Although $\dim \Vt=2$ in the current bosonic model example, the present discussion applies to cases with general $\dim \Vt$.} In this case, the multi-strand partition function 
\eqref{Z_sum_lambda-1} is
\begin{align}
    \cZ(x,\xt)=\sum_\lambda S_\lambda(x)\St_\lambda(\xt)\ ,
    \label{Z=sum_S_lambda-S_lambdat}
\end{align}
where $S_\lambda(x)=\tr_{V_\lambda}(g)$ is the $GL(\infty)$-character of $V_\lambda$ on $g$, which is given by a Schur function, and similarly for $\St_\lambda(\xt)$.

The Cauchy identity for Schur functions (see \eqref{eq:Cauchy_id}) says that \eqref{Z=sum_S_lambda-S_lambdat} is in fact equal to
\begin{align}
    \cZ(x,\xt)={1\over\prod_{i,\itilde}(1-x_i \xt_{\itilde})}\ .
    \label{Cauchy(x,xt)}
\end{align}
The indices $i$ and $\itilde$ label the left- and right-moving states of a strand.  One sees from Eq.~\eqref{Cauchy(x,xt)} that $\cZ(x,\xt)$ is nothing but the grand-canonical partition function of strands regarded as independent bosons, whose internal states are labelled by $(i,\itilde)$.
This is essentially the so-called DMVV formula \cite{Dijkgraaf:1996xw} which gives the grand partition of symmetric orbifold CFTs by relating it to the multi-string partition function of second-quantized strings.  We will see this repeatedly in explicit examples below.

To apply the above abstract expressions to the present example, let us write the left-moving single-strand partition function \eqref{z_bosonic} in the form
\begin{align}
z(p,q)=\sum_{k,r}c(k,r)p^k q^r\ ,
\end{align}
where $c(k,r)$ is the number of states with quantum numbers $(k,r)$.
This is an ``unpacked'' form of the condensed expression \eqref{z(x)_gen}, where $g=p^{\hat{k}} q^{L_0}$ and
$c(k,r)$ gives the multiplicity of the eigenvalues $x_i$ taking the same value $p^k q^r$.  The right-moving single-strand partition function is still given by \eqref{zt_bosonic}, giving two eigenvalues $\xt_{\itilde}=\yt,\yt^{-1}$.  In this case, the Cauchy identity
\eqref{Cauchy(x,xt)} gives a DMVV-form \cite{Dijkgraaf:1996xw} of the multi-strand partition function:
\begin{align}
  \cZ(p,q,\yt)
  ={1\over\prod_{k,r}
  [(1- p^k q^r \yt)(1- p^k q^r \yt^{-1})]^{c(k,r)}
  }\ .
\end{align}
The power $c(k,r)$ is because $c(k,r)$ gives the number of different bosons with the same left-moving quantum numbers $(k,r)$.  The bosons also have two possible right-moving states, leading to two factors with $\yt$ and $\yt^{-1}$.


\subsection{Adding fermions} \label{ssec:fermionEx}

The Schur-Weyl formalism discussed above can be straightforwardly extended to the case with both bosonic and fermionic states. This will allow for the direct application to holographically relevant theories such as the one discussed in Section~\ref{ssec:SW_T4}.

We again start with the Hilbert space on a single strand, which is assumed to factorize into left- and right-moving parts.  Let $V_k$ be the left-moving Hilbert space on a strand of length $k$. The left-moving single-strand Hilbert space is $V=\oplus_{k\ge 1} V_k$, which is now assumed to contain infinitely many bosonic as well as fermionic states. If we regard $V$ as the space of vectors $T^I$, where $I=i,i'$ runs over bosonic ($i$) and fermionic ($i'$) states, an operator $g$ in $V$ acts on $T^I$ as a $GL(\infty|\infty)$ matrix $g^I{}_J$; namely, $V$ is a representation space of the fundamental representation $\YTnormalsize\ydiagram{1}$ of $GL(\infty|\infty)$.
On the other hand, the right-moving Hilbert space $\Vt$ is assumed to be $k$-independent and contains $b$ bosonic and $f$ fermionic states.  $\Vt$ is the representation space of the fundamental representation $\ydiagram{1}$ of $GL(b|f)$.  The total single-strand Hilbert space is then $\cH_1=V\otimes \Vt$.

%
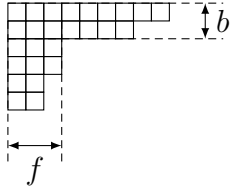
\begin{figure}
\begin{center}
\begin{tikzpicture}[scale=1.3]
\draw (0  ,0) rectangle +(1ex,1ex);
\draw (1ex,0) rectangle +(1ex,1ex);
\draw (2ex,0) rectangle +(1ex,1ex);
\draw (3ex,0) rectangle +(1ex,1ex);
\draw (4ex,0) rectangle +(1ex,1ex);
\draw (5ex,0) rectangle +(1ex,1ex);
\draw (6ex,0) rectangle +(1ex,1ex);
\draw (7ex,0) rectangle +(1ex,1ex);
\draw (8ex,0) rectangle +(1ex,1ex);
\draw (0  ,-1ex) rectangle +(1ex,1ex);
\draw (1ex,-1ex) rectangle +(1ex,1ex);
\draw (2ex,-1ex) rectangle +(1ex,1ex);
\draw (3ex,-1ex) rectangle +(1ex,1ex);
\draw (4ex,-1ex) rectangle +(1ex,1ex);
\draw (5ex,-1ex) rectangle +(1ex,1ex);
\draw (6ex,-1ex) rectangle +(1ex,1ex);

\draw (0,-2ex) rectangle +(1ex,1ex);
\draw (1ex,-2ex) rectangle +(1ex,1ex);
\draw (2ex,-2ex) rectangle +(1ex,1ex);
\draw (0,-3ex) rectangle +(1ex,1ex);
\draw (1ex,-3ex) rectangle +(1ex,1ex);
\draw (2ex,-3ex) rectangle +(1ex,1ex);
\draw (0,-4ex) rectangle +(1ex,1ex);
\draw (1ex,-4ex) rectangle +(1ex,1ex);
\draw (0,-5ex) rectangle +(1ex,1ex);
\draw (1ex,-5ex) rectangle +(1ex,1ex);

\draw[densely dashed] (0,1ex) -- +(12ex,0);
\draw[densely dashed] (0,-1ex) -- +(12ex,0);
\draw[densely dashed] (0,0) --   +(0,-8ex);
\draw[densely dashed] (3ex,0) -- +(0,-8ex);

\draw[latex-latex] (11ex,1ex) -- +(0,-2ex) node [midway,right] {$b$};
\draw[latex-latex] (0,-7ex) -- +(3ex,0) node [midway,below] {$f$};

\end{tikzpicture}
    \caption{The form of a Young diagram $\lambda$ satisfying the $(b|f)$-hook condition in \eqref{eq.bfHook}.}
    \label{fig:(b|f)-hook_cond}
\end{center}
\end{figure}

The construction of the $n$-strand Hilbert space follows just as in the bosonic case, using the super Schur-Weyl duality \cite{469926,BERELE1985225,BERELE1987118} which takes exactly the same form as the bosonic relation used in \eqref{SchurWeyl_bosonic}.  In this relation $V_\lambda$ can again be regarded as the space of tensors $T^{I_1\dots I_n}$ with symmetry dictated by $\lambda$, except that exchanging two fermionic indices yields an extra minus sign. The same is true for the right-moving space, $\Vt_{\lambdat}$.  Imposing $S_n$ invariance (including possible fermionic signs) again enforces the condition $\lambda=\lambdat$, leading to the same expression \eqref{eq:Hn_LxR} for the $n$-strand Hilbert space, $\cH_n$, and summing $\cH_n$ over $n$ gives the covering space $\cH$ just as in \eqref{eq:H_LxR}.  One thing to note, however, is that the right-moving space $\Vt_{\lambda}$ now vanishes unless the Young diagram $\lam$ is of the $(b|f)$-hook type, which we denote by $\lam\in H(b|f)$ (see Figure~\ref{fig:(b|f)-hook_cond}). We define such Young diagrams by the $(b|f)$-hook condition
\begin{align} \label{eq.bfHook}
    \lambda_{b+1}\le f \ .
\end{align}
This condition is due to the fact that one cannot antisymmetrise more than $b$ bosonic indices and symmetrize more than $f$ fermionic indices in the tensor $T^{I_1\dots I_n}$.  Therefore, the multi-strand Hilbert space can be written with this explicit restriction of the tensor sum:
\begin{align}
    \cH = \bigoplus_{\lambda\in H(b|f)}
    V_\lambda\otimes \Vt_{\lambda} \ .
    \label{H=sum_VV_super}
\end{align}

Turning to the partition function, because we have both bosonic and fermionic states, the natural generalization of the left-moving single-strand partition function \eqref{z(x)_gen} is the supertrace ($\bbZ_2$ graded trace)
\begin{align}
    z(x|x')=\tr_V\!\big[(-1)^F g\big]=\sum_i x_i - \sum_{i'} x'_{i'}\ ,
\end{align}
where $g\in GL(\infty|\infty)$ is an arbitrary operator and $F$ is the left-moving fermion number operator.  The indices $i$ and $i'$ run over bosonic and fermionic states respectively in $V$, and $x_i$ and $x'_{i'}$ are the eigenvalues of $g$.   Analogously, the right-moving single-strand partition function, assumed to be $k$-independent, is
\begin{align}
    \zt(\xt|\xt')=\tr_{\Vt}\!\big[(-1)^{\tilde{F}}\tilde{g}\big]=\sum_{\itilde}\xt_{\itilde}- \sum_{\itilde'}\xt'_{\itilde'}\ ,
    \label{ztilde_condensed}
\end{align}
where $\tilde{g}\in GL(b|f)$ is an arbitrary operator, $\tilde{F}$ is the right-moving fermion number operator, and $\xt_{\itilde}$ and $\xt'_{\itilde'}$ are the eigenvalues of $\tilde{g}$.

The super Schur-Weyl decomposition \eqref{H=sum_VV_super} leads to the expression for the multi-strand partition function
\begin{align}
    \cZ(x,\xt|x',\xt') = \sum_{\lambda\in H(b|f)}
    S_\lambda(x|x')\, \St_\lambda(\xt|\xt') \ ,
    \label{Z=sum_SS_super}
\end{align}
where $S_\lambda(x|x')$ and  $\St_\lambda(\xt|\xt')$ are the characters of the irreducible representation subspaces $V_\lambda,\Vt_\lambda$ and are given by super Schur functions (see \eqref{eq:Super_Schur_power_sum}). 

The Cauchy identity for super Schur functions says that the multi-strand partition function \eqref{Z=sum_SS_super} can be written in DMVV form as
\begin{equation} \label{eq:Cauchy_id_general}
    \cZ(x,\xt|x',\xt') = \frac{\prod_{i,\tilde{\imath}'}(1-x_i \tilde{x}'_{\tilde{\imath}'})
    \prod_{i',\tilde{\imath}}(1-x'_{i'}\tilde{x}_{\tilde{\imath}})
    }{
    \prod_{i,\tilde{\imath}}(1-x_i \tilde{x}_{\tilde{\imath}})
    \prod_{i',\tilde{\imath}'}(1-x'_{i'}\tilde{x}'_{\tilde{\imath}'})
    }\ .
\end{equation}
Just as in the bosonic case, this can be interpreted as the grand partition function of strands regarded as independent bosons and fermions.
The strand is bosonic if left- and right-moving internal states are both bosonic, namely, for $(i,\itilde)$ or $(i',\itilde')$.  On the other hand, the strand is fermionic if one side is bosonic and the other side is fermionic, namely, for $(i',\itilde)$ or $(i,\itilde')$.  So, the Schur-Weyl formulation is very well suited for studying the symmetric orbifold CFT, which is nothing but a theory of multiple strings, both bosonic and fermionic.

\subsection{Symmetric orbifold of $T^4$}
\label{ssec:SW_T4}


We now specify to the symmetric product orbifold theory of $T^4$, introduced in Section~\ref{sec:background}, which is the holographic CFT for the D1-D5 system on $\cM=T^4$.
We will first cast its BPS spectrum in the language of the Schur-Weyl formalism  laid out above
and then use this to study supersymmetric indices for this theory.

In order to decompose the Hilbert space of this theory in the Schur-Weyl form \eqref{H=sum_VV_super} and write its grand partition function in the form \eqref{Z=sum_SS_super}, we first have to specify the relevant left- and right-moving single-strand Hilbert spaces $V_k$ and $\Vt$. Working in the Ramond-Ramond sector, these are given by
\begin{equation} \label{eq:Vk_Vt_def}
    V_k = \mathrm{span}\Big\{\cO \ket{\alpha}_k\, ,\ \cO \ket*{\Ad}_k\Big\} _\cO
    \quad,\quad \Vt = \mathrm{span}\Big\{\ket{\alphad}\, ,\ \ket*{\Ad}\Big\} \ ,
\end{equation}
where $\cO$ represents any generators of the holomorphic contracted large $\mathcal{N}=4$ algebra. As discussed in Section~\ref{sec:background}, the spaces of Ramond ground states on strands of any length are all isomorphic and so we do not include the strand-length label for the right-moving Hilbert space $\Vt$. 

In our conventions, the two right-moving ground states $\ket{\alphad}$ are bosonic and the two $\ket*{\Ad}$ are fermionic, and hence $b=f=2$ here. 
Therefore, operators on the right-moving Hilbert space $\Vt$ can be thought of as $GL(b|f)=GL(2|2)$ matrices, and $\Vt$ is a representation space of the fundamental representation $\YTnormalsize\ydiagram{1}$ of $GL(2|2)$.  
We will give explicit representations of the generators of this $GL(2|2)$ in Appendix \ref{ss:gl(2|2)_and_A-alg}.
The right-moving partition function over $\Vt$ is
\begin{equation} \label{eq.gl22_char_def1}
    \tilde{z}(\yt) \equiv \tr_{\Vt} [\tilde{g}] = \yt^{-1}-2+\yt \ ,
\end{equation}
where $\tilde{g} = (-1)^{F_R}\tilde{y}^{2\tilde{J}^3_0}\in GL(2|2)$.
The terms in the above polynomial represent the Ramond ground states $\ket*{\dot{-}}_k$, $\ket*{\Ad}_k$, and $\ket*{\dot{+}}_k$ respectively. This is a specific example of the general expression \eqref{ztilde_condensed}.  As discussed before, mathematically, $\zt(\yt)$ is the $GL(2|2)$-character of the fundamental representation $\ytableausetup{centertableaux}\YTnormalsize\lam=\ydiagram{1}$ evaluated on $\tilde{g}$.
Then the character $\St_\lambda(\yt)$ of the generic irreducible  $G(2|2)$-representation labeled by a Young diagram $\lambda$ is equal to the (super) Schur function constructed using the formula \eqref{eq:Super_Schur_power_sum} with \eqref{eq.gl22_char_def1} as the (super) power sum functions $P_{\alpha}(\yt)=\tilde{z}(\yt^{\alpha})$, as defined in \eqref{eq:def_super_power_sum}.

By way of example, the right-moving Ramond ground states for $N=2$ fall into three irreducible representations labelled by the Young diagrams $\YTnormalsize\lam=\ydiagram{1}, \ydiagram{2},\ydiagram{1,1}$.  Here,
$\lambda=\ydiagram{1}$ corresponds to the twisted sector where there is a single strand of length two, while
$\lambda=\ydiagram{2},\ydiagram{1,1}$ correspond to the untwisted sector where there are two strands of length one.
In the Schur-Weyl formulation, the states in $\lam$-sectors have definite symmetry under the permutation of  strands, and so this $\lam$-sector organisation of states is a different basis from the more conventional, twist-sector basis described in Section \ref{sec:background}. The characters of these $N=2$ representations are given in terms of \eqref{eq.gl22_char_def1} as
\begin{subequations} \label{eq.RightSchurPolysN=2}
\begin{align}\YTscriptsize
    \St_{\ydiagram{1}}(\ty) &= \tilde{z}(\ty) = \yt^{-1} - 2 + \yt \ ,\label{eq.RightSchurPolysN=2_1}\\
    \St_{\ydiagram{2}}(\ty) &= \frac12 \big[\tilde{z}(\ty)^2 + \tilde{z}(\ty^2)\big] = \ty^{-2} -2\ty^{-1} + 2 - 2\ty + \ty^{2}\ ,\label{eq.RightSchurPolysN=2_2}\\
    \St_{\ydiagram{1,1}}(\ty) &= \frac12 \big[\tilde{z}(\ty)^2 - \tilde{z}(\ty^2)\big] = -2\ty^{-1} + 4 -2\ty \ .\label{eq.RightSchurPolysN=2_3}
\end{align}
\end{subequations}
The terms in \eqref{eq.RightSchurPolysN=2_1} represent the standard twisted-sector ground states, while those in \eqref{eq.RightSchurPolysN=2_2} represent the strand-symmetric Ramond ground states\footnote{Here, for example, $\ket*{\dot{-}}_1\ket*{\dot{A}}_1 + \ket*{\dot{A}}_1\ket*{\dot{-}}_1$ in the second line means $\ket*{\dot{-}}_1^{[1]}\ket*{\dot{A}}_1^{[2]} + \ket*{\dot{A}}_1^{[1]}\ket*{\dot{-}}_1^{[2]}$ where the label $[I]$ denotes the strand number.  The permutation $S_2$ acts on the strand labels; see \eqref{eq:Sn_Vn_action_strand}.
}
\begin{equation} \label{eq.Rgs_N=2_symm}
    \begin{aligned}
    +\ty^{-2} &:\quad \ket*{\dot{-}}_1\ket*{\dot{-}}_1\ ,\\
    -2\ty^{-1} &:\quad \ket*{\dot{-}}_1\ket*{\dot{A}}_1 + \ket*{\dot{A}}_1\ket*{\dot{-}}_1 \ ,\\
    +2&:\quad \ket*{\dot{-}}_1\ket*{\dot{+}}_1 + \ket*{\dot{+}}_1\ket*{\dot{-}}_1 \ , \ \ \ep_{\Ad\Bd}\big(\ket*{\dot{A}}_1\ket*{\dot{B}}_1 - \ket*{\dot{B}}_1\ket*{\dot{A}}_1\big)  \ ,\\
    -2\ty &:\quad \ket*{\dot{+}}_1\ket*{\dot{A}}_1 + \ket*{\dot{A}}_1\ket*{\dot{+}}_1 \ ,\\
    +\ty^2 &: \quad \ket*{\dot{+}}_1\ket*{\dot{+}}_1\ ,
\end{aligned}
\end{equation}
and those in \eqref{eq.RightSchurPolysN=2_3} represent the strand-antisymmetric Ramond ground states
\begin{equation} \label{eq.Rgs_N=2_asymm}
    \begin{aligned}
    -2\ty^{-1} &:\quad \ket*{\dot{-}}_1\ket*{\dot{A}}_1 - \ket*{\dot{A}}_1\ket*{\dot{-}}_1 \ ,\\
    +4&:\quad \ket*{\dot{-}}_1\ket*{\dot{+}}_1 - \ket*{\dot{+}}_1\ket*{\dot{-}}_1 \ , \ \ \sigma^{a}_{\Ad\Bd}\big(\ket*{\dot{A}}_1\ket*{\dot{B}}_1 + \ket*{\dot{B}}_1\ket*{\dot{A}}_1\big)  \ ,\\
    -2\ty &:\quad \ket*{\dot{+}}_1\ket*{\dot{A}}_1 - \ket*{\dot{A}}_1\ket*{\dot{+}}_1 \ .
\end{aligned}
\end{equation}
See Appendix \ref{app:conventions} for the definition of $\epsilon$ and $\sigma^a$ matrices, which respectively span the bases of antisymmetric and symmetric matrices.
While for these simple cases the explicit form of states with a particular strand symmetry is intuitive, for a general symmetry sector labelled by $\lam$ the states captured by the character $\St_{\lam}(\ty)$ are \textit{a priori} less obvious. The technique for finding the explicit form of such states is that of Young symmetrizers, which we discuss in Appendix~\ref{app:S-W++}.

Returning to the general case, for the left-moving covering space of single-strand Hilbert spaces, $V=\bigoplus_{k}V_k$, the single-strand partition function (or characters, defined with respect to the operator $g=(-1)^{F_L}p^{\hat{k}}q^{L_0-\frac{c}{24}}y^{2J^3_0}$) is given by 
\begin{equation} \label{eq:1-strand_z(p,q,y)}
    z_1(p,q,y) \equiv \tr_{V} [g] = \sum_{k,r,\ell} c(k,r,\ell)\, p^k q^r y^{\ell} \ ,
\end{equation}
where in symmetric orbifold CFTs the expansion coefficients $c(k,r,\ell)$ are related to those for the seed theory, $c(r,\ell)$, via \cite{Dijkgraaf:1996xw}
\begin{align}
c(k,r,\ell)=c(kr,\ell)\ .
\label{eq:c(k,r,l)=c(kr,l)}
\end{align}
In the present case of the $T^4$ theory, the seed-theory left-moving  (signed) partition function is given by
\begin{align}
    z_{\mathrm{seed}}(q,y) &\equiv 
    (y+y^{-1}-2)
    \prod_{r=1}^\infty {(1-yq^{r})^2(1-y^{-1}q^{r})^2\over (1-q^r)^4}
    \notag\\
    &=
    -\left(\frac{\vartheta_1(\nu,\tau)}{\eta(\tau)^3}\right)^{2} 
    = \sum_{r,\ell} c(r,\ell)\,q^r y^{\ell} \ ,
    \label{eq:seed_z(q,y)}
\end{align}
where $y=e^{2\pi i\nu}$, $q=e^{2\pi i\tau}$ and the definitions for the Jacobi theta and Dedekind eta functions are given in \eqref{eq.theta1etaDef}. 
The coefficients are actually a function of only one variable \cite{Maldacena:1999bp},
\begin{align}
    c(r,\ell)=c(4r-\ell^2),
    \label{eq:c(r,l)=c(4r-l^2)}
\end{align}
with the first few nonzero values given by $c(-1)=1$, $c(0)=-2$, $c(3)=8$, $c(4)=-12$.
From \eqref{eq:1-strand_z(p,q,y)} a general $n$-strand character $S_{\lam}(p,q,y)$ for an irreducible representation of $GL(\infty|\infty)$ labelled by $\lam\vdash n$ can be defined analogously to the right-moving case discussed above.  

As we discussed generally, by the Schur-Weyl duality, the multi-strand Hilbert space of the $T^4$ theory can be decomposed into $\lambda$-sectors as in \eqref{H=sum_VV_super}.  The Young diagram $\lambda\vdash n$ simultaneously labels the $S_n$ representation under permutation of strands, the $GL(\infty|\infty)$ representation of the left-moving sector, and the $GL(2|2)$ representation of the right-moving sector.
As discussed in \eqref{eq.bfHook}, $\lambda$ is restricted to be of the $(2|2)$-hook type, satisfying $\lambda_3\le 2$.
The grand partition function, \textit{i.e.}\ the character for the multi-strand Hilbert space \eqref{H=sum_VV_super}, is given in the Schur-Weyl form \eqref{Z=sum_SS_super}. Specifically, in this case it is given by
\begin{equation} \label{eq:Z=sum_SS_T4}
    \cZ(p,q,y,\ty) \equiv \sum_{N=1}^{\infty} p^N Z_N(q,y,\ty) = \sum_{\lam\in H(2|2)} S_{\lam}(p,q,y) \St_{\lam}(\ty) \ .
\end{equation}

We can also apply the Cauchy identity \eqref{eq:Cauchy_id_general}
to the Schur-Weyl form of the grand partition function, \eqref{eq:Z=sum_SS_T4}.
In this case $x_i, x'_{i'}$ are respectively the bosonic and fermionic (with respect to $F_L$) eigenvalues of the operator $g=(-1)^{F_L}p^{\hat{k}}q^{L_0-\frac{c}{24}}y^{2J^3_0}$ and $\tilde{x}_{\itilde},\tilde{x}'_{\itilde'}$ are the bosonic and fermionic (with respect to $F_R$) eigenvalues of $\tilde{g}=(-1)^{F_R}\ty^{2\tilde{J}^3_0}$. This immediately yields a DMVV form \cite{Dijkgraaf:1996xw} of the grand partition function,
\begin{equation}
    \cZ(p,q,y,\ty) = \prod_{k,r,\ell}\left[{(1-p^k q^r y^{\ell})^2 \over (1-p^k q^r y^{\ell}\yt)(1-p^k q^r y^{\ell} \yt^{-1})}\right]^{c(4kr-\ell^2)} \ ,
    \label{eq:Z_T4_DMVV-form}
\end{equation}
where we used \eqref{eq:c(k,r,l)=c(kr,l)} and \eqref{eq:c(r,l)=c(4r-l^2)}.  Note that the original DMVV formula \cite{Dijkgraaf:1996xw} is not just the statement that the generating function of symmetric orbifold CFTs can be written in a product form with exponents determined by the coefficients $c(k,r,\ell)$.  It also includes the fact that the multiply-wound strand coefficients $c(k,r,\ell)$ can be written in terms of the singly-wound (seed theory) ones $c(r,\ell)$, as in \eqref{eq:c(k,r,l)=c(kr,l)}.  In the Schur-Weyl formulation, the first statement follows immediately from the Cauchy identity.

\subsection{The modified elliptic genus}
\label{ssec:T4_MEG}

The supersymmetric index for the D1-D5 CFT on $T^4$ with central charge $c=6N$ is the modified elliptic genus (MEG) \cite{Maldacena:1999bp}, defined by the trace over the Ramond-Ramond sector Hilbert space
\begin{equation}
    \cE_N(q,y) \equiv \frac12 \Tr_{\cH_{\mathrm{CFT}}^{\mathrm{R}}} \Big[(-1)^{F} q^{L_0-{c\over 24}} y^{2J^3_0} \tq^{\tilde{L}_0-{c\over 24}} \big(2\tilde{J}^3_0\big)^2\Big] \ ,
    \label{eq:MEG_T^4}
\end{equation}
where $(-1)^F = (-1)^{2(J^3_0-\tilde{J}^3_0)}$ provides a $\mathbb{Z}_2$ grading by spin statistics. This quantity, in fact, only receives contributions from BPS states with right-moving part being a Ramond ground state ($\tilde{h}=\frac{c}{24}$), and so the $\tq$-dependence drops out in the final expression.

The ordinary elliptic genus, defined as the similar trace $\Tr_{\cH_{\mathrm{CFT}}^{\mathrm{R}}} [(-1)^{F} q^{L_0-{c\over 24}} y^{2J^3_0} \tq^{\tilde{L}_0-{c\over 24}}]$, vanishes identically in the $T^4$ theory. This is due to the presence of the right-moving fermion zero modes $\psit^{\alphad \Ad}_0$ which generate an $SO(4)$ Clifford algebra,\footnote{From the hermiticity \eqref{eq:hermiticity} it follows that this is an $SO(4)$ Clifford algebra.}
\begin{align}
    \big\{\psit_0^{\alphad \Ad},\psit_0^{\betad \Bd}\big\}=-N\epsilon^{\alphad\betad}\epsilon^{\Ad\Bd}\ .
    \label{eq:cl_4^psit}
\end{align}
We will refer to this algebra as $cl_4^{\psit}$. The zero modes
$\psit^{\alphad \Ad}_0$ are total modes in the sense defined in \eqref{eq.total_mode_def_strand}, as reflected in the factor $N$ on the right-hand side.
This algebra implies that the contribution to the elliptic genus from any left-moving state comes with a right-moving ``$\psit$-quartet'', generated by the action of the zero modes $\psit^{\alphad \Ad}_0$.  As a result, the total contribution from the quartet of states vanishes. 

The MEG, meanwhile, is defined such that these $\psit$-quartets built on a state with quantum numbers $h,\tilde{h}=\frac{c}{24},m,\mt$, (the eigenvalues of $L_0,\tilde{L}_0,J^3_0,\Jt^3_0$) contribute to the trace as
\begin{align} \label{eq:MEGcontributionsShort}
    \cE_{N}(q,y) &\supset \frac12(-1)^{2m-2\mt}q^{h-{c\over 24}} y^{2m} \Big[(2\mt)^2 -2(2\mt+1)^2 + (2\mt+2)^2\Big] \nonumber\\
    &= (-1)^{2m-2\mt}q^{h-\frac{c}{24}} y^{2m} \ .
\end{align}
This modification to the elliptic genus therefore removes the right-moving degeneracy coming from the $\psit$-quartets present in the $T^4$ theory. The only other fermionic zero modes in the symmetry algebra of this theory are the right-moving supercharge zero modes $\tilde{G}^{\alphad A}_0$.  BPS states are annihilated by these and therefore contribute to the MEG as in \eqref{eq:MEGcontributionsShort}. Non-BPS states, on the other hand, sit in long supersymmetry multiplets, and so are not annihilated by half of the zero modes $\tilde{G}^{\alphad A}_0$. The contribution to the MEG of a long multiplet is equivalent to that of four short multiplets \eqref{eq:MEGcontributionsShort} with quantum numbers related by the action of $\tilde{G}^{\alphad A}_0$; \textit{i.e.},
\begin{equation} \label{eq:MEGcontributionsLong}
    \cE_{N}(q,y) \supset (-1)^{2m}q^{h-\frac{c}{24}} y^{2m} \Big[ (-1)^{-2\mt} +2 (-1)^{-2\mt+1} + (-1)^{-2\mt+2}\Big] = 0 \ .
\end{equation}
In other words, non-BPS states do not contribute to the MEG.

Although the 1/4-BPS spectrum of the D1-D5 CFT is not invariant under turning on couplings for the 20 exactly marginal operators of this theory, the MEG is protected. Thus, we can compute \eqref{eq:MEG_T^4} at the free point of the moduli space of the D1-D5 CFT and, despite its BPS spectrum being massively enhanced relative to a generic point, the index captures only this generic subspace of states. The lifting of the symmetric orbifold's BPS spectrum is a much studied phenomenon, yet many of its general features remain mysterious. We will discuss the representation theory of this process further in Section~\ref{ssec:Lifting} and so for now we simply state that this lifting mechanism proceeds by the joining of four free-theory short multiplets, by the deformed right-moving supercharges, into a long multiplet of the deformed theory which contributes zero to the MEG as in \eqref{eq:MEGcontributionsLong}.

As an alternative to the definition in \eqref{eq:MEG_T^4}, the MEG can also be written as the differential operator ${\mathcal D}[\,\cdot\,] \equiv \frac12 (\tilde{y}\partial_{\tilde y})^2[\,\cdot\,]|_{\ty=1}$ acting on the signed partition function
\begin{equation} \label{eq.ZRCFT}
    Z_{N}(q,y,\tq,\ty) \equiv \Tr_{\cH^{\mathrm{CFT}}_R}\!\Big[(-1)^{F} q^{L_0-\frac{c}{24}} y^{2J^3_0} \tq^{\tilde{L}_0-\frac{c}{24}} \ty^{2\tilde{J}^3_0} \Big]\ ,
\end{equation}
namely as,
\begin{align}
    \cE_N(q,y)=\cD Z_N(q,y,\qt,\yt)\ .
    \label{cE_N=cD Z_N}
\end{align}
If we apply this to the DMVV form of the grand partition function, \eqref{eq:Z_T4_DMVV-form}, we find
a more standard form \cite[Eq.~(5.6)]{Maldacena:1999bp} of the MEG:
\begin{align} \label{eq.MMS_generating}
    \cE(p,q,y) = \cD \cZ(p,q,y,\ty) = \sum_{k,r,\ell} \frac{c(k,r,\ell)\,p^k q^r y^{\ell}}{(1-p^k q^r y^{\ell})^2} \ .
\end{align}
This can be most easily derived by instead considering $\log \cZ$ and noting that $\cD\log\cZ = \cD\cZ$ due to the fact that $\cZ|_{\ty=1}=1$ and $\ty\pd_{\ty}\log\cZ|_{\ty=1}=0$. 

Expanding \eqref{eq.MMS_generating} yields \cite[Eq.~(5.8)]{Maldacena:1999bp}
\begin{equation} \label{eq:MEG_id_strand_form}
    \cE(p,q,y) = \sum_{s,k,r,\ell} s\big(p^kq^{r}y^{\ell}\big)^s c(k,r,\ell) \ .
\end{equation}
This form of the MEG, along with the discussion of \cite[Sec.~6]{Maldacena:1999bp}, implies that the only contributing states are those consisting of $s$ strands, all in the same state: they not only have the same quantum numbers $(k,r,\ell)$, but are in exactly the same internal state. In other words, only ``identical-strand states'' contribute to the MEG\@. We will return to this intuition in Section~\ref{ssec:Aspects_of_REG}, where a different picture will be proposed.

\subsubsection*{A Schur-Weyl expression of the MEG}

Using the expression \eqref{cE_N=cD Z_N} for the MEG, we can connect it with the Schur-Weyl form of the grand partition function \eqref{eq:Z=sum_SS_T4} and derive a ``Schur-Weyl'' expression of the MEG:
\begin{equation}
     \cE(p,q,y) \equiv \sum_{N=1}^{\infty} p^N\cE_N(q,y) = \sum_{\lam\in H(2|2)} S_{\lam}(p,q,y)\, \cD\St_{\lam}(\ty) \ .
     \label{eq:MEG_as_H(2|2)_sum}
\end{equation}
This expression shows how the MEG can be decomposed into contributions from different sectors of states labeled by $\lambda$.  Recall that $\lambda$ simultaneously labels $S_n$, $GL(\infty|\infty)$ and $GL(2|2)$ representations.

\begin{figure}[tb]
\begin{center}
\begin{tikzpicture}[scale=1.25]
\draw (0,0) rectangle +(0.2,-0.2);
\draw (0.2,0) rectangle +(0.2,-0.2);
\draw (0.4,0) rectangle +(0.2,-0.2);
\draw (0.6,0) rectangle +(0.2,-0.2);
\draw (0.8,0) rectangle +(0.2,-0.2);
\draw (1.0,0) rectangle +(0.2,-0.2);
\draw (1.2,0) rectangle +(0.2,-0.2);
\draw (0,-0.2) rectangle +(0.2,-0.2);
\draw (0,-0.4) rectangle +(0.2,-0.2);
\draw (0,-0.6) rectangle +(0.2,-0.2);
\draw[decorate,decoration={brace,mirror,amplitude=6pt}]
       (-0.05,0) -- (-0.05,-0.8)
       node[midway, left=5pt]{$\rho_\lambda$};
\draw[decorate,decoration={brace,amplitude=6pt}]
       (0,0.05) -- (1.4,0.05)
       node[midway, above=5pt]{$n_\lambda-\rho_\lambda+1$};
\end{tikzpicture}
\caption{\raggedright\sl A single-hook Young diagram $\lambda\in H(1|1)$ is uniquely defined from its number of boxes $n_\lambda$ and number of rows $\rho_\lambda$.
\label{fig:hook_diag}}
\end{center}
\end{figure}
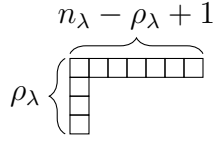%

Interestingly, while a right-moving character $\St_\lambda(\ty)$ is non-vanishing if and only if $\lambda_3\le 2$, its contribution to the MEG, $\cD \St_\lambda$, is non-vanishing if and only if $\lambda_2\le 1$, namely when $\lambda$ is a single-hook Young diagram, $\lambda\in H(1|1)$ (see Figure~\ref{fig:hook_diag}). A character labelled by a single hook Young diagram $\lambda$ contributes
\begin{equation} \label{eq:GL22MEGcontribution}
    \cD \St_\lambda = (-1)^{\rho_\lambda-1}n_\lambda\ ,
\end{equation}
where $\rho_\lambda$ is the number of rows and $n_\lambda$ is the number of boxes in $\lambda$ (see Table~\ref{tab:1} for the relevant diagrams to the case of $N=4$). We derive this result in Appendix \ref{app:GL22MEG} (for a slightly more general $GL(2|2)$ character which includes an additional fugacity $\te$). 
Therefore, we arrive at an alternative expression for the MEG:
\begin{equation}
   \cE(p,q,y)=\sum_{\lambda\in H(1|1)} S_\lambda(p,q,y) 
    \,\, (-1)^{\rho_\lambda-1} n_\lambda\ .
    \label{eq:MEG_as_hook_sum}
\end{equation}
The first terms in this expansion are
\begin{align} \label{eq:MEG_exp}
    \YTscriptsize
    \cE &= S_{\ydiagram{1}} +2(S_{\ydiagram{2}} -S_{\ydiagram{1,1}}) +3(S_{\ydiagram{3}} - S_{\ydiagram{2,1}} + S_{\ydiagram{1,1,1}}) \notag\\
    &\qquad + 4(S_{\ydiagram{4}} - S_{\ydiagram{3,1}} + S_{\ydiagram{2,1,1}} - S_{\ydiagram{1,1,1,1}}) +\cdots.
\end{align}

An advantage of the Schur-Weyl form of the partition function \eqref{eq:Z=sum_SS_T4} and the MEG \eqref{eq:MEG_as_hook_sum} is that 
 it makes it manifest
how the relatively simple right-moving Ramond ground states constrain the representations of the much more complicated left-moving states which can contain arbitrary excitations. In the MEG \eqref{eq:MEG_as_hook_sum}, not only is the set of contributing representations much reduced -- with sectors labelled by non-single-hook diagrams such as $\YTnormalsize\ydiagram{2,2}$ simply not contributing -- but it also makes the cancellations between sectors more transparent \eqref{eq:MEG_exp}. We will return in Section~\ref{ssec:Aspects_of_REG} to comment on the precise form of these cancellations between $\lam$-sectors.

\begin{table}[htb]
\begin{adjustbox}{center,scale=1}
    \def\arraystretch{1.25}
    \begin{NiceTabular}{|c|cccc|}
        \hline\CYon\YTnormalsize
        &~$n_{\lambda}=4$~&~$n_{\lambda}=3$~&~$n_{\lambda}=2$~&~$n_{\lambda}=1$~~\\
        \hline
        ~$\rho_{\lambda}=1$~~&\ydiagram{4}&\ydiagram{3}&\ydiagram{2}&\ydiagram{1}\\
         ~$\rho_{\lambda}=2$~~&\hspace{-6pt}\ydiagram{3,1}&\hspace{-6pt}\ydiagram{2,1}&\hspace{-6pt}\ydiagram{1,1}\\[0.8ex]
        ~$\rho_{\lambda}=3$~~&\hspace{-13pt}\ydiagram{2,1,1}&\hspace{-12.5pt}\ydiagram{1,1,1}\\[1.2ex]
        ~$\rho_{\lambda}=4$~~&\hspace{-19pt}\ydiagram{1,1,1,1}\\[1.7ex]
        \hline
    \end{NiceTabular}
\end{adjustbox}
\caption{\raggedright\sl Organization of single-hook Young diagrams $\lam\in H(1|1)$, by number of rows $\rho_{\lam}$ and number of boxes $n_{\lam}$, relevant to the MEG for $N=4$.\label{tab:1}}
\end{table}

As a side remark, note that, if we use the explicit form of the Schur functions 
\eqref{Schur_poly_example}, Eq.~\eqref{eq:MEG_exp} gives\footnote{Strictly speaking, we need to use the super version of these Schur functions~\eqref{eq:Super_Schur_power_sum}, though this does not change the essence of the current remark.}
\begin{align}
    \YTscriptsize
    \cE &= \sum_i x_i + 2\sum_{i} x_i^2 + 3\sum_{i} x_i^3 + \cdots
    =\sum_n n \sum_i x_i^n.
    \label{eq:MEG_id_strand_abstract}
\end{align}
Recall that, in the abstract formulation in \eqref{z(x)_gen},
$x_i$ are the eigenvalues of an arbitrary operator $g$ acting on single-strand states. 
So, $\sum_i x_i$ is the trace of $g$ on the single-strand Hilbert space, while $\sum_i x_i^n$ can be interpreted as the trace of the same operator $g$ on the $n$-strand Hilbert space, with the restriction that all strands are in exactly the same state (the factor $n$ comes from the right-moving sector). Therefore, \eqref{eq:MEG_id_strand_abstract} can be interpreted as another manifestation of the ``identical-strand'' nature of the MEG\@. This reduces to
\eqref{eq:MEG_id_strand_form} if we take the arbitrary operator to be $g=(-1)^{F_L}p^{\hat{k}}q^{L_0-{c\over 24}}y^{2J^3_0}$.

\section{The structure of lifting}
\label{ssec:Lifting}


In the previous section, we reviewed the standard supersymmetry index of the D1-D5 CFT for $T^4$ -- the modified elliptic genus (MEG) \eqref{eq:MEG_T^4} -- which receives contributions only from BPS states.
From the perspective of the free symmetric-orbifold theory, the contribution vanishes for non-BPS states, and also for a set of BPS states that combine and lift once the theory is deformed away from the free point. 
With the Schur-Weyl decomposition of the MEG, in the form \eqref{eq:MEG_as_H(2|2)_sum} or \eqref{eq:MEG_as_hook_sum}, we now have a detailed picture of the sectors of free-theory states that contribute to the index.
It is therefore interesting to investigate how the representation theory involved in the lifting process fits together with this sector decomposition of the MEG\@.   In particular, if there is a superselection rule that forbids interaction between certain sectors, one can define a new protected index by summing the MEG separately over each resulting superselection sector.

\subsection{Lifted quartets}
\label{sss:struct_lifting}

Let us start by reviewing how free-theory BPS states can interact and lift once a deformation is turned on.
As previously mentioned, the BPS spectrum of the free $\mathrm{Sym}^N(T^4)$ theory consists of states in right-moving Ramond ground states, having $\tilde{h}=\frac{N}{4}$. The overall state can be either 1/2-BPS or 1/4-BPS, depending on the number of left-moving supersymmetries it preserves. Let us denote free-theory BPS states by $\ket{\phi_\mu}$, satisfying the shortening condition
\begin{equation} \label{eq:free_theory_short}
    \tilde{G}^{\alphad A}_{0}\ket*{\phi_\mu}=0 \ ,
\end{equation}
where $\tilde{G}^{\alphad A}_{0}$ are right-moving supercharges in the free theory. While the 1/2-BPS spectrum is protected under deformations away from the free orbifold locus of moduli space, the 1/4-BPS spectrum is not: states that satisfy the condition \eqref{eq:free_theory_short} with respect to the free theory supersymmetry generators do not necessarily satisfy it in the deformed theory. The deformed right-moving supercharges can thus generate a long multiplet from four appropriate free-theory short multiplets. Such free-theory BPS states gain anomalous contributions to their conformal dimensions starting at second order in the coupling: \textit{i.e.}, $(h,\frac{N}{4}) \to (h+\frac{1}{2}E^{(2)},\frac{N}{4}+\frac{1}{2}E^{(2)})$, where the left- and right-moving anomalous dimensions are equal since the spin $h-\tilde{h} \in \mathbb{Z}$ cannot vary continuously.
At the level of explicit states, this process is complicated due to the large degeneracies in the free theory and the fact that both the supercharges and the states get deformed (see \cite{Hampton:2018ygz,Gava:2002xb,Gaberdiel:2023lco} for approaches to this in conformal perturbation theory).

In \cite{Gava:2002xb,Guo:2019pzk}, it was shown that, from the perspective of the free theory's BPS spectrum, the so-called Gava-Narain operator, which we denote by $\tilde{\cG}^{\alphad A}$, is responsible for generating lifted long multiplets (see Appendix \ref{app:GN_op} for some details of the Gava-Narain operator). 
Let us focus on free-theory BPS states $\{\ket{\phi_\mu}\}$ with conformal dimensions $(h,\frac{N}{4})$ for some given $h$.\footnote{In the Ramond sector of the free theory, $h\in\bbZ_{\ge 0}+{N\over 4}$.}  Then the anti-commutator of $\cGt^{\alphad A}$ computes the second-order lifting matrix $E^{(2)}_{\mu\nu}$ via \cite{Guo:2019ady, Guo:private}
\begin{equation}
    2\,\big\langle \phi_{\mu} \big| \big\{ \tilde{\cG}^{\alphad A}, \tilde{\cG}^{\betad B} \big\} \big|\phi_{\nu}\big> = -\epsilon^{\alphad\betad}\epsilon^{AB} E^{(2)}_{\mu\nu} \ .
    \label{eq:anom_dim}
\end{equation}
When diagonalized, $E^{(2)}_{\mu\nu}$ gives the anomalous dimension $E^{(2)}_\mu$ of the state $\ket{\phi_\mu}$, which can be written as\footnote{See \eqref{eq:GN_hermiticity} for the hermiticity of the Gava-Narain operator.}
\begin{align} \label{eq:anom_dim_diag}
    E^{(2)}_\mu 
    =    \mel{\phi_\mu}{\hat{E}^{(2)}}{\phi_\mu} \ ,\quad
\hat{E}^{(2)}
\equiv 
       2\big\{\cGt^{\dot{+}1},(\cGt^{\dot{+}1})^\dagger\big\}
    =   2\big\{\cGt^{\dot{+}2},(\cGt^{\dot{+}2})^\dagger\big\} \ .
\end{align}

If $E^{(2)}_\mu=0$, the state $\ket{\phi_\mu}$ is annihilated by the Gava-Narain operator and remains BPS in the deformed theory to second order in the coupling.
On the other hand, if $E^{(2)}_\mu>0$, the algebra \eqref{eq:anom_dim} satisfied by $\cGt^{\alphad A}$ is a Clifford algebra on the free BPS Hilbert space and $\cGt^{\alphad A}$ generate a lifted quartet (``$\cGt$-quartet''), whose members can be written as
\begin{equation} \label{eq:GN_quartet}
    \ket{\phi_\mu} \quad,\quad \tilde{\cG}^{\dot{+} A} \ket{\phi_\mu} \quad,\quad\tilde{\cG}^{\dot{+} 1} \tilde{\cG}^{\dot{+} 2} \ket{\phi_\mu} \ ,
\end{equation}
where $\ket{\phi_\mu}$ is annihilated by $\cGt^{\dot{-}A}$.
We note that the second-order anomalous dimensions are non-negative since the lift, for example from the diagonalised form \eqref{eq:anom_dim_diag}, can be written as a sum of manifestly non-negative quantities as
\begin{equation}
    E^{(2)}_{\mu} = 2\big| \tilde{\mathcal{G}}^{\dot{+}1}\ket{\phi_{\mu}}\!\big|^2 + 2\big| \tilde{\mathcal{G}}^{\dot{-}2}\ket{\phi_{\mu}}\!\big|^2 \ ,
\end{equation}
with the second term vanishing if $\ket{\phi_{\mu}}$ is the bottom member of a lifted quartet.
Each state in \eqref{eq:GN_quartet} has the same second-order anomalous dimension $E^{(2)}_\mu$ because the Gava-Narain operator commutes with the lifting operator $\hat{E}^{(2)}$ (see \eqref{eq:[cGt,E(2)]=0}),
\begin{equation}
    \big[\cGt^{\alphad A},\hat{E}^{(2)}\big]=0 \ . 
\end{equation}

While the precise form of the Gava-Narain operator is not critical to our discussion, we give its form in order to point out some helpful features.
The Gava-Narain operator acts on free-theory states as
\begin{equation} \label{eq:GN_action}
    \tilde{\cG}^{\alphad A} \equiv \pi \cP G^{+A}_{-\frac12} \sigma_2^{-\alphad}\cP \ ,
\end{equation}
where $\cP$ is a projection operator onto the space of states with free conformal dimensions $(h,\frac{N}{4})$. Due to the presence of the twist operator $\sigma_2$, in spite of the tilde, the Gava-Narain operator acts both on the left and the right, as well as mapping between twist sectors (see for example, \cite{Carson:2014ena,Guo:2022and} for details of how the $\sigma_2$ operator acts on states).   Also, its $A$ index is left-moving in the sense that it comes from the left-moving mode $G^{+A}_{-\half}$.

\subsection{Symmetries of the free theory}
\label{ss:sym_free_thy}

We saw above that the Gava-Narain operator $\cGt$ connects free-theory BPS states that lift under deformation.  One might then naively wonder if $\cGt$ actually connects entire $\lambda$-sectors in the Schur-Weyl decomposition \eqref{eq:MEG_as_H(2|2)_sum} to each other, grouping them into quartets of $\lambda$-sectors that lift as a whole.  This might seem plausible since the boxes of the Young diagram $\lambda$ correspond to strands, and different $\lambda$-sectors can be connected due to the twist operator $\sigma_2$ in \eqref{eq:GN_action} which changes the number of strands.  However, the situation is not as simple as that for multiple reasons.
Firstly, because $\sigma_2$ both splits and joins strands, acting with $\cGt$ on an $n$-strand state generally produces a linear combination of $(n+1)$- and $(n-1)$-strand states.  A $\cGt$-quartet therefore cannot generally be identified with a quartet of $\lambda$-sectors.
Secondly, although a $\lambda$-sector furnishes a representation of $GL(2|2)$, or its algebra $gl(2|2)$, the actual symmetry of the interacting theory is smaller (as we will discuss below). Therefore, states within a $\lambda$-sector need not behave uniformly under deformation; in particular, they need not have the same lifting.
In fact, we will find below that states lift or remain BPS at the level of the multiplets of a subalgebra of $gl(2|2)$, which we call the ``$\scrA$-algebra''.
Namely, each $\scrA$-multiplet either lifts as a whole or remains BPS as a whole.

To better understand the situation, let us briefly recall the symmetry structure of the free theory.  First of all, we have the contracted large $\cN=(4,4)$ superconformal algebra, both left- and right-moving, which is defined everywhere on the D1-D5 CFT's moduli space.  
In addition, the free theory has extra $su(2)$ algebras (see \eqref{eq:SO(4)_E}) whose representations can be used to further organise states.\footnote{We are working in the sector of zero momentum and winding on the $T^4$, on which the free boson zero modes vanish: $\alpha^{\Ad A}_0=\alphat^{\Ad A}_0=0$.}   
Due to the left- and right-moving modes commuting, the actions of these $su(2)$'s split accordingly in the free theory:\footnote{Such separate left/right actions of these $su(2)$ have previously been used in, for example, \cite{Gaberdiel:2025smz,Chang:2025wgo}.}
\begin{align}
    \underbrace{(su(2)_1\oplus su(2)_2)}_{\rm left}
    \oplus
    \underbrace{(\widetilde{su}(2)_1\oplus \widetilde{su}(2)_2)}_{\rm right}~.
\end{align}
In Table \ref{tbl:su(2)s}, we list the currents and doublet indices associated with each $su(2)$ algebra.  The left-moving free fields $\alpha^{\Ad A},\psi^{\alpha \Ad}$ transform under left-moving $su(2)$'s as doublets according to their left-moving doublet indices, while the right-moving fields $\tilde{\alpha}^{\Ad A},\psit^{\alphad \Ad}$ transform under right-moving $su(2)$'s according to their right-moving indices.
Note that we are using the same letters $A,\Ad$ for left and right doublet indices,\footnote{In other words, these $A,\Ad$ are indices of the diagonal left times right $su(2)$'s.}
to avoid cluttering the notation.
The free-theory supercurrent modes $G_n^{\alpha A}$ and $\tilde{G}_n^{\alphad A}$ are also charged under $su(2)_1$ and $\widetilde{su}(2)_1$, respectively.

 
\begin{table}[tb]
\centering
\begin{math}
\renewcommand{\arraystretch}{1.3}
\begin{array}{|c||cc|cc|cc|}
\hline
\text{symmetry algebra} &
su(2)_L & su(2)_R & su(2)_1 & su(2)_2 & \widetilde{su}(2)_1 & \widetilde{su}(2)_2 \\
\hline
\hline
\begin{minipage}{0.2\textwidth}\centering
\raisebox{1ex}{\strut}
valid away from the free point?
\raisebox{-1ex}{\strut}
\end{minipage} &
\multicolumn{2}{c|}{\text{Yes}}
&
\multicolumn{4}{c|}{\text{No}}
\\
\hline
\text{left/right moving} & L & R & L & L & R & R\\
\text{current} & J^a & \Jt^a & K_1^a &  K_2^a & \Kt_1^a & \Kt_2^a \\
\text{quantum numbers}&j,m & \jt,\mt& j_1,m_1 & j_2,m_2 & \jt_1,\mt_1& \jt_2,\mt_2\\
\hline
\text{doublet index}&
\alpha & \alphad & A & \Ad & A & \Ad\\
\hline
\end{array}
\end{math}
\caption{\label{tbl:su(2)s} \sl Various $su(2)$ symmetry algebras of the free theory and their associated conserved currents.  We also show the convention for the quantum numbers and the doublet index for each $su(2)$ algebra.  Quantum numbers shown are the Casimirs and the third component of the zero modes of the current.  For example, for $su(2)_L$,  $j\in \bbZ_{\ge 0}/2$ and  $m=-j,-j+1,\dots,j$ is the eigenvalue of $J^3_0$.
}
\end{table}

Furthermore, we have an additional ``symmetry'' in the subspace of right-moving Ramond ground states.  The four single-strand right-moving ground states -- the two bosonic $\ket{\alphad}$ and the two fermionic $\ket*{\Ad}$ in $\Vt$ of \eqref{eq:Vk_Vt_def} -- all have the same energy $\tilde{h}=\frac{k}{4}$.  Hence any unitary change of basis mixing them, \textit{i.e.}\ any $U(2|2)$ transformation is a symmetry leaving the Hamiltonian invariant.
This $U(2|2)$ is the unitary subgroup of $GL(2|2)$ that appeared  in Section \ref{ssec:SW_T4} in the Schur-Weyl decomposition into irreducible $GL(2|2)$-representations labeled by $\lambda$. This ``symmetry'' is rather trivial: restricted to the ground-state subspace, the Hamiltonian is proportional to the identity.
The corresponding Lie superalgebra is $u(2|2)$.  The zero-mode algebras $su(2)_R$ and $\widetilde{su}(2)_2$ discussed above are two bosonic (non-maximal) sub-algebras of this $u(2|2)$, while the fermionic zero modes $\psit_0^{\alphad \Ad}$ constitute half of its eight fermionic generators (recall that the $\psit_0^{\alphad \Ad}$ generate a Clifford algebra $cl_4^{\psit}$).  For later convenience, let us denote the right-moving zero-mode algebra generated by these subalgebras by
\begin{align} \label{eq:A_alg_def}
    \scrA\equiv (su(2)_R\oplus \widetilde{su}(2)_2)\loplus cl_4^{\psit}
    ~~\subset~~ u(2|2)
    ~~\subset~~ gl(2|2) \ ,
\end{align}
where $h \loplus g $ means a semidirect sum with $g$ being the ideal (namely, the commutator $[h,g]\subset g$).
Although $\scrA$ is a symmetry of the free theory, the full $u(2|2)$ is not: the free boson modes $\tilde{\alpha}_n^{\Ad A}$ have an $\Ad$ index but no $\alphad$ index, so general $u(2|2)$ rotations would not preserve the spectrum beyond the ground state sector.

The Gava-Narain operator $\cGt^{\alphad A}$, which represents the supercharge in the deformed theory, preserves much of the above symmetry structure.  It commutes with all of the left-moving generators of the contracted large $\cN=4$ superconformal algebra (as shown in Appendix~\ref{app:GN_alg}\@).  
As for the free-theory $su(2)$ algebras (see Table \ref{tbl:su(2)s}), it commutes with all the current zero modes except for $K^a_{1,0}$, for which it transforms covariantly according to its $A$ index (recall that the $A$ index of $\cGt^{\alphad A}$ is left-moving as we mentioned below \eqref{eq:GN_action}).
For the right-moving $\cN=4$ generators, it anticommutes with $\psit_0^{\alphad \Ad}$ and covariantly transforms under $\Jt_0^a$  (see \eqref{eq:[At,cGt]}).  However, it does not fully respect the right-moving ground-state $u(2|2)$ symmetry; namely, it only respects\footnote{Note that the $\cGt^{\alphad A}$ do not commute with $\scrA$.  It does commute with $\widetilde{su}(2)_2$ and $cl_4^{\psit}$ but it is charged under $su(2)_R$.  Hence the use of the word ``respect'' in this context.} the $\scrA$-algebra \eqref{eq:A_alg_def}.  
The fact that $\cGt^{\alphad A}$ breaks $u(2|2)$ to $\scrA$ can be confirmed by explicit computations and we will expand on this in Section \ref{sec:summary}.
This means that the right-moving ground-state symmetry is broken once we deform away from the free theory, and that states lift or remain BPS at the level of $\scrA$-representations rather than in full $u(2|2)$ (or $gl(2|2)$) representations.

\subsection{$\scrA$-algebra representations (diamonds)}
\label{sssec:diamonds}

We found that lifting and the preservation of BPS states occur at the level of multiplets of the free-theory right-moving zero-mode algebra $\scrA$:
each $\scrA$-multiplet either lifts as a whole or remains BPS as a whole.
This leads us to consider the representation theory of the $\scrA$-algebra, which we now turn to.

The $\scrA$-algebra is the right-moving zero-mode algebra generated by $\psit^{\alphad \Ad}_0$, $\Jt^a_0$, $\Kt^a_{2,0}$.
More precisely, these are the total modes as defined in \eqref{eq.total_mode_def_strand}.  Namely, at the free point, each strand has generators defined on it, and 
the $\scrA$ generators are the diagonal sum of these strand-wise generators:
\begin{align}
\psit^{\alphad \Ad}_0&=\sum_{I=1}^n \psit^{\alphad \Ad[I]}_0,
\quad
\Jt^a_0=\sum_{I=1}^n \Jt^{a[I]}_0,
\quad 
\Kt^a_{2,0}=\sum_{I=1}^n \Kt^{a[I]}_{2,0} \ ,
\end{align}
where $I$ labels the strand number.
In this section, henceforth, we will suppress the mode-number subscript 0 on $\psit^{\alphad \Ad},\Jt^a,\Kt_2^{a}$ to avoid cluttering (the subscript 2 on $\Kt^a_2$ is not the mode number). 

The $\scrA$-algebra is defined by the following (anti)commutation relations among these total generators:
\begin{subequations} \label{A-alg_comm}
 \begin{gather}
 \big\{\psit^{\alphad\Ad},\psit^{\betad\Bd}\big\}=-N\epsilon^{\alphad\betad}\epsilon^{\Ad\Bd}\ ,
 \label{A-alg_comm_psit}
 \\
 \big[\Jt^a,\Jt^b\big]=i\epsilon^{ab}{}_{c}\Jt^c\quad,\quad
 \big[\Jt^a,\psit^{\alphad \Ad}\big]=\thalf (\sigma^{a})^\alphad{}_\betad \psit^{\betad\Ad}\ ,
  \label{A-alg_comm_Jt}
\\
 \big[\Kt_2^a,\Kt_2^b]=i\epsilon^{ab}{}_{c}\Kt_2^c\quad,\quad
 \big[\Kt_2^a,\psit^{\alphad \Ad}\big]=\thalf (\sigma^{a})^{\Ad}{}_{\Bd} \psit^{\alphad\Bd}\ .
  \label{A-alg_comm_Kt2}
 \end{gather}
\end{subequations}

Below, we discuss the representations of this algebra in general, but it is also instructive to start instead with concrete representations in simple examples.
Readers interested in such examples may wish to first read Appendix \ref{app:A-alg_rep_simple_cases} and then return to the general discussion below.

Let us now discuss the representation theory of the algebra \eqref{A-alg_comm} in general.
We begin with the algebra \eqref{A-alg_comm_psit} satisfied by the total modes $\psit^{\alphad \Ad}$.  This is an $SO(4)\cong SU(2)_R\times \widetilde{SU}(2)_2$ Clifford algebra $cl_4^{\psit}$ and its unique irreducible representation, the Dirac representation, is spanned by a quartet of states with $(\mt,\mt_2)=(\pm \half,0),(0,\pm\half)$, where $(\mt,\mt_2)$ are the eigenvalues of $(\Jt^3,\Kt^3_2)$; see Table \ref{tbl:su(2)s}.

Next, we turn to the $su(2)_R\oplus \widetilde{su}(2)_2$ part of the algebra, \eqref{A-alg_comm_Jt} and \eqref{A-alg_comm_Kt2}.  The $su(2)$ generators $\Jt^a,\Kt^a_2$ can be decomposed into a part expressible in terms of the total modes $\psit^{\alphad \Ad}$, which we denote by $\Jt^a_{(\psit)}$ and $\Kt^a_{2(\psit)}$, and the remaining parts, which we denote by $\Jt'^a$ and $\Kt'^a_2$.  Explicitly, we define the total-$\psit$-mode part by
\begin{align}
 \Jt_{(\psit)}^a
 \equiv -\tfrac{1}{4N}(\sigma^a)_\alphad{}^\betad \psit^{\alphad\Ad}\psit_{\betad\Ad} \quad ,
\quad
\Kt_{2(\psit)}^a\equiv 
  -\tfrac{1}{4N}(\sigma^a)_{\Ad}{}^{\Bd} \psit^{\alphad\Ad}\psit_{\alphad\Bd}\ .
\end{align}
It is straightforward to show that these satisfy the
$su(2)_R\oplus \widetilde{su}(2)_2$ relations on their own,
\begin{align} 
\begin{aligned}
\big[\Jt_{(\psit)}^a,\Jt_{(\psit)}^b\big]&=i\epsilon^{ab}{}_c \Jt_{(\psit)}^c\quad,&\quad
 \big[\Jt_{(\psit)}^a,\psit^{\alphad \Ad}\big]&=\thalf (\sigma^{a})^\alphad{}_\betad \psit^{\betad\Ad}\ ,
\\
 \big[\Kt_{2(\psit)}^a,\Kt_{2(\psit)}^b\big]&=i\epsilon^{ab}{}_c \Kt_{2(\psit)}^c\quad,&
 \big[\Kt_{2(\psit)}^a,\psit^{\alphad \Ad}\big]&=\thalf (\sigma^{a})^{\Ad}_{\ \Bd}\, \psit^{\alphad\Ad}\ .
 \end{aligned}
\end{align}
We then define the remaining parts by
\begin{align}
 \Jt'^a   \equiv \Jt^a- \Jt_{(\psit)}^a\ ,\qquad
 \Kt'^a_2 \equiv \Kt^a_2 - \Kt_{2(\psit)}^a\ .
\end{align}
We can show that $\Jt'^a$ and $\Kt'^a_2$ satisfy the $su(2)_R\oplus \widetilde{su}(2)_2$ algebra on their own,
\begin{align}
\big[\Jt'^a,\Jt'^b\big]&=i\epsilon^{ab}{}_c \,\Jt'^c\ ,\qquad
 \big[\Kt'^a_2,\Kt'^b_2\big]=i\epsilon^{ab}{}_c \,\Kt'^c_2\ ,
\end{align}
and are independent of the total mode $\psit$:
\begin{align}
 \big[\Jt'^a,\Jt_{(\psit)}^b\big]=
 \big[\Kt'^a_2,\Kt_{2(\psit)}^b\big]=
 \big[\Jt'^a,\psit^{\alphad \Ad}\big]=
 \big[\Kt'^a_2,\psit^{\alphad \Ad}\big]=0\ .
\end{align}

This means that the $\scrA$-algebra \eqref{A-alg_comm} is in fact a direct sum of the $su(2)'_R\oplus \widetilde{su}(2)_2'$ algebra generated by $\Jt'^a$, $\Kt'^a_2$ and the Clifford algebra $cl_4^\psit$ generated by $\psit$:
\begin{align}
    \scrA\cong su(2)_R'\oplus \widetilde{su}(2)_2'\oplus cl_4^\psit \ .
\end{align}
Therefore, an irreducible representation of the $\scrA$-algebra is obtained by taking an irreducible representation of $su(2)_R'\oplus \widetilde{su}(2)_2'$, namely the spin-$(\jt,\jt_2)$ representation with $\tj,\tj_2\in\half\bbZ_{\ge 0}$, and tensoring it with the Dirac quartet of the total mode $\psit$.  Let us denote the resulting $\scrA$-representation by $\cR^\scrA_{\tj,\tj_2}$.
The $SU(2)_R\times \widetilde{SU}(2)_2$ character of $\cR^{\scrA}_{\tj,\tj_2}$ is then
\begin{align} 
    \chi^{\scrA}_{\tj,\tj_2}(\yt,\etat) 
    &\equiv
    \tr_{\cR^{\scrA}_{\tj,\tj_2}}
    \bigl[(-1)^{2\Kt^3_2}\yt^{2\Jt^3}\etat^{2\Kt^3_2}\bigr]\notag\\
    &=
    (-1)^{2\tj_2}\bigl[\chi_{\half}(\yt)-\chi_{\half}(\etat)\bigr]
    \,\chi_{\tj}(\yt)\,\chi_{\tj_2}(\etat)\ .
    \label{eq:A_alg_chars}
\end{align}
where $\chi_{\half}(\yt)-\chi_{\half}(\etat)$ is the character of the Dirac representation.  The factor $(-1)^{2\Kt^3_2}$ is because $\ket*{\Ad}$ are fermionic, and we define $\tilde{\chi}^{\scrA}_{\tj,\tj_2}$ to be zero if either $\tj<0$ or $\tj_2<0$.

To graphically present the state content of $\scrA$-algebra representations, \emph{diamond diagrams} are useful.  This is obtained by first starting with a representation of $su(2)'_R\oplus \widetilde{su}(2)'_2$ as dots on the $(\mt,\mt_2)$ plane and then replacing each dot with a \emph{diamond}, representing a $\psit$-quartet.

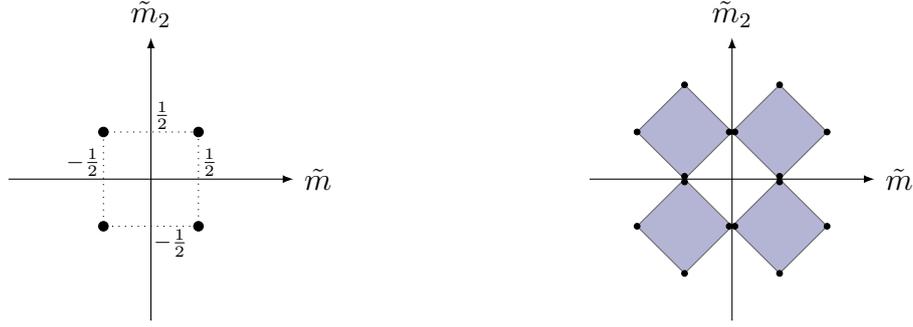
\begin{figure}[tb]
\centering
\subcaptionbox{\sl Spin-$(\half,\half)$ representation of the $su(2)_R'\oplus \widetilde{su}(2)_2'$ algebra. Dots represent states in the representation.}[0.4\textwidth]{
\begin{tikzpicture}[scale=1.25]
\draw[-latex] (-1.5,0) -- (1.5,0) node [right] {$\mt$};
\draw[-latex] (0,-1.5) -- (0,1.5) node [above] {$\mt_2$};
 \draw[fill=black] (1/2,1/2) circle (0.05);
 \draw[fill=black] (1/2,-1/2) circle (0.05);
 \draw[fill=black] (-1/2,1/2) circle (0.05);
 \draw[fill=black] (-1/2,-1/2) circle (0.05);
 \draw[dotted] (1/2,1/2) -- (-1/2,1/2) -- (-1/2,-1/2) -- (1/2,-1/2) -- cycle;
 \node at ( 1/2,0) [above right,font=\scriptsize,xshift=-3,yshift=-3] {$\tfrac12$};
 \node at (-1/2,0) [above left, font=\scriptsize,xshift=3,yshift=-3] {$-\tfrac12$};
 \node at (0, 1/2) [above right,font=\scriptsize,xshift=-3,yshift=-3] {$\tfrac12$};
 \node at (0,-1/2) [below right,font=\scriptsize,xshift=-3,yshift=3] {$-\tfrac12$};
\end{tikzpicture}
}
\qquad
\subcaptionbox{\sl The diamond diagram for the $\scrA$-algebra representation $\cR^\scrA_{\half,\half}$. Dots represent individual states in the representation.  At the four points $(\mt,\mt_2)=(\pm\half,0),(0,\pm \half)$ where two diamonds touch, there are two states.
}[0.4\textwidth]{
\begin{tikzpicture}[scale=1.25]
\draw[-latex] (-1.5,0) -- (1.5,0) node [right] {$\mt$};
\draw[-latex] (0,-1.5) -- (0,1.5) node [above] {$\mt_2$};
\diam{1/2}{1/2}
\diam{1/2}{-1/2}
\diam{-1/2}{1/2}
\diam{-1/2}{-1/2}
\draw[fill=black] (0.5,0.03) circle (0.03);
\draw[fill=black] (1.0,0.5) circle (0.03);
\draw[fill=black] (0.5,1.0) circle (0.03);
\draw[fill=black] (0.03,0.5) circle (0.03);
\draw[fill=black] (-0.5,0.03) circle (0.03);
\draw[fill=black] (-1.0,0.5) circle (0.03);
\draw[fill=black] (-0.5,1.0) circle (0.03);
\draw[fill=black] (-0.03,0.5) circle (0.03);
\draw[fill=black] (0.5,-0.03) circle (0.03);
\draw[fill=black] (1.0,-0.5) circle (0.03);
\draw[fill=black] (0.5,-1.0) circle (0.03);
\draw[fill=black] (0.03,-0.5) circle (0.03);
\draw[fill=black] (-0.5,-0.03) circle (0.03);
\draw[fill=black] (-1.0,-0.5) circle (0.03);
\draw[fill=black] (-0.5,-1.0) circle (0.03);
\draw[fill=black] (-0.03,-0.5) circle (0.03);
\end{tikzpicture}
}
\caption{\label{fig:A-alg_reps0} 
\sl $su(2)_R'\oplus \widetilde{su}(2)_2'$ representations and
$\scrA$-algebra representations
}
\end{figure}

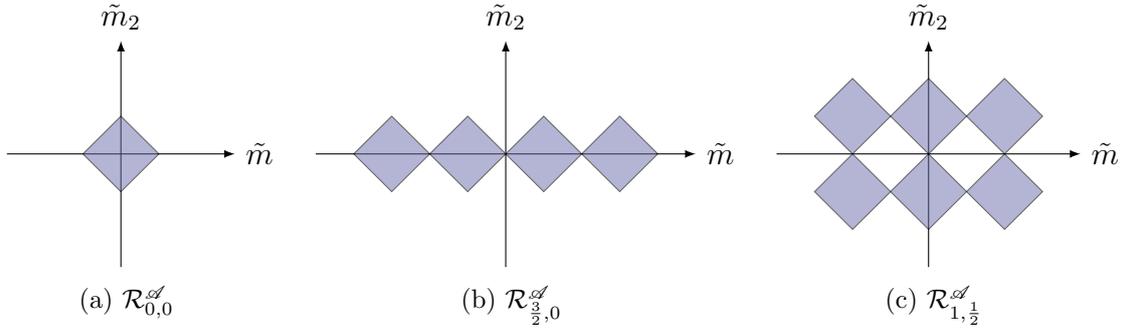
\begin{figure}[tb]
\centering
\subcaptionbox{$\cR^\scrA_{0,0}$~~~~}{
\begin{tikzpicture}
\draw[-latex] (-1.5,0) -- (1.5,0) node [right] {$\mt$};
\draw[-latex] (0,-1.5) -- (0,1.5) node [above] {$\mt_2$};
\diam{0}{0}
\end{tikzpicture}
}
\subcaptionbox{$\cR^\scrA_{\frac32,0}$~~~~}{
\begin{tikzpicture}
\draw[-latex] (-2.5,0) -- (2.5,0) node [right] {$\mt$};
\draw[-latex] (0,-1.5) -- (0,1.5) node [above] {$\mt_2$};
\diam{-3/2}{0}
\diam{-1/2}{0}
\diam{1/2}{0}
\diam{3/2}{0}
\end{tikzpicture}
}
\subcaptionbox{$\cR^\scrA_{1,\half}$~~~~}{
\begin{tikzpicture}
\draw[-latex] (-2,0) -- (2,0) node [right] {$\mt$};
\draw[-latex] (0,-1.5) -- (0,1.5) node [above] {$\mt_2$};
\diam{-1}{1/2} \diam{0}{1/2} \diam{1}{1/2}
\diam{-1}{-1/2} \diam{0}{-1/2} \diam{1}{-1/2}
\end{tikzpicture}
}
\caption{\label{fig:A-alg_reps_ex} 
\sl Examples of diamond diagrams for various $\scrA$-representations.  We have omitted the dots at the vertices representing states.}
\end{figure}

As an example of an $\scrA$-representation, let us consider $\cR^{\scrA}_{\half,\half}$.  This is based on the spin-$(\half,\half)$ representation of $su(2)'_R\oplus \widetilde{su}(2)'_2$, whose member states have $(\mt,\mt_2)=(\half,\pm \half)$, $(-\half,\pm \half)$ as shown in Figure~\ref{fig:A-alg_reps0}(a).  Multiplying this representation by the $\psit$-quartet means to replace each dot in Figure~\ref{fig:A-alg_reps0}(a) by a $\psit$-quartet with charges $(\mt,\mt_2)=(\pm\half,0),(0,\pm\half)$, which we represent by a diamond as in Figure~\ref{fig:A-alg_reps0}(b).  The dots at the vertices of diamonds represent states.  This means that when two diamonds touch at a point, as happens at four points in
Figure~\ref{fig:A-alg_reps0}(b), there are two states with the same $(\mt,\mt_2)$ values there. We call diagrams of diamonds on the $(\mt,\mt_2)$-plane representing  $\scrA$-algebra representations diamond diagrams.
For more examples of diamond diagrams, see Figure~\ref{fig:A-alg_reps_ex}, where we omitted the dots for states.

\subsection{$\scrG$-algebra representations (garnets)}
\label{sssec:garnets}

In this previous section, we saw that the free-theory BPS states can be organized into $\scrA$-representations, or equivalently diamond diagrams.  On the other hand, the Gava-Narain operator $\cGt^{\alphad A}$ connects a quartet of free-theory short multiplets that form a lifted long multiplet of the deformed theory.
This implies that, upon deformation, diamond diagrams containing lifted states get connected to one another by $\cGt^{\alphad A}$,  forming a larger structure. We now examine this structure.

Let us recall that $\cGt^{\alphad A}$ (anti)commutes with $\psit^{\alphad\Ad}$ and $\Kt^a_2$, and is charged under $su(2)_R\times su(2)_1$ as its ($\alphad,A$) indices indicate. Namely,
\begin{align}
    \big[\Jt^a,\cGt^{\alphad A}\big]
    =\half (\sigma^{a})^\alphad{}_\betad\, \cGt^{\betad A}\quad,\quad
    \big[K_1^a,\cGt^{\alphad A}\big]
    =\half (\sigma^{a})^A{}_B\, \cGt^{\alphad B}\ .
    \label{JtcGcomm}
\end{align}

A BPS state $\ket*{\phi_\mu}$ of the free theory, when deformed, can remain BPS, or otherwise get lifted.
First consider the case where a state $\ket{\phi}$ remains BPS, namely when $\cGt^{\alphad A}\ket{\phi}=0$ for all $\alphad,A$.  Then the entire $\scrA$-representation that $\ket{\phi}$ belongs to is annihilated by $\cGt^{\alphad A}$.  This is because $\psit,\Kt_2$ (anti)commute with $\cGt$ and
\begin{align}
    \cGt^{\alphad A}\big(\Jt^a \ket{\phi}\big)
    =
    \Jt^a \cGt^{\alphad A} \ket{\phi}
    -\big[\Jt^a, \cGt^{\alphad A}\big] \ket{\phi}
    =0 \ ,
\end{align}
where the second term vanishes due to \eqref{JtcGcomm}.
Therefore, unlifted states come in $\scrA$-represen\-tations of the free theory.  In other words, unlifted states form ``isolated'' diamond diagrams that are not connected to each other by $\cGt$.\footnote{These multiplets of unlifted states can, however, come in nontrivial $SU(2)_1$ representations, as considered in \cite{Chang:2025wgo,Guo:2020gxm}.}

Next, consider the case where $\ket{\phi}$ is lifted once the theory is deformed; namely,  when $\cGt^{\alphad A}\ket{\phi}\neq 0$ for some $\alphad,A$.  Then $\ket{\phi}$ must be a part of a lifted $\cGt$-quartet \eqref{eq:GN_quartet}, and its nonzero lift is determined by the lifting operator $\{\cGt^{\alphad A}, \cGt^{\betad B}\}$ in  \eqref{eq:anom_dim}.  If the lift of $\ket{\phi}$ is given by $E^{(2)}_\phi>0$, the states that can be related to it by the zero-mode generators $\psit,\Jt,\Kt_2$ also have the same lift $E^{(2)}_\phi$.  This is obvious for $\psit,\Kt_2$ which (anti)commute with $\cGt$ and thus with $\{\cGt^{\alphad A}, \cGt^{\betad B}\}$.  For $\Jt$, one can show
\begin{align}
    \ev*{\phi_b| \big[\Jt^a,\{\cGt^{\alphad A}, \cGt^{\betad B}\} \big]  |\phi_a }=0 \ ,
\end{align}
using \eqref{JtcGcomm} and the fact that $(\sigma^{a})^{\alphad\betad}=(\sigma^{a})^{\betad\alphad}$, which means that the states obtained by the action of $\Jt^a$ also have the same lift.
Therefore, each member of the $\cGt$-quartet \eqref{eq:GN_quartet} is embedded into a full $\scrA$-representation, all lifting by the same amount.

If we restrict to the eigenspace of the lifting operator with a fixed eigenvalue $E_\phi>0$ in the free-theory BPS Hilbert space, we can talk about the representation of the zero-mode algebra generated by the $\scrA$-algebra generators ($\Jt^a,\Kt_2^a,\psit^{\alphad \Ad}$) and $K_1^a,\cGt^{\alphad A}$. In this space, from \eqref{eq:anom_dim}, 
\begin{equation}
    \big\{ \cGt^{\alphad A}, \cGt^{\betad B} \big\} 
    = -\epsilon^{\alphad\betad}\epsilon^{AB} E^{(2)}_\phi/2 \ ,
\end{equation}
which means that $\cGt$ generates another Clifford algebra, $cl_4^{\cGt}$, in addition to $cl_4^{\psit}$.  Let us denote the full algebra by
\begin{align}
    \scrG
    \equiv
        (su(2)_R\oplus su(2)_1 \oplus \widetilde{su}(2)_2)\loplus (cl_4^{\psit}\oplus cl_4^{\cGt})\ .
\end{align}
Just as we did for the $\scrA$-algebra, we can show that this can be rewritten as
\begin{align}
    \scrG
    \cong
        (su(2)_R'\oplus su(2)_1' \oplus \widetilde{su}(2)_2')\oplus (cl_4^{\psit}\oplus cl_4^{\cGt})\ ,
\end{align}
where $su(2)'_R$ is obtained from $su(2)_R$ by subtracting the contribution to it from the total $\psit$ and $\cGt$ modes.  Likewise, $su(2)'_1$ is obtained from $su(2)_1$ by subtracting the total $\cGt$ mode contribution and $\widetilde{su}(2)'_2$ is obtained from $\widetilde{su}(2)_2$ by subtracting the total $\psit$ mode contribution.  This means that an irreducible representation of $\scrG$, denoted by $\cR^{\scrG}_{\tj,j_1,\tj_2}$ (where $\tj,j_1,\tj_2\in \frac12\bbZ_{\ge 0}$), is a direct product of a representation of $su(2)'_R\oplus su(2)'_1\oplus \widetilde{su}(2)'_2$ and the two Dirac quartets, one generated by $\psit$ and another generated by $\cGt$.
The character of the $\scrG$-algebra representation $\cR^{\scrG}_{\tj,j_1,\tj_2}$ is
\begin{align}
    \chi^{\scrG}_{\tj,j_1,\tj_2}(\yt,\etat_1,\etat)
    &\equiv
    \tr_{\cR^{\scrG}_{\tj,j_1,\tj_2}}
    \bigl[(-1)^{2\Kt^3_2}\yt^{2\Jt^3}\etat_1^{2K^3_1}\etat^{2\Kt^3_2}\bigr]
    \notag\\
    &=(-1)^{2\tj_2}
    \big[\chi_\half(\yt)-\chi_\half(\etat_1)\big]
    \big[\chi_\half(\yt)-\chi_\half(\etat)\big]
    \chi_{\tj}(\yt)\,\chi_{j_1}(\etat_1)\chi_{\tj_2}(\etat)\,.
    \label{eq:garnet_char}
\end{align}

The product of the $\psit$- and $\cGt$-quartets contains 16 states whose quantum numbers $(\mt,m_1,\mt_2)$, defined as the eigenvalues of $(\Jt^3,K^3_1,\Kt^3_2)$, are
\begin{align}
    &(\pm1,0,0)\,,\ 2(0,0,0)\,,\ (\thalf,\pm\thalf,0)\,,\ (-\thalf,\pm\thalf,0)\,,\notag\\
    &\qquad (\thalf,0,\pm\thalf)\,,\ (-\thalf,0,\pm\thalf)\,,\ (0,\thalf,\pm\thalf)\,,\ (0,-\thalf,\pm\thalf)\ .
    \label{garnet_pts}
\end{align}
In the $(\mt,m_1,\mt_2)$ space, the polyhedron with vertices at these points is a variation of the rhombic dodecahedron (the points in \eqref{garnet_pts} include ones that are inside the polyhedron's outer surface).  This is the generalization of the diamond for the $\scrA$-algebra to the $\scrG$-algebra.
We call this structure a \emph{garnet}, because garnet is known to have a rhombic dodecahedral crystal habit.\footnote{The actual rhombic dodecahedron of the garnet crystal is a topologically equivalent variant of the one represented by \eqref{garnet_pts} or Figure~\ref{fig:garnets}(a) \cite{wiki:rhombicdodecahedron}.} In Figure~\ref{fig:garnets}(a), we present a diagram of a garnet in the $(\mt,m_1,\mt_2)$ space.  It contains four $\psit$-diamonds (which are blue vertical diamonds in the figure) and four $\cGt$-diamonds (yellowish horizontal diamonds).
\begin{figure}[tb]
    \centering
    \subcaptionbox{}{
    \includegraphics[height=4.0cm,trim={2cm 3.5cm 0 0},clip]{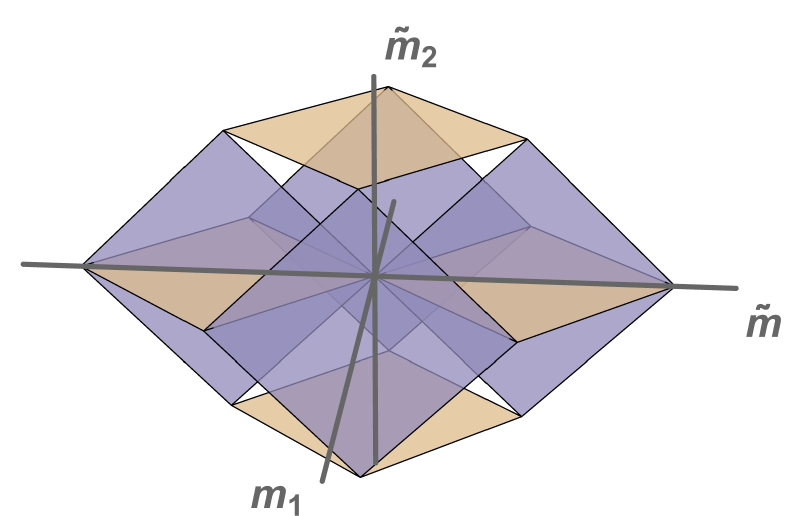}
    }
    \subcaptionbox{}{
    \includegraphics[height=4.0cm,trim={2cm 3.5cm 1cm 0},clip]{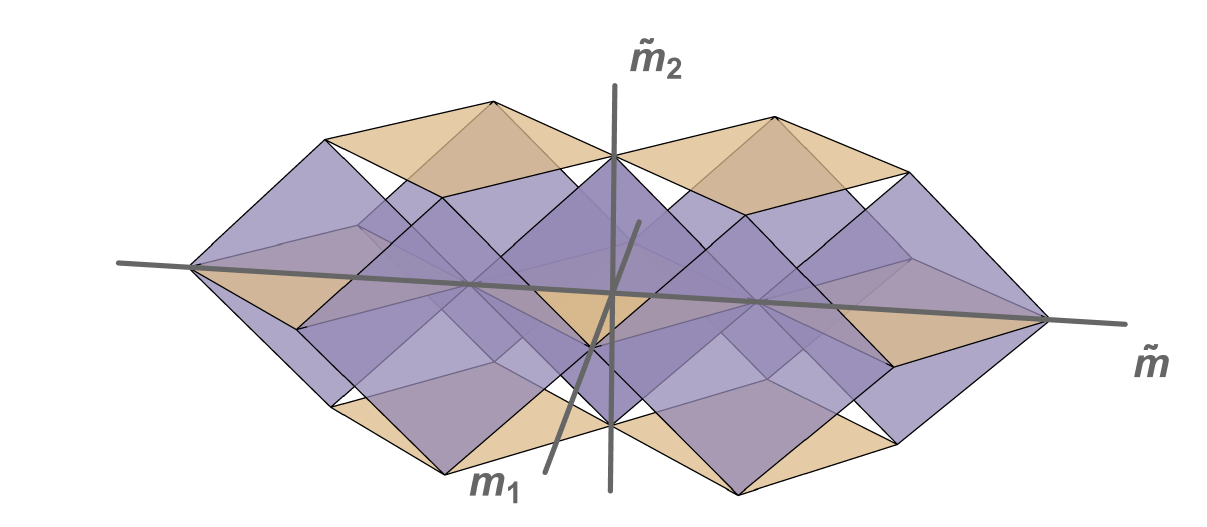}
    }
\caption{\sl Examples of garnet diagrams.  (a)~A garnet, which represents the product of $\psit$- and $\cG$-quartets.  This can also be interpreted as the garnet diagram for $\cR^{\scrG}_{0,0,0}$.
  (b)~The garnet diagram for $\cR^{\scrG}_{\half,0,0}$.  Two garnets are arrayed along the $\mt$ direction.  Two vertical diamonds (bluish) centered at the origin overlap, and two horizontal diamonds (yellowish) centered at the origin overlap.}
    \label{fig:garnets}
\end{figure}
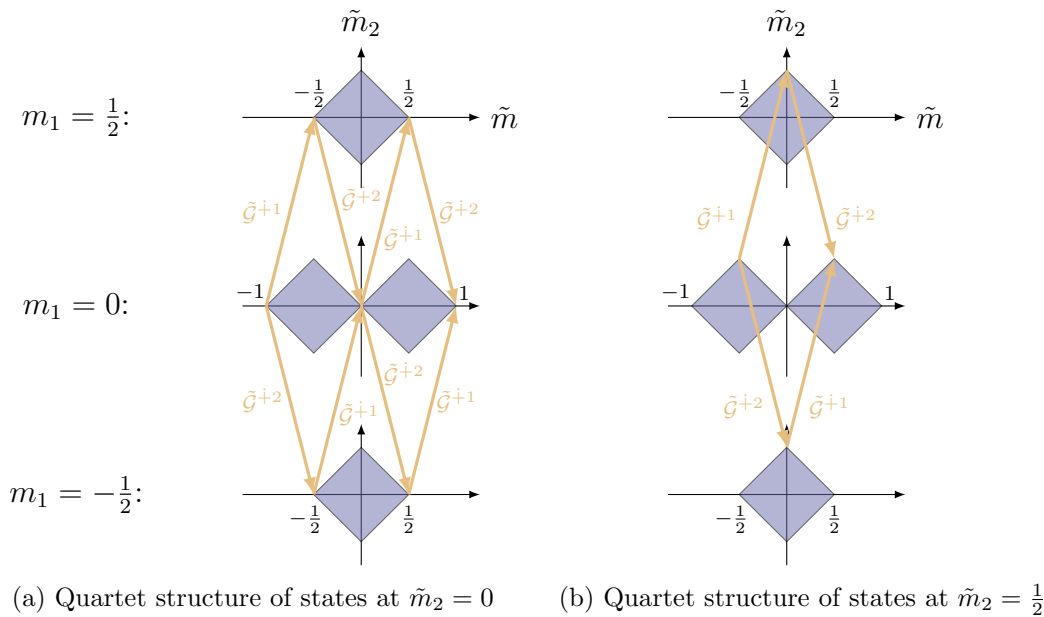
\begin{figure}[tb]
\centering
\subcaptionbox{Quartet structure of states at $\mt_2=0$}[0.4\textwidth]{
\begin{tikzpicture}[scale=1.25]
\node at (-3, 2) {$m_1=\thalf$:};
\node at (-3, 0) {$m_1=0$:};
\node at (-3,-2) {$m_1=-\thalf$:};
\draw[-latex] (-1.25,2)  -- +(2.5,0) node [right] {$\mt$};
\draw[-latex] (-1.25,0)  -- +(2.5,0);
\draw[-latex] (-1.25,-2) -- +(2.5,0);
\draw[-latex] (0,1.25)   -- +(0,1.5) node [above] {$\mt_2$};
\draw[-latex] (0,-0.75)  -- +(0,1.5);
\draw[-latex] (0,-2.75)  -- +(0,1.5);
\node[above,font=\scriptsize] at (0.5,2)  {$\thalf$};
\node[above,font=\scriptsize] at (-0.5,2) {$-\thalf~$};
\node[above,font=\scriptsize,xshift=3,yshift=-2] at (1,0)  {$1$};
\node[above,font=\scriptsize,xshift=-6,yshift=-2] at (-1,0) {$-1$};
\node[below,font=\scriptsize] at (0.5,-2)  {$\thalf$};
\node[below,font=\scriptsize] at (-0.5,-2) {$-\thalf~~$};
\diam{0}{2}
\diam{-1/2}{0} \diam{1/2}{0}
\diam{0}{-2}
\draw[-latex,color=garrow1,very thick,font=\scriptsize] (-1,0) -- (-0.5, 2) node [midway,left,xshift=2] {$\cGt^{\dot{+}1}$};
\draw[-latex,color=garrow1,very thick,font=\scriptsize] (-1,0) -- (-0.5,-2) node [midway,left,xshift=2] {$\cGt^{\dot{+}2}$};
\draw[-latex,color=garrow1,very thick,font=\scriptsize] (-0.5, 2) -- (0,0) node [midway,right,xshift=-3,yshift=5] {$\cGt^{\dot{+}2}$};
\draw[-latex,color=garrow1,very thick,font=\scriptsize] (-0.5,-2) -- (0,0) node [midway,right,xshift=-4,yshift=-5] {$\cGt^{\dot{+}1}$};
\draw[-latex,color=garrow1,very thick,font=\scriptsize] (0,0) -- (0.5, 2) node [midway,right,xshift=-5,yshift=-10] {$\cGt^{\dot{+}1}$};
\draw[-latex,color=garrow1,very thick,font=\scriptsize] (0,0) -- (0.5,-2) node [midway,right,xshift=-5,yshift=10] {$\cGt^{\dot{+}2}$};
\draw[-latex,color=garrow1,very thick,font=\scriptsize] (0.5, 2) -- (1,0) node [midway,right,xshift=-2] {$\cGt^{\dot{+}2}$};
\draw[-latex,color=garrow1,very thick,font=\scriptsize] (0.5,-2) -- (1,0) node [midway,right,xshift=-3] {$\cGt^{\dot{+}1}$};
\end{tikzpicture}}
\quad
\subcaptionbox{Quartet structure of states at $\mt_2=\thalf$}[0.4\textwidth]{
\begin{tikzpicture}[scale=1.25]
\draw[-latex] (-1.25,2)  -- +(2.5,0) node [right] {$\mt$};
\draw[-latex] (-1.25,0)  -- +(2.5,0);
\draw[-latex] (-1.25,-2) -- +(2.5,0);
\draw[-latex] (0,1.25)   -- +(0,1.5) node [above] {$\mt_2$};
\draw[-latex] (0,-0.75)  -- +(0,1.5);
\draw[-latex] (0,-2.75)  -- +(0,1.5);
\node[above,font=\scriptsize] at (0.5,2)  {$\thalf$};
\node[above,font=\scriptsize] at (-0.5,2) {$-\thalf~$};
\node[above,font=\scriptsize,xshift=3,yshift=-2] at (1,0)  {$1$};
\node[above,font=\scriptsize,xshift=-6,yshift=-2] at (-1,0) {$-1$};
\node[below,font=\scriptsize] at (0.5,-2)  {$\thalf$};
\node[below,font=\scriptsize] at (-0.5,-2) {$-\thalf~~$};
\diam{0}{2}
\diam{-1/2}{0} \diam{1/2}{0}
\diam{0}{-2}
\draw[-latex,color=garrow1,very thick,font=\scriptsize] (-1/2,0.5) -- (0, 2.5) node [midway,left,xshift=-5,yshift=-20] {$\cGt^{\dot{+}1}$};
\draw[-latex,color=garrow1,very thick,font=\scriptsize] (-1/2,0.5) -- (0,-1.5) node [midway,left,xshift= 5,yshift=-20] {$\cGt^{\dot{+}2}$};
\draw[-latex,color=garrow1,very thick,font=\scriptsize] (0, 2.5) -- (1/2,0.5) node [midway,right,xshift= 5,yshift=-20] {$\cGt^{\dot{+}2}$};
\draw[-latex,color=garrow1,very thick,font=\scriptsize] (0,-1.5) -- (1/2,0.5) node [midway,right,xshift=-5,yshift=-20] {$\cGt^{\dot{+}1}$};
\end{tikzpicture}}
\caption{\label{fig:diam_gnt_1} 
\sl How diamond diagrams  within the garnet diagram $\cR^{\scrG}_{0,0,0}$ are connected by the action of the Gava-Narain operator $\cGt^{\alphad A}$, for different values of $\mt_2$.  The yellowish arrows indicate how states are mapped into each other by the indicated Gava-Narain operator. The diamonds outlined by four arrows exactly represent the horizontal yellowish diamonds in Figure~\ref{fig:garnets}(a).
In (a), there are two states at $(\mt,m_1,\mt_2)=(0,0,0)$, and the one at which two arrows end and the one from which two different arrows start are distinct states.  The quartet structure at $\mt_2=-\half$ is similar to~(b).}
\end{figure}

In the $(\mt,m_1,\mt_2)$ space, the states of the $\scrG$-representation $\cR^{\scrG}_{\tj,j_1,\tj_2}$ can be obtained as follows:
\begin{itemize}
    \item[(i)]
Put dots at points $(\mt,m_1,\mt_2)$ that correspond to the states of spin-$(\tj,j_1,\tj_2)$ representation of $su(2)'_R\oplus su(2)'_1\oplus \widetilde{su}(2)'_2$. 
\item[(ii)]Replace each dot by a garnet, by adding 
$(\mt,m_1,\mt_2)$ charges given in \eqref{garnet_pts} to it.
This corresponds to multiplying by $\psit$- and $\cGt$-quartets.
\end{itemize}
We call the array of garnets obtained by this procedure a \emph{garnet diagram}.  In a garnet diagram, neighboring garnets overlap in the $\mt$ direction, because a garnet has length two in the $\mt$ (or $\Jt^3$) direction while the lattice spacing is one.  In the $m_1,\mt_2$ (or $K^3_1,\Kt^3_2$) directions a garnet has length one, and neighboring garnets intersect at a face in those directions, sharing a $\psit$-diamond or a $\cGt$-diamond.  For example, in the garnet diagram for $\cR^{\scrG}_{\half,0,0}$, there are two garnets arrayed along the $\mt$ axis; see Figure~\ref{fig:garnets}(b). 
We see that there are two $\psit$-diamonds (bluish vertical diamonds centered at the origin) on top of each other and two $\cGt$-diamonds (yellowish horizontal diamonds centered at the origin) on top of each other.

\subsection{Diamonds, garnets, and superselection} \label{ssec:superselection}

Diamond diagrams connected by the Gava-Narain operator $\cGt$ form a garnet diagram representing lifted states.   Here let us discuss how the $\cGt$-quartet structure \eqref{eq:GN_quartet} is embedded in diamond and garnet diagrams, representing a set of states that have the same nonzero lift.

First, note that the $\scrG$-character \eqref{eq:garnet_char} can be decomposed into $\scrA$-characters \eqref{eq:A_alg_chars} as \begin{align}
    \chi^{\scrG}_{\tj,j_1,\tj_2}(\yt,\eta_1,\etat)
    =
\chi_{j_1}(\eta_1)\,\big(\chi^{\scrA}_{\jt+\half,\tj_2}+\chi^{\scrA}_{\jt-\half,\tj_2}\big)-\big(\chi_{j_1+\half}(\eta_1)+\chi_{j_1-\half}(\eta_1)\big)\,\chi^{\scrA}_{\jt,\jt_2}\ .
\end{align}
All $\scrA$-characters on the right-hand side have the same $\widetilde{su}(2)_2$ spin $\tj_2$.
Diagrammatically, this means that only diamond diagrams with the same size along the $\mt_2$-axis (measured by the number of diamonds along the $\mt_2$-axis, $2\tj_2+1$) can combine to form a garnet diagram.  Physically, this implies a \emph{$\tj_2$-superselection rule:} $\scrA$-representations with different values of $\tj_2$ belong to different sectors, and no lifting (or mixing) occurs between them.

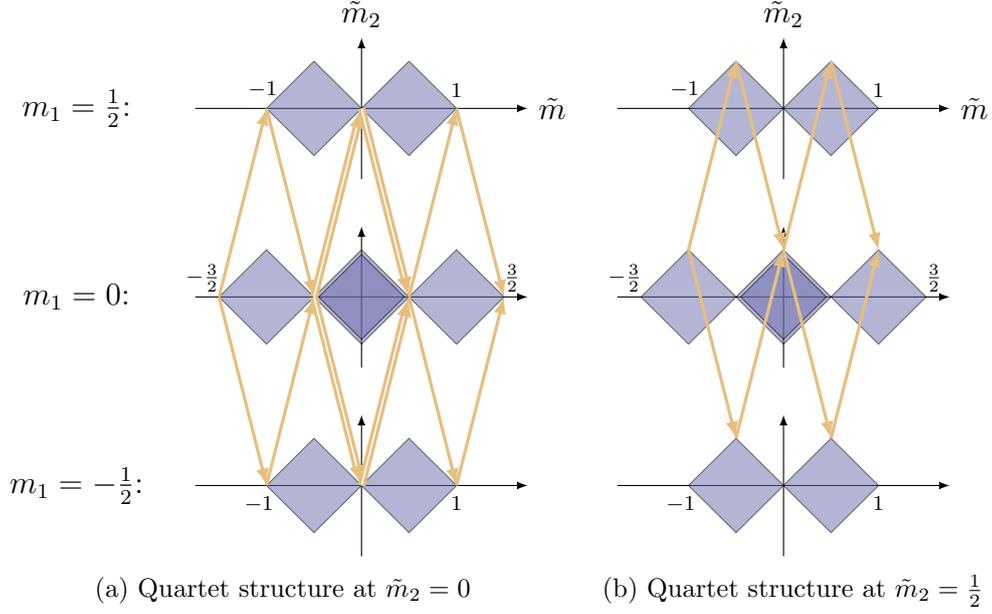
\begin{figure}[tb]
\centering
\subcaptionbox{Quartet structure at $\mt_2=0$}{
\begin{tikzpicture}[scale=1.25]
\node at (-3,2) {$m_1=\thalf$:};
\node at (-3,0) {$m_1=0$:};
\node at (-3,-2) {$m_1=-\thalf$:};
\draw[-latex] (-1.75,2)  -- +(3.5,0) node [right] {$\mt$};
\draw[-latex] (-1.75,0)  -- +(3.5,0);
\draw[-latex] (-1.75,-2) -- +(3.5,0);
\draw[-latex] (0,1.25)   -- +(0,1.5) node [above] {$\mt_2$};
\draw[-latex] (0,-0.75)  -- +(0,1.5);
\draw[-latex] (0,-2.75)  -- +(0,1.5);
\node[above,font=\scriptsize] at (1,2)  {$1$};
\node[above,font=\scriptsize] at (-1,2) {$-1~$};
\node[above,font=\scriptsize,xshift=3,yshift=-2] at (1.5,0)  {$\frac32$};
\node[above,font=\scriptsize,xshift=-6,yshift=-2] at (-1.5,0) {$-\frac32$};
\node[below,font=\scriptsize] at (1,-2)  {$1$};
\node[below,font=\scriptsize] at (-1,-2) {$-1~~$};
\diam{-1/2}{2} \diam{1/2}{2}
\diam{-1}{0} \diam{0}{0} \diam[0.45]{0}{0} \diam{1}{0}
\diam{-1/2}{-2} \diam{1/2}{-2}
\draw[-latex,color=garrow1,very thick,font=\scriptsize] (-1.5,0) -- (-1, 2);
\draw[-latex,color=garrow1,very thick,font=\scriptsize] (-1.5,0) -- (-1,-2);
\draw[-latex,color=garrow1,very thick,font=\scriptsize] (-1, 2) -- (-0.5,0);
\draw[-latex,color=garrow1,very thick,font=\scriptsize] (-1,-2) -- (-0.5,0);
\draw[-latex,color=garrow1,double,very thick,font=\scriptsize] (-0.5,0) -- (0, 2);
\draw[-latex,color=garrow1,double,very thick,font=\scriptsize] (-0.5,0) -- (0,-2);
\draw[-latex,color=garrow1,double,very thick,font=\scriptsize] (0, 2) -- (0.5,0);
\draw[-latex,color=garrow1,double,very thick,font=\scriptsize] (0,-2) -- (0.5,0);
\draw[-latex,color=garrow1,very thick,font=\scriptsize] (0.5,0) -- (1, 2);
\draw[-latex,color=garrow1,very thick,font=\scriptsize] (0.5,0) -- (1,-2);
\draw[-latex,color=garrow1,very thick,font=\scriptsize] (1, 2) -- (1.5,0);
\draw[-latex,color=garrow1,very thick,font=\scriptsize] (1,-2) -- (1.5,0);
\end{tikzpicture}}
\subcaptionbox{Quartet structure at $\mt_2=\half$}{
\begin{tikzpicture}[scale=1.25]
\draw[-latex] (-1.75,2)  -- +(3.5,0) node [right] {$\mt$};
\draw[-latex] (-1.75,0)  -- +(3.5,0);
\draw[-latex] (-1.75,-2) -- +(3.5,0);
\draw[-latex] (0,1.25)   -- +(0,1.5) node [above] {$\mt_2$};
\draw[-latex] (0,-0.75)  -- +(0,1.5);
\draw[-latex] (0,-2.75)  -- +(0,1.5);
\node[above,font=\scriptsize] at (1,2)  {$1$};
\node[above,font=\scriptsize] at (-1,2) {$-1~$};
\node[above,font=\scriptsize,xshift=3,yshift=-2] at (1.5,0)  {$\frac32$};
\node[above,font=\scriptsize,xshift=-6,yshift=-2] at (-1.5,0) {$-\frac32$};
\node[below,font=\scriptsize] at (1,-2)  {$1$};
\node[below,font=\scriptsize] at (-1,-2) {$-1~~$};
\diam{-1/2}{2} \diam{1/2}{2}
\diam{-1}{0} \diam{0}{0} \diam[0.45]{0}{0} \diam{1}{0}
\diam{-1/2}{-2} \diam{1/2}{-2}
\draw[-latex,color=garrow1,very thick,font=\scriptsize] (-1,0.5) -- (-0.5, 2.5);
\draw[-latex,color=garrow1,very thick,font=\scriptsize] (-1,0.5) -- (-0.5,-1.5);
\draw[-latex,color=garrow1,very thick,font=\scriptsize] (-0.5, 2.5) -- (0,0.5);
\draw[-latex,color=garrow1,very thick,font=\scriptsize] (-0.5,-1.5) -- (0,0.5);
\draw[-latex,color=garrow1,very thick,font=\scriptsize] (0,0.5) -- (0.5, 2.5);
\draw[-latex,color=garrow1,very thick,font=\scriptsize] (0,0.5) -- (0.5,-1.5);
\draw[-latex,color=garrow1,very thick,font=\scriptsize] (0.5, 2.5) -- (1,0.5);
\draw[-latex,color=garrow1,very thick,font=\scriptsize] (0.5,-1.5) -- (1,0.5);
\end{tikzpicture}}
\caption{\label{fig:diam_gnt_2} 
\sl How diamond diagrams within the garnet diagram $\cR^{\scrG}_{\half,0,0}$ are connected by the action of the Gava-Narain operator $\cGt^{\alphad A}$, for different values of $\mt_2$.  There are two degenerate $\psit$-diamonds (drawn in blue) centered at $\mt=\mt_2=0$ for $m_1=0$.  Double-lined arrows mean that there are two states connected by $\cGt$ (two at the head and two at the tail). The quartet structure at $\mt_2=-\half$ is similar to (b).}
\end{figure}

Let us study how diamond diagrams with the same $\tj_2$ value constitute a garnet diagram.
The simplest example is $\cR^{\scrG}_{0,0,0}$.  In this case, step (i) above puts a single dot at the origin and step (ii) replaces it with a single garnet.  So, the garnet diagram in this case is exactly the one in Figure~\ref{fig:garnets}(a).  In that figure, we see four vertical $\psit$-diamonds (bluish) representing $\psit$-quartets: one lying in the $m_1=\half$ plane, two in the $m_1=0$ plane, and one in the $m_1=-\half$ plane.  They are connected by horizontal $\cGt$-diamonds (yellowish), representing $\cGt$-quartets.  To make the structure more visible, in Figure~\ref{fig:diam_gnt_1}, we drew three diamond diagrams lying in the $m_1=- \half,0,\half$ plane separately and connected by arrows states that are mapped into each other by $\cGt$.  We can identify the diamonds outlined by yellowish arrows in that figure with the horizontal yellowish diamonds in Figure~\ref{fig:garnets}(a).  In Figure~\ref{fig:diam_gnt_2}, we do the same for the garnet diagram $\cR^{\scrG}_{\half,0,0}$.  This way we see how all states in diamond diagrams at different values of $m_1$, each of which is an irreducible $\scrA$-representation, are connected to each other by $\cGt$ in a consistent way, forming a garnet diagram, which is an irreducible $\scrG$-representation.

Lastly, we make the simple observation that for any $N$ the totally antisymmetric sector, labelled by $\lam=\{1,\dots,1\}\vdash N$, contains only one $\scrA$-representation and that this representation has the maximal $\tj_2$ charge for that value of $N$. At the level of characters, from \eqref{eq.gl22_to_A_Def} we have
\begin{equation}
    \tilde{S}_{\{1,\dots,1\}} = \tilde{\chi}^{\scrA}_{0,\frac{N-1}{2}} \ .
\end{equation}
The fact that this is the only representation with this value of $\tj_2$ charge at fixed $N$ means that this entire sector of states remains unlifted. This is a generalisation of an observation made in \cite{Guo:2020gxm} about states in the $N=2$ and $N=3$ theories with antisymmetric right-moving part not lifting.

\subsection{Summary}

To summarize, 
free-theory BPS states that remain BPS when deformed can be organized into diamond diagrams.  All states in a diamond diagram are related to one another by the generators of the $\scrA$-algebra, which is a symmetry algebra of the free theory and of the second-order perturbation theory.  A diamond diagram gives a nonzero contribution to the MEG (which is equal to the number of diamonds up to sign, as we will see in \eqref{eq.A_char_to_MEG}).
Diamond diagrams (or $\scrA$-representations) with different values of $\widetilde{su}(2)_2$ spin $\tj_2$ belong to different superselection sectors and lifting does not happen between them.

On the other hand, free-theory BPS states that get lifted when deformed can be organized into garnet diagrams, which are made of multiple diamond diagrams with the same $\tj_2$ value. All states in a garnet diagram have the same lift, and are related to one another by the $\scrG$-algebra generators in a way consistent with the $\cGt$-quartet structure \eqref{eq:GN_quartet}.  Garnet diagrams contribute zero to the MEG\@.

\section{The resolved elliptic genus (REG)}
\label{sec:REG}

\subsection{Constructing the REG}
\label{ssec:REG}

We now return to the supersymmetry index of this theory, namely the modified elliptic genus (MEG)\@. While the Schur-Weyl form of the MEG \eqref{eq:MEG_as_H(2|2)_sum}, or equivalently \eqref{eq:MEG_as_hook_sum}, is naturally expressed as a sum of $\lam$-sectors, the superselection sectors with respect to the Gava-Narain operator discussed in Section~\ref{ssec:Lifting} are not. Instead, these sectors are naturally described in terms of $\scrA$-algebra representations. 

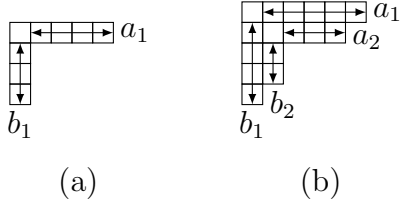
\begin{figure}[tb]
    \centering
\begin{tabular}{c@{\hspace{5ex}}c}
\begin{tikzpicture}[scale=1.5]
 \draw (0  ,0) rectangle +(1ex,1ex);
 \draw (1ex,0) rectangle +(1ex,1ex);
 \draw (2ex,0) rectangle +(1ex,1ex);
 \draw (3ex,0) rectangle +(1ex,1ex);
 \draw (4ex,0) rectangle +(1ex,1ex);
 \draw (0,-1ex) rectangle +(1ex,1ex);
 \draw (0,-2ex) rectangle +(1ex,1ex);
 \draw (0,-3ex) rectangle +(1ex,1ex);
\draw[latex-latex] (1ex,0.5ex) -- +(4ex,0) node [right,xshift=-0.3ex] {$a_1$};
\draw[latex-latex] (0.5ex,0) -- +(0,-3ex) node [below,yshift=0.3ex] {$b_1$};
\end{tikzpicture}
&
\begin{tikzpicture}[scale=1.5]
 \draw (0  ,0) rectangle +(1ex,1ex);
 \draw (1ex,0) rectangle +(1ex,1ex);
 \draw (2ex,0) rectangle +(1ex,1ex);
 \draw (3ex,0) rectangle +(1ex,1ex);
 \draw (4ex,0) rectangle +(1ex,1ex);
 \draw (5ex,0) rectangle +(1ex,1ex);
 \draw (0  ,-1ex) rectangle +(1ex,1ex);
 \draw (1ex,-1ex) rectangle +(1ex,1ex);
 \draw (2ex,-1ex) rectangle +(1ex,1ex);
 \draw (3ex,-1ex) rectangle +(1ex,1ex);
 \draw (4ex,-1ex) rectangle +(1ex,1ex);
 \draw (0,-2ex) rectangle +(1ex,1ex);
 \draw (1ex,-2ex) rectangle +(1ex,1ex);
 \draw (0,-3ex) rectangle +(1ex,1ex);
 \draw (1ex,-3ex) rectangle +(1ex,1ex);
 \draw (0,-4ex) rectangle +(1ex,1ex);
\draw[latex-latex] (1ex,0.5ex) -- +(5ex,0) node [right,xshift=-0.3ex] {$a_1$};
\draw[latex-latex] (2ex,-0.5ex) -- +(3ex,0) node [right,xshift=-0.3ex,yshift=-0.5ex] {$a_2$};
\draw[latex-latex] (0.5ex,0) -- +(0,-4ex) node [below,yshift=0.3ex] {$b_1$};
\draw[latex-latex] (1.5ex,-1ex) -- +(0,-2ex) node [below,yshift=0.3ex,xshift=0.7ex] {$b_2$};
\end{tikzpicture}
\\
(a)&(b)
\end{tabular}
    \caption{\sl Frobenius coordinates for Young diagrams in the case of (a) diagrams with $d_\lambda=1$, $\lambda=(a_1|a_2)\in H(1|1)$ and (b) diagrams with $d_\lambda=2$, $\lambda=(a_1,a_2|b_1,b_2)\in H(2|2)$ with $a_1>a_2$ and $b_1>b_2$.}
    \label{fig:frobenius_coords}
\end{figure}

Returning to the Schur-Weyl form \eqref{eq:MEG_as_H(2|2)_sum}, we are then motivated to decompose the $gl(2|2)$ (or equivalently $u(2|2)$) irreducible representations (irreps) labelled by Young diagrams $\lam\in H(2|2)$ into $\scrA$-algebra representations. In terms of characters, we want to decompose the right-moving character $\tilde{S}_{\lam}$ into $\scrA$-algebra characters $\tilde{\chi}^{\scrA}_{\tj,\tj_2}$, defined in \eqref{eq:A_alg_chars}. To make this decomposition, it is convenient to use so-called Frobenius coordinates for the Young diagram $\lam$, \textit{i.e.}, $\lam = (a_1|b_1)$ for $d_\lambda=1$ and $\lam = (a_1,a_2|b_1,b_2)$ for $d_\lambda=2$, where $d_{\lam}$ is the number of hooks in the diagram $\lam$, or in other words the largest integer $s$ for which $\lam_s\geq s$ (see Figure~\ref{fig:frobenius_coords}\@). 
These coordinates satisfy $a_1,b_1\ge 0$ for $d_\lambda=1$, while
$a_1>a_2\ge 0$ and $b_1>b_2\ge 0$ for $d_\lambda=2$.\footnote{The Frobenius coordinates for $d_\lambda>2$ are defined similarly, although we only need $d_\lambda=1,2$ in this paper.}
We then find the following character decomposition:\footnote{This formula has been checked for a large range of cases, however, a derivation is currently not known to us.}
\begin{align} \label{eq.gl22_to_A_Def}
    \tilde{S}_{\lam} = \begin{cases}
        \ \tilde{\chi}^{\scrA}_{\frac{a_1}{2},\frac{b_1}{2}} + \tilde{\chi}^{\scrA}_{\frac{a_1-1}{2},\frac{b_1-1}{2}} & \text{if } d_{\lam} = 1\\[2ex]
        \ \tilde{\chi}^{\scrA}_{\frac{a_{12}-1}{2},\frac{b_{12}-2}{2}} + \tilde{\chi}^{\scrA}_{\frac{a_{12}-2}{2},\frac{b_{12}-1}{2}} + \tilde{\chi}^{\scrA}_{\frac{a_{12}}{2},\frac{b_{12}-1}{2}} + \tilde{\chi}^{\scrA}_{\frac{a_{12}-1}{2},\frac{b_{12}}{2}} & \text{if } d_{\lam} = 2
    \end{cases} \ ,
\end{align}
where $a_{12}=a_1-a_2$ and $b_{12}=b_1-b_2$ and $\tilde{\chi}^{\scrA}_{\tj,\tj_2}=0$ if either $\tj<0$ or $\tj_2<0$. As an example, schematically the $gl(2|2)$ irreps labelled by Young diagrams with up to $n_{\lam}=4$ boxes decompose into $\scrA$-algebra irreps (represented by diamond diagrams) as
\begin{equation} \label{eq.lam_to_diamonds_ex}
\begin{aligned}
\ytableausetup{boxsize=1.25ex}
    \ydiagram{1} &= \vtikz{\diam[0.2]{0}{0}} \ , \\
    \ydiagram{2} &= \vtikz{\diam[0.2]{-0.2}{0}\diam[0.2]{0.2}{0}} \ ,& 
    \ydiagram{1,1} &= \vtikz{\diam[0.2]{0}{0.2}\diam[0.2]{0}{-0.2}} \ ,&\\
    \ydiagram{3} &= \vtikz{\diam[0.2]{-0.4}{0}\diam[0.2]{0}{0}\diam[0.2]{0.4}{0}} \ ,& 
    \ydiagram{2,1} &= \vtikz{\diam[0.2]{-0.2}{0.2}\diam[0.2]{0.2}{0.2}\diam[0.2]{-0.2}{-0.2}\diam[0.2]{0.2}{-0.2}} + \vtikz{\diam[0.2]{0}{0}} \ ,& 
    \ydiagram{1,1,1} &= \vtikz{\diam[0.2]{0}{-0.4}\diam[0.2]{0}{0}\diam[0.2]{0}{0.4}} \ ,\\
    \ydiagram{4} &= \vtikz{\diam[0.2]{-0.6}{0}\diam[0.2]{-0.2}{0}\diam[0.2]{0.2}{0}\diam[0.2]{0.6}{0}} \ ,& 
    \ydiagram{3,1} &= \vtikz{\diam[0.2]{-0.4}{0.2}\diam[0.2]{0}{0.2}\diam[0.2]{0.4}{0.2}\diam[0.2]{-0.4}{-0.2}\diam[0.2]{0}{-0.2}\diam[0.2]{0.4}{-0.2}} + \vtikz{\diam[0.2]{-0.2}{0}\diam[0.2]{0.2}{0}} \ ,\\
    \ydiagram{2,2} &= \vtikz{\diam[0.2]{-0.2}{0}\diam[0.2]{0.2}{0}} + \vtikz{\diam[0.2]{0}{0.2}\diam[0.2]{0}{-0.2}}\ ,& 
    \ydiagram{2,1,1} &= \vtikz{\diam[0.2]{-0.2}{0.4}\diam[0.2]{-0.2}{0}\diam[0.2]{-0.2}{-0.4}\diam[0.2]{0.2}{0.4}\diam[0.2]{0.2}{0}\diam[0.2]{0.2}{-0.4}} + \vtikz{\diam[0.2]{0}{0.2}\diam[0.2]{0}{-0.2}} \ ,&
    \ydiagram{1,1,1,1} &= \vtikz{\diam[0.2]{0}{0.6}\diam[0.2]{0}{0.2}\diam[0.2]{0}{-0.2}\diam[0.2]{0}{-0.6}}\ .
\end{aligned}
\end{equation}
%

The Schur-Weyl form of the MEG \eqref{eq:MEG_as_H(2|2)_sum} therefore can be broken down into $\scrA$-irrep sectors as
\begin{align}
    \cE(p,q,y) = \sum_{\lambda\in H(2|2)} \sum_{(\tj,\tj_2)\subset\lambda} S_\lambda(p,q,y)\,\cD \tilde{\chi}^{\scrA}_{\tj,\tj_2} \ ,
\label{MEGdecompA}
\end{align}
where here by $\sum_{(\tj,\tj_2)\subset\lambda}$ we mean a summation over $\scrA$-representations contained in the $gl(2|2)$ representation labeled by $\lam$, as given by the decomposition \eqref{eq.gl22_to_A_Def}. According to the superselection sectors of different $\tj_2$ charges at the level of the $\scrA$-algebra character (or equivalently the number of rows of diamonds in the corresponding diamond diagram), we can define a ``resolved'' version of the MEG by summing only over terms in \eqref{MEGdecompA} with a particular value of $\tj_2$. This yields what we refer to as the \emph{resolved elliptic genus (REG)}:
\begin{align} \label{eq.REG1}
    \mathcal{E}_{\tj_2}(p,q,y) &\equiv \!\!\sum_{\substack{\lambda\in H(2|2)\\\tj_2\subset\lam}}\sum_{\tj_{\lam}^{}} S_\lambda(p,q,y)\, \cD \tilde{\chi}^{\scrA}_{\tj_{\lam}^{},\tj_2^{}}(\ty,\te) \notag\\
    &\,= \!\!\sum_{\substack{\lambda\in H(2|2)\\\tj_2\subset\lam}}\!\! S_\lambda(p,q,y) (-1)^{\tj_2}(2\tj_2+1) \sum_{\tj_{\lam}} (2\tj_{\lam}+1) \ ,
\end{align}
where by the condition ``$\tj_2\subset \lambda$'' we mean that the $gl(2|2)$-representation labeled by $\lam$ must contain $\scrA$-representations with $\widetilde{su}(2)_2$ charge $\tj_2$, and $\tj_\lambda$ runs over values of the $su(2)_R$ label $\tj$ that appear in the $gl(2|2)$ representation $\lambda$. In the second line of \eqref{eq.REG1} we used the result
\begin{equation} \label{eq.A_char_to_MEG}
    \mathcal{D} \tilde{\chi}^{\scrA}_{\tj,\tj_2} = \frac12 \big(\ty\pd_{\ty}\big)^2 \tilde{\chi}^{\scrA}_{\tj,\tj_2}(\ty,\te)\big|_{\ty=\te=1} = (-1)^{2\tj_2}(2\tj+1)(2\tj_2+1) \ .
\end{equation}
In other words, the REG \eqref{eq.REG1} is a sum of contributions of left-moving $\lam$-sectors weighted (up to a sign) by the number of diamonds in the contributing $\scrA$-irrep's diamond diagram.

Explicitly, the first terms in \eqref{eq.REG1} are
\begin{align} \label{eq:REG_exp}
\begin{split}
\YTscriptsize
\cE_{\tj_2=0}&=S_{\ydiagram{1}}
+2S_{\ydiagram{2}}
+(3S_{\ydiagram{3}}
+S_{\ydiagram{2,1}})
+(4S_{\ydiagram{4}}
+2S_{\ydiagram{3,1}}
+2S_{\ydiagram{2,2}})
+\cdots,\\
\cE_{\tj_2=\frac12}
&=-2S_{\ydiagram{1,1}}
-4S_{\ydiagram{2,1}}
-(6S_{\ydiagram{3,1}}
+2S_{\ydiagram{2,2}}
+2S_{\ydiagram{2,1,1}})
+\cdots,\\
\cE_{\tj_2=1}&=
3S_{\ydiagram{1,1,1}}
+6S_{\ydiagram{2,1,1}}
+\cdots,\\
\cE_{\tj_2=\frac32}&=
-4S_{\ydiagram{1,1,1,1}}+\cdots,\\[-2ex]
& \dots
\end{split}
\end{align}
where we showed only terms with up to four boxes. By summing the REG in \eqref{eq:REG_exp} we recover the expansion of the MEG \eqref{eq:MEG_exp} as expected and, in particular, the contributions from non-single-hook $\lam$-sectors such as $S_{\ydiagram{2,2}}$ cancel out.

Since $\cE_{\tj_2}(p,q,y)$ is a generating function for theories with different values of $N$, the REG for fixed $N$, which we denote by $\cE_{N,\tj_2}(q,y)$, can be found from
\begin{equation}
    \cE_{\tj_2}(p,q,y) = \sum_{N=1}^{\infty} p^N \cE_{N,\tj_2}(q,y) \quad,\quad \tj_2 = 0,\frac12,\dots,\frac{N-1}{2} \ .
\end{equation}

One may wonder if the REG is equivalent to a simple refinement of the MEG, obtained by introducing an additional fugacity for $\widetilde{su}(2)_2$ charge inside the index's trace \eqref{eq:MEG_T^4}. This is not the case.  In the MEG, a single short multiplet (or a diamond) makes a non-vanishing contribution as in \eqref{eq:MEGcontributionsShort}, while a long multiplet (or a garnet) contributes zero \eqref{eq:MEGcontributionsLong}, which is necessary for the MEG to be a protected supersymmetry index.
If we include $\etat^{2\Kt^3_{2,0}}$ in the MEG's trace \eqref{eq:MEG_T^4}, a short multiplet containing states with charges $(\mt-\half,\mt_2)$, $(\mt,\mt_2\pm \half)$, $(\mt+\half,\mt_2)$, namely a diamond centered at $(\mt,\mt_2)$, now contributes (here we focus just on the fugacities of the right-moving charges)
\begin{align}
 I_{\vtikz{\diam[0.1]{0}{-0.1}}}(\mt,\mt_2) &\equiv 
 \frac12 \tr_{\vtikz{\diam[0.1]{0}{-0.1}}}\!\Big[(2\Jt^3_0)^3(-1)^{2\Jt^3_0}\etat^{2\Kt^3_{2,0}}\Big]\notag\\
 &=\frac12  (-1)^{2\mt} 
 \Bigl[ (2\mt-1)^2 \etat^{2\mt_2}
 -(2\mt)^2  (\etat^{2\mt_2+1}+\etat^{2\mt_2-1})
 +(2\mt+1)^2  \etat^{2\mt_2}\Bigr]
 \notag\\
 &=\frac12 (-1)^{2\mt} \etat^{2\mt_2}
 \Bigl[8\mt^2+2
 -4\mt^2(\etat+\etat^{-1})\Bigr]\ .
\end{align}
However, a long multiplet (a garnet) made of four short multiplets contributes a total of
\begin{align}
 I_{\vtikz{\diam[0.1]{0}{-0.1}}}(\mt-\thalf,\mt_2) 
+2I_{\vtikz{\diam[0.1]{0}{-0.1}}}(\mt,\mt_2) 
+I_{\vtikz{\diam[0.1]{0}{-0.1}}}(\mt+\thalf,\mt_2)
&=
(-1)^{2\mt}\etat^{2\mt_2}
(\etat-\etat^{-1}-2)\ .
\end{align}
Since this does not vanish, a simple refinement by inserting $\etat^{2\Kt^3_{2,0}}$ in the MEG does not yield a protected index.

Although a simple fugacity for the $\widetilde{su}(2)_2$ charge does not lead to a protected index, we can define the following generating function for the protected REG:
\begin{align}
    \cE(p,q,y,\etatb) 
&\equiv\sum_{\tj_2}\mathcal{E}_{\tj_2}(p,q,y)\frac{\etatb^{2\tj_2+1}+\etatb^{-2\tj_2-1}}{2} \ ,
\end{align}
where $\etatb$ is not a fugacity for the $\widetilde{su}(2)_2$ charge, but is a bookkeeping parameter that records the $\tj_2$ label of the $\scrA$-multiplets.  Having
both $\etatb^{2\tj_2+1}$ and $\etatb^{-2\tj_2-1}$ is for mathematical convenience which makes $\cE(p,q,y,\etatb)$ invariant under $\etatb\to \etatb^{-1}$. Thus, in a sense, we are inserting a fugacity at the level of $\scrA$-multiplets, not at the level of individual states.
We can derive a closed-form expression for this generating function
(see Appendix \ref{app:genFuncREG}):
\begin{align}
    \cE(p,q,y,\etatb) 
&={1\over (\etatb^{\half}-\etatb^{-\half})^2}
 \left(1-{\etatb-\etatb^{-1}\over 2}\etatb\partial_{\etatb}\right)\cZ(p,q,y,1,\etatb)
\notag\\
&={\cZ(p,q,y,1,\etatb)\over (\etatb^{\half}-\etatb^{-\half})^2}
\left[
 1+\half(\etatb-\etatb^{-1})^2
 \sum_{k,r,\ell}{c(k,r,\ell)\, p^k q^r y^{\ell}
 \over (1-p^k q^r y^{\ell} \etatb)(1-p^k q^r y^{\ell} \etatb^{-1})}
\right],
\label{genFuncREG}
\end{align}
where the $c(k,r,\ell)$ are defined from the seed-theory partition function in \eqref{eq:seed_z(q,y)}, and satisfies \eqref{eq:c(k,r,l)=c(kr,l)}  and \eqref{eq:c(r,l)=c(4r-l^2)}, while
\begin{align}
 \cZ(p,q,y,\yt,\etat)
&= \sum_N p^N
\Tr_N \left[(-1)^F q^{L_0-{c\over 24}}y^{2J^3_0}\tilde{q}^{\tilde{L}_0-{c\over 24}}\yt^{2\Jt^3_0}\etat^{2\Kt^3_{2,0}}\right]\notag\\
&= \prod_{k,r,\ell}\left[{(1-p^k q^r y^{\ell} \etat)(1-p^k q^r y^{\ell} \etat^{-1})
\over (1-p^k q^r y^{\ell}\yt)(1-p^k q^r y^{\ell}\yt^{-1})}\right]^{c(k,r,\ell)} \ ,
\end{align}
is a refinement of the grand partition function \eqref{eq:Z_T4_DMVV-form} by the additional fugacity $\etat$ (this quantity is not protected).
In the $\etatb \to 1$ limit, \eqref{genFuncREG} reduces to the MEG \eqref{eq.MMS_generating} (up to some irrelevant divergent terms at order $p^0$) as we show in Appendix \ref{app:genFuncREG}\@.
The formula \eqref{genFuncREG} allows us to find the REG without using the Schur-Weyl expansion.

\subsection{Matching}

Given the REG defined in the previous section for the symmetric orbifold CFT of $T^4$, one can also apply it to the space of states dual to supergravitons in AdS$_3\times S^3$ (or for large enough conformal dimensions these states will be dual  to superstrata backgrounds \cite{Shigemori:2020yuo}). Supergraviton states are naturally defined in the Neveu-Schwarz (NS) sector of the CFT and their left-moving single-strand Hilbert space is a restriction of the full NS Hilbert space to
\begin{equation}
    V_{\mathrm{SG}} = \mathrm{span}\Big\{\mathcal{O}_g\ket{\alpha}_k\ ,\ \mathcal{O}_g\ket*{\dot{A}}_k\Big\}_{k,\mathcal{O}_g} \ ,
\end{equation}    
where the $\mathcal{O}_g$ represent single-strand modes of the $su(1,1\,|\,2)$ global (anomaly-free) subalgebra of the full $\mathcal{N}=4$ symmetry algebra \cite{Maldacena:1999bp}, generated by $L_{-1},G^{- A}_{-\half},J^-_0$. Our conventions for spectral flow are given in \eqref{eq.symmModeDeform} and \eqref{eq.SF_hm}.

For convenience we choose to consider mixed NS-R sector, with the right-moving single-strand Hilbert space $\tilde{V}$ for supergraviton states being the same as the CFT's Ramond ground-state space
\footnote{The mixed NS-R sector that we consider here allows both an easy comparison to the supergraviton spectrum as well as preserving the simplicity of the right-moving characters in the R sector that we took advantage of in Section~\ref{sec:S-W}.}.

The REG generating function for supergraviton states is again given by \eqref{eq.REG1}, but now $S_{\lambda}(p,q,y)$ is the Schur function for the single-strand supergraviton character
\begin{align} \label{eq.SGchar}
    z_{\mathrm{SG}}^{}(p,q,y) &= \sum_{k}p^k\Big[\phi^{(s)}_{\frac{k-1}{2}}(q,y) -2\phi^{(s)}_{\frac{k}{2}}(q,y) + \phi^{(s)}_{\frac{k+1}{2}}(q,y)\Big] \notag\\
    &\equiv \sum_{k,r,\ell} c_{\mathrm{SG}}^{}(k,r,\ell)\, p^kq^ry^{\ell}\ ,
\end{align}
where $\phi^{(s)}_{j}$ are characters of short representations of $SU(1,1\,|\,2)$ as given in \eqref{eq.su112short}. Equivalently, one can use the REG generating function \eqref{genFuncREG} with the $c(k,r,\ell)$ replaced by the coefficients $c_{\mathrm{SG}}^{}(k,r,\ell)$. 

The supergraviton spectrum was compared with the full CFT spectrum using the MEG \eqref{eq:MEG_T^4} in  \cite{Maldacena:1999bp} and agreement was found below the black-hole threshold, \textit{i.e.}\ for
\begin{equation}
    h < h_{\rm BH} = \frac{N}{4} \ .
\end{equation}
This agreement, however, was rather empty since both the CFT and supergraviton MEGs in fact vanish below this threshold, apart from the global NS vacuum contribution at order $q^0y^0$. By instead using the REG \eqref{eq.REG1} we see that this ``$0=0$'' statement turns into a meaningful comparison. By resolving each side of the comparison into a sum of non-vanishing $\tj_2$-sector contributions, $\cE_{\tj_2}$, with $\tj_2 = 0,\frac12,\dots,\frac{N-1}{2}$. 
To make this comparison we have to spectral flow the CFT left-moving character \eqref{eq:1-strand_z(p,q,y)} from the R-sector to the NS-sector via~\cite{Schwimmer:1986mf}
\begin{equation} \label{eq:z_R_to_NS}
    z_{\rm NS}^{}(p,q,y)=z_{\rm  R}^{}\big(pq^{\frac12}y,q,yq^{\frac12}\big)
    =\sum_{k,r,\ell} c(4kr-k^2-\ell^2)\,p^k q^r y^\ell
    \ .
\end{equation}

In the case of $N=6$, for example, the black hole bound is $h_{\rm BH} = \frac32$ and the CFT and supergraviton REGs have the $q$-expansions\footnote{We acknowledge the use of the GAP system for computational discrete algebra \cite{GAP4}, which we used for the computation of symmetric group characters.}

\begin{align} \label{eq:REG6}
    \cE^{\rm CFT}_{6,0}&= 6 + \sqrt{q} (-12y-\cdots) + q(6y^2 +\textcolor{blue}{54}+\cdots) + q^{\frac32}(3y^3-95y-\cdots)+\cdots\notag\\
    \cE^ {\rm SG}_{6,0}&= 6 + \sqrt{q} (-12y-\cdots) + q(6y^2 +\textcolor{blue}{54}+\cdots) + q^{\frac32}(3y^3-92y-\cdots)+\cdots\notag\\[1ex]
    \cE^{\rm CFT}_{6,\frac12}&= \sqrt{q}(12y + \cdots) + q(-24y^2\textcolor{blue}{-72}-\cdots) + q^{\frac32}(12y^3+180y+\cdots) + \cdots\notag\\
    \cE^ {\rm SG}_{6,\frac12}&= \sqrt{q}(12y + \cdots) + q(-24y^2\textcolor{blue}{-72}-\cdots) + q^{\frac32}(12y^3+176y+\cdots)+\cdots\notag\\[1ex]
    \cE^{\rm CFT}_{6,1}&= q(18y^2+\textcolor{blue}{18}+\cdots) + q^{\frac32}(-36y^3\textcolor{red}{-108}y-\cdots) + q^2(18y^4+252y^2+342+\cdots)+\cdots\notag\\
    \cE^ {\rm SG}_{6,1}&= q(18y^2+\textcolor{blue}{18}+\cdots) + q^{\frac32}(-36y^3\textcolor{red}{-108}y-\cdots) + q^2(18y^4+264y^2+378+\cdots)+\cdots \notag\\[1ex]
    \cE^{\rm CFT}_{6,\frac32}&= \textcolor{blue}{0}q + q^{\frac32}(24y^3+\textcolor{red}{24}y+\cdots) + q^2(-48y^4-160y^2-176-\cdots)+\cdots\notag\\
    \cE^ {\rm SG}_{6,\frac32}&= \textcolor{blue}{0}q + q^{\frac32}(24y^3+\textcolor{red}{24}y+\cdots) + q^2(-48y^4-144y^2-132-\cdots)+\cdots\notag\\[1ex]
    \cE^{\rm CFT}_{6,2}&= \textcolor{blue}{0}q + \textcolor{red}{0}q^{\frac32} + q^2(30y^4+\textcolor{red}{30}y^2+\textcolor{red}{30}+\cdots) + q^{\frac52}(-60y^5-120y^3-45y-\cdots)+\cdots\notag\\
    \cE^ {\rm SG}_{6,2}&= \textcolor{blue}{0}q + \textcolor{red}{0}q^{\frac32} + q^2(30y^4+\textcolor{red}{30}y^2+\textcolor{red}{30}+\cdots) + q^{\frac52}(-60y^5-205y^3-210y-\cdots)+\cdots \notag\\[1ex]
    \cE^{\rm CFT}_{6,\frac52}&= \textcolor{blue}{0}q + \textcolor{red}{0}q^{\frac32} +\textcolor{red}{0}q^{2} + q^{\frac52}(36y^5+\textcolor{red}{60}y^3+\textcolor{red}{72}y+\cdots)
    \notag\\
    &\hspace{30ex}
    +q^3(-72y^6-360y^3-558y^2-624-\cdots)+\cdots\notag\\
    \cE^ {\rm SG}_{6,\frac52} &= \textcolor{blue}{0}q + \textcolor{red}{0}q^{\frac32} +\textcolor{red}{0}q^{2} + q^{\frac52}(36y^5+\textcolor{red}{60}y^3+\textcolor{red}{72}y+\cdots)
    \notag\\
    &\hspace{30ex}
    +q^3(-72y^6-270y^3-414y^2-462-\cdots)+\cdots\notag\\[1ex]
\end{align}
where we see not only detailed non-trivial matching up to the expected order $O(q^{1})$ (terms in \textcolor{blue}{blue}), but also an enhanced matching (terms in \textcolor{red}{red}) for the $\tj_2=1,\frac32,2,\frac52$ sectors. We have explicitly checked that these features (matching below the threshold and enhanced matching in certain $\tj_2$-sectors) of the REG hold up to $N=12$. This matching with the supergraviton spectrum provides evidence that the REG are protected, \textit{i.e.}\ that lifted states cancel from the REG separately in each $\tj_2$-sector.

\begin{figure}[h]
\begin{adjustbox}{center}
    \begin{subfigure}[h]{0.67\linewidth}
        \includegraphics[height=7.5cm,trim={0.5cm 0 0 0},clip]{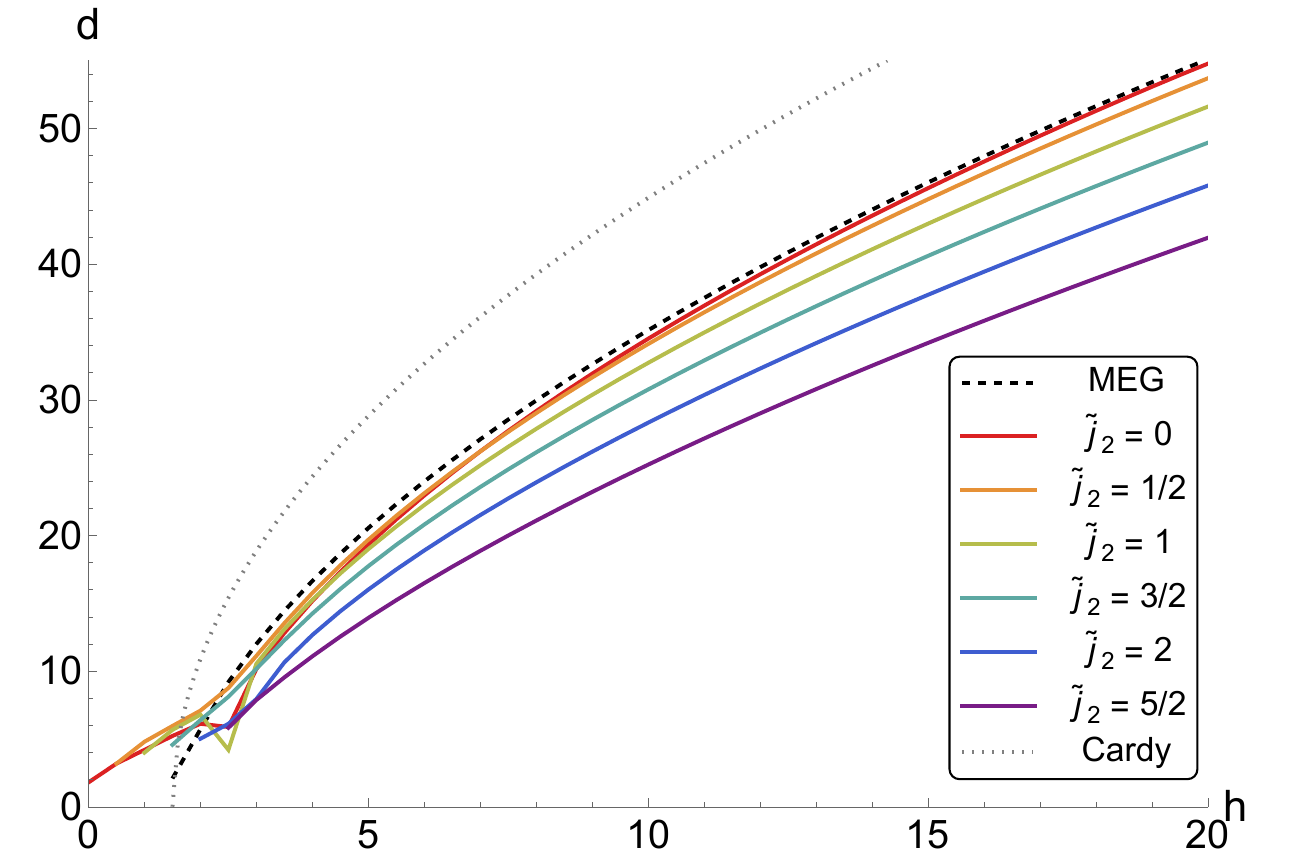}
        \caption{\label{fig:REG1a}}
    \end{subfigure}
    \begin{subfigure}[h]{0.33\linewidth}
        \includegraphics[width=\linewidth]{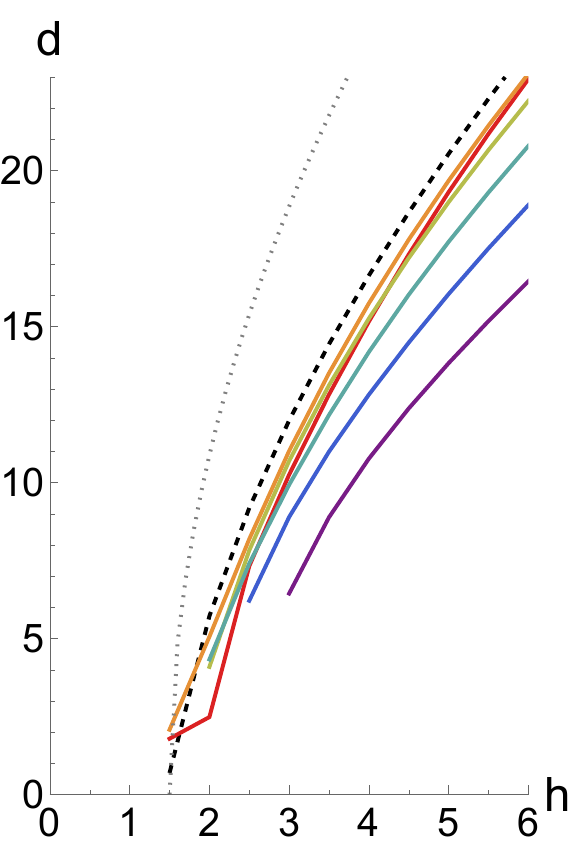}
        \caption{\label{fig:REG1b}}
    \end{subfigure}
\end{adjustbox}
\caption{\raggedright\sl Plots of the logarithmic degeneracies \eqref{eq:log_degens_def} from the REG $d^{\mathrm{CFT}}_{N=6,\tj_2}$ in (a) and $d^{\mathrm{BH}}_{N=6,\tj_2}$ in (b). For comparison we display the analogous quantities obtained from the MEG, as well as the universal Cardy growth. The first contributing states to sectors of the REG with larger $\tj_2$ are at higher $h$, as discussed in Section~\ref{ssec:Aspects_of_REG}.\label{fig:REG1}}
\end{figure}

In order to study the behavior of the REG well above the black-hole threshold, we consider the logarithmic degeneracies of the CFT REG (working here in the NS sector using \eqref{eq:z_R_to_NS}) as well as of the difference between the CFT and supergraviton REGs which we define as
\begin{subequations} \label{eq:log_degens_def}
    \begin{align}
        d^{\mathrm{CFT}}_{N,\tj_2}(h) &\equiv \log\abs{\rule{0pt}{2.5ex}
        \mathcal{E}_{N,\tj_2}^{\mathrm{CFT}}(q,1)|_{q^h}} \ ,\\
        d^{\mathrm{BH}}_{N,\tj_2}(h) &\equiv \log\abs{\rule{0pt}{2.5ex}
        \mathcal{E}_{N,\tj_2}^{\mathrm{CFT}}(q,1)|_{q^h}-\mathcal{E}_{N,\tj_2}^{\mathrm{SG}}(q,1)|_{q^h}} \ ,
    \end{align}
\end{subequations}
where setting $y=1$ has the effect of summing over values of $J^3_0$. From these logarithmic degeneracies, which we show for the case of $N=6$ in Figure~\ref{fig:REG1}, we observe that the leading-order growth of states in each $\tj_2$-sector of the REG appears to be the same as the full MEG in the regime $h\gg N$, namely Cardy growth \cite{Cardy:1986ie}. Due to $\tj_2$-sectors containing contributions from a mixture of different twist sectors, we are motivated to conclude that black hole states are distributed among twist sectors. The growths of the REG at the subleading level should, however, depend on $\tj_2$. In the next section we briefly consider what the more refined structure of the REG tells us about the regime $h\sim N\gg1$, in which the ``long-string sector'' (the sector labelled by $\lam=\YTnormalsize\ydiagram{1}$ and contained within the $\tj_2=0$ sector of the REG) is believed to be dominant~\cite{Maldacena:1996ds,Das:1996wn}.

\subsection{Comments on the REG}
\label{ssec:Aspects_of_REG}

\subsubsection*{First non-vanishing terms in the $q$-expansion}

In the $q$-expansions of REGs (in the NS sector) for $N=6$ presented in \eqref{eq:REG6}, we see that the $q$-expansion does not generally start at $q^0$, being zero up to some order in $q$.  In fact, $\cE_{N,\tj_2}(q,y)$ for both CFT and supergravity starts at $q^{\tj_2}$, where they agree, irrespective of~$N$.

This phenomenon has a simple origin. Eq.~\eqref{eq.gl22_to_A_Def} implies that $\cE_{N,\jt_2}$, counting states in $\scrA$-representations with $\widetilde{su}(2)_2$ spin $\jt_2$, comes from $\lambda$-sectors whose number of rows is at least $2\jt_2+1$.  Because $\lambda$ also labels the $S_n$-representation, if all strands were bosonic, this would mean that we need at least $2\jt_2+1$ strands in different states in order for the column antisymmetrization imposed by the Young symmetrizer not to kill the total state.  However, in the present case, we have left-moving fermionic states such as $\ket*{\Ad}_1$ and $J_0^-\ket*{\Ad}_1 \propto \psi^{-\Ad}_{-\half}\ket*{-}_1$.  Any number of left-moving strands can be put in these states while being antisymmetric under $S_n$. Excluding $\ket{-}_1$, these are states with the smallest possible conformal weight, $h=\half$, and are common to CFT and supergravity.  Therefore, these states are responsible for the first terms in the $q$-expansions of the REG sectors.

As an example, consider the REG sector $\cE_{6,\frac52}=6S_\lambda$ with $\lambda=\YTscriptsize\ydiagram{1,1,1,1,1,1}\,$, whose $q$-expansion is given in \eqref{eq:REG6}. Because of the column antisymmetrization, the state with the lowest conformal weight $h$ appears at $q^{\frac52}$ and its left-moving part is
$\ket{-}_1\ket*{\Ad_1}_1
 \cdots
 \ket*{\Ad_5}_1$, 
where $\Ad_1,\dots,\Ad_5=\dot{1},\dot{2}$. Total antisymmetry under permutation of the fermionic strands implies that the $\Ad_i$ indices must be symmetrized under exchange.  This gives a contribution of $-6$, where the minus sign is because the overall state is fermionic. Multiplied by the right-moving contribution $\cD \tilde{S}_\lambda=-6$, we obtain $36$ in the last two lines of \eqref{eq:REG6}. 
Note that the states involved  here are supergraviton states, which explains the agreement between CFT and supergravity REGs at this level.  

The leading terms in other $\jt_2$ sectors can be explained using similar arguments, by including $\ket{-}_2$ and other states when the number of strands is less than $N$.

\subsubsection*{Caveats about the identical-strand and long-string pictures}

Below \eqref{eq:MEG_id_strand_form}, we noted that the MEG can be written as a sum of contributions from ``identical-strand states'' in which all strands have the same length and internal state. Based on this one might think that, upon turning on the deformation, only identical-strand states of the symmetric orbifold theory survive and all other states lift.  However, the REG says that this picture is naive.

To demonstrate this, let us consider the CFT MEG in the NS sector for the case of $N=2$: from \eqref{eq:MEG_as_hook_sum} and using the spectral flow \eqref{eq:z_R_to_NS}
\begin{align}
\YTscriptsize
  \cE_{N=2}(q,y)\big|_{q^4} =
    (S_{\ydiagram{1}}+
    2S_{\ydiagram{2}}- 2S_{\ydiagram{1,1}})\big|_{q^4} =
   - 192y^4 - 5616y^2 - 14514 - \cdots\ ,
   \label{eq:MEG_N=2_q6}
\end{align}
where to simplify the presentation, we are focusing on the $q^4$ term\footnote{We chose $q^4$ so that we are well above the black-hole threshold $h_{\rm BH}=N/4$, where the MEG is expected to be dominated by black-hole states.} in the $q$ expansion.  We are also suppressing the terms with negative powers of $y$ because of the $y\to y^{-1}$ symmetry.  The individual contributions in \eqref{eq:MEG_N=2_q6} are
\begin{subequations}
\begin{align}
 S_{\ydiagram{1}}   |_{q^4}&=-208y^4 - 5616y^2 - 14592-\cdots\ ,\label{eq:S1}\\
2S_{\ydiagram{2}}   |_{q^4}&=+248y^4 + 4128y^2 + \phantom{0}9654+\cdots\ ,\label{eq:S2}\\
-2S_{\ydiagram{1,1}}|_{q^4}&=-232y^4 - 4128y^2 - \phantom{0}9576-\cdots\ .\label{eq:S11}
\end{align}
\end{subequations}
First, let us rewrite the MEG \eqref{eq:MEG_N=2_q6} as the sum of REGs
\begin{align}
    \cE_{N=2}\big|_{q^4}
    =
    \underbrace{(S_{\ydiagram{1}}+2S_{\ydiagram{2}})\big|_{q^4}}
    _{\cE_{2,0}:\, 40y^4 - 1488y^2 - 4938 -\cdots}
    ~~+~~
    \underbrace{(-2S_{\ydiagram{1,1}})\big|_{q^4}}
    _{\cE_{2,\frac12}:\, -232y^4-4128y^2-9576- \cdots},
    \label{eq:MEG_as_sum_REG_N=2}
\end{align}
where the first and second terms (respectively $\cE_{2,0}$ and $\cE_{2,\frac12}$) are separately protected.  Within the first term, $\cE_{2,0}$,  we see that contributions \eqref{eq:S1} and \eqref{eq:S2} partly cancel, indicating the combining and lifting of states.  Indeed, in \cite{Guo:2020gxm}, it was shown that for $N=2$, all states that are not affine descendants of 1/2-BPS states in the $\YTnormalsize\ydiagram{2}$ sector combine and lift with states in the $\ydiagram{1}$ sector up to $h=4$.  The second term, $\cE_{2,\frac12}$, is protected on its own (as was also shown in \cite{Guo:2020gxm}). Note that the two terms contribute to the MEG at a comparable order of magnitude.\footnote{We are not very quantitative here in comparing the magnitudes of terms, but the point is that the difference in the magnitude between the two terms in \eqref{eq:MEG_as_sum_REG_N=2} is not as much as that in \eqref{eq:MEG_as_id_str_N=2}.\label{ftnt:comparable_order}}

Alternatively, we can write the MEG \eqref{eq:MEG_N=2_q6} in a way that makes manifest the identical-strand picture:
\begin{align}
\YTscriptsize
    \cE_{N=2}\big|_{q^4}
    =
    \underbrace{~~~~(S_{\ydiagram{1}})\big|_{q^4}~~~~}
    _{- 208y^4 - 5616y^2 - 14592 -\cdots}
    \!\!+\ \underbrace{(2S_{\ydiagram{2}}-2S_{\ydiagram{1,1}})\big|_{q^4}}
    _{16y^4+0y^2+78+    \cdots} \ ,
    \label{eq:MEG_as_id_str_N=2}
\end{align}
which groups contributions by the twist sector. The first term is the contribution from one strand of length two and, by \eqref{eq:MEG_id_strand_form}, we can interpret the second term as the contribution from two identical strands.  Within the second term we see an almost perfect cancellation between \eqref{eq:S2} and \eqref{eq:S11}, making it much smaller than the first term. Based on this, one might naively conclude that there is substantial combining and lifting happening between the $\YTnormalsize\ydiagram{2}$ and $\ydiagram{1,1}$ sectors. This is purely illusory, however, since these $\lam$-sectors belong to different superselection sectors and no such mixing can occur.  So, the massive cancellations within the second term is simply due to the definition of the MEG and is therefore kinematical rather than dynamical.

Note also that the first term in \eqref{eq:MEG_as_id_str_N=2} is the contribution from the so-called ``long-string'' or maximally-twisted sector
\cite{Maldacena:1996ds,David:2002wn}, in which all copies are glued into one strand (or string).  From \eqref{eq:MEG_as_id_str_N=2} it may appears that the long-string sector (the $\ydiagram{1}$ sector) completely dominates the MEG, but that is misleading because many states in the $\ydiagram{1}$ sector combine with states in the $\ydiagram{2}$ sector and lift. Indeed, we saw in \eqref{eq:MEG_as_sum_REG_N=2} that the non-long-string sector with $\YTnormalsize\ydiagram{1,1}$ contributes at a comparable order of magnitude.\footnote{Although comparable, numerical results suggest that the long-string sector contribution will dominate over the non-long-string sector contribution in the large $h\gg N$ limit.    However, as mentioned in footnote \ref{ftnt:comparable_order}, this is irrelevant to the point that we are making here.}

In \eqref{eq:MEG_as_sum_REG_N=2} for $N=2$, as well as in all other cases we have checked, we observe that for quantum numbers sufficiently inside the black hole regime -- that is, well within the parabola $h=\frac{m^2}{N}+\frac{N}{4}$ \cite{Breckenridge:1996is} -- the coefficients of $q^hy^{2m}$ for fixed $h$ and $m$ which appear in each $\tj_2$-sector of the REG all have the same sign. This is also visible from the different terms in \eqref{eq:MEG_as_sum_REG_N=4}. While not proof of anything, we would argue that it does lend credibility to the idea that the states contributing to the REG remain BPS finitely away from the free point. 
If different REG sectors were to come in different signs, this would suggest the possiblity of combining and lifting between those sectors and the protection of the REG could be called into question.

All of these comments and observations persist for larger values of $N$. For example, for $N=4$ and focusing on the $q^8$ term, the MEG is
\begin{align}
    \cE_{N=4}(q,y=1)\big|_{q^8}=-44\,112\,587\,940 \ ,
\end{align}
where we set $y=1$ so that all values of $J^3_0$ are summed over.  The MEG decomposed into REG contributions, using \eqref{eq.REG1}, is
\begin{align}
    \label{eq:MEG_as_sum_REG_N=4}
\begin{split}
\YTscriptsize
\cE_{N=4}(q,y=1)\big|_{q^8}
&=
\underbrace{(S_{\ydiagram{1}}
+2S_{\ydiagram{2}}
+3S_{\ydiagram{3}}
+S_{\ydiagram{2,1}}
+4S_{\ydiagram{4}}
+2S_{\ydiagram{3,1}}
+2S_{\ydiagram{2,2}})\big|_{q^8, y=1}}_{\cE_{4,0}:\, -23\,956\,050\,240}
\\
&\quad
+\underbrace{(-2S_{\ydiagram{1,1}}
-4S_{\ydiagram{2,1}}
-6S_{\ydiagram{3,1}}
-2S_{\ydiagram{2,2}}
-2S_{\ydiagram{2,1,1}})\big|_{q^8,y=1}}_{\cE_{4,\frac12}:\, -15\,895\,262\,740}
\\
&\quad
+\underbrace{(3S_{\ydiagram{1,1,1}}
+6S_{\ydiagram{2,1,1}})\big|_{q^8,y=1}}_{\cE_{4,1}:\, -3\,802\,922\,148}
+\underbrace{(-4S_{\ydiagram{1,1,1,1}})\big|_{q^8,y=1}}_{\cE_{4,\frac32}:\, -458\,352\,812} \ ,
\end{split}
\end{align}
where the numbers under underbraces are the coefficients of $q^8$.
We see that different REG sectors contribute at a comparable\footnote{See footnote \ref{ftnt:comparable_order} again.} order of magnitude.
On the other hand, in the twist-sector decomposition, which manifests the identical-strand structure, the MEG is
\begin{align}
\begin{split}
\YTscriptsize
\cE_{N=4}(q,y=1)\big|_{q^8}
&=
\underbrace{~~(S_{\ydiagram{1}})\big|_{q^8,y=1}~~}_{-44\,112\,535\,680}
+\underbrace{2(S_{\ydiagram{2}}-S_{\ydiagram{1,1}})\big|_{q^8,y=1}}_{-52\,480}
+\underbrace{3(S_{\ydiagram{3}}-S_{\ydiagram{2,1}}+S_{\ydiagram{1,1,1}})\big|_{q^8,y=1}}_{0} \\
&\quad+\underbrace{4(S_{\ydiagram{4}}-S_{\ydiagram{3,1}}+S_{\ydiagram{2,1,1}}-S_{\ydiagram{1,1,1,1}})\big|_{q^8,y=1}}_{220}\ .
\end{split}
\end{align}
We again see an apparent dominance of the long-string sector $\YTnormalsize\ydiagram{1}\,$; the non-long-string sectors' contributions are many orders of magnitude smaller (the third term contributes zero because we cannot divide the total length $N=4$ into three equal-length strands).
However, again, this grouping of sectors is misleading because it does not take into account how combining and lifting actually occurs; namely that it does not occur within twist sectors.

\subsubsection*{Caveats to the protection of the REG}

Since the arguments made in Section \ref{ssec:superselection} about the nature of lifting and superselection sectors were made with reference to the second-order deformation of the symmetric orbifold theory by the twisted-sector modulus \eqref{eq:deformation_op_def}, the second-order BPS spectrum needs to be exact for the REG to be a fully protected index. 
While this has been long thought to be true in the D1-D5 CFT, as well as in a wide number of other theories, recent work~\cite{Chang:2025mqp,Choi:2025bhi} casts doubt on this for $\mathcal{N}=4$ super Yang-Mills, where higher-order effects invalidate a similar extrapolation.

We also note that the representations of the $\widetilde{su}(2)_2\subset\scrA$ algebra which we used in defining superselection sectors for the first-order deformed supercharge is only natural from the free orbifold point. Away from the free point, the $\widetilde{su}(2)_2$ charge is not a good quantum number; the currents for $\widetilde{su}(2)_2$ defined at the free point are known to get lifted once deformed \cite{Benjamin:2021zkn,Guo:2019ady}.
Therefore, it is likely that at a generic point in moduli space
the division of states into REG sectors by the values $\jt_2$ is not a very useful labeling.\footnote{Note, however, that in the K3 theory the $SU(2)_1$ outer-automorphism bundle over moduli space can be studied using the $tt^*$ equations \cite{deBoer:2008ss, deBoer:2008qe}.  It would be interesting to investigate whether the same method can be generalized to the $T^4$ case to study the $\widetilde{SU}(2)_1\times \widetilde{SU}(2)_2$ outer-automorphism bundle and track the evolution of the REG sectors as we deform away from the free orbifold point.
}
Despite this, the REG does provide a meaningful separation of the free orbifold's BPS spectrum, one that is consistent with second-order lifting and demonstrates detailed matching with the bulk Kaluza-Klein spectrum.

\section{Discussion and outlook}
\label{sec:summary}

In this follow-up work to \cite{Hughes:2025oxu}, we have presented the resolved elliptic genus (REG); a modification of the standard supersymmetry index of the D1-D5 CFT for $T^4$. This is based on superselection sectors of the free-theory BPS spectrum under the action of the first-order deformed supersymmetry generators. The REG finds a natural description using the Schur-Weyl formalism, developed in Section~\ref{sec:S-W}, which decomposes the free-theory BPS Hilbert space into left- and right-moving sectors transforming covariantly under the symmetric group acting on the strand structure.

We provided two expressions for the generating function of REG: one \eqref{eq.REG1} in terms of a sum of Schur functions, over Young diagrams satisfying a hook condition; the other \eqref{genFuncREG} a DMVV-style product formula, with the only input being the seed theory partition function expansion coefficients \eqref{eq:seed_z(q,y)}. Using the REG, we demonstrated a detailed agreement between the full CFT Hilbert space and the supergraviton sector for conformal dimensions below the black hole bound, \textit{i.e.}\ for $h<\frac{N}{4}$. This constitutes a dramatic improvement over the previously observed agreement made using the modified elliptic genus \cite{Maldacena:1999bp}, whose only non-zero contribution below the black hole bound is from the global NS vacuum.

\bigskip
In Section \ref{ss:sym_free_thy}, we stated that turning on the deformation breaks the $u(2|2)$ symmetry algebra of the free theory to the $\scrA$-algebra.  Let us now explain this in more detail.
First of all, $cl_4^\psit,su(2)_R\subset \scrA$ are symmetry algebras of the deformed theory and $\widetilde{su}(2)_2\subset\scrA$ is respected by the Gava-Narain operators.  However, other generators of $u(2|2)$ are not part of the symmetry algebra (see Appendix \ref{ss:gl(2|2)_and_A-alg}) and on general grounds are expected to be broken by the deformation.  Furthermore, in fact, one can directly show that the generators of $u(2|2)$ that are not part of $\scrA$ do not commute with the Gava-Narain operator \cite{Guo:private}.
We can also see the breaking $u(2|2)\to \scrA$ in a somewhat more indirect way as follows.  If the full $u(2|2)$ were a symmetry, all states in a given $\lambda$-sector of the Schur-Weyl decomposition would have the same lift, because $\lambda$ labels irreducible $u(2|2)$ representations. Then the MEG would receive nonzero contribution only from hook diagrams, as we saw in Section \ref{ssec:T4_MEG}.  Then, based on \eqref{eq:GL22MEGcontribution}, we would be able to prove that the Gava-Narain operator $\cGt$ links only $\lambda$-sectors with the same number of rows, $\rho_\lambda$ (``$\rho$-protection'').  However, one can explicitly show that $\cGt$ links $\lambda=\ydiagram{2,1}$ and $\ydiagram{2}$, contradicting with the $\rho$-protection.\footnote{We thank Bin Guo for communication on this matter.}  This means that states in a given $\lambda$-sector can have different lifting, implying that $u(2|2)$ is broken.

If this $\rho$-protection were true, $\lambda$-sectors with the same value of $\rho_\lambda$ would form a superselection sector, and summing the Schur-Weyl expansion of the MEG \eqref{eq:MEG_as_hook_sum} only over such a $\rho$-sector would give a protected index, which one might call the ``$\rho$-sector REG''\@.  Because $u(2|2)$ is broken, one expects that such a $\rho$-sector REG is not protected.  Nevertheless, surprisingly, numerical computations show that the $\rho$-sector REG agrees between the CFT and supergravity below the black-hole threshold, just as the $\widetilde{su}(2)_2$-sector REG does in \eqref{eq:REG6}.  We find this both intriguing and puzzling, but leave a fuller understanding for future investigation.

One technical matter to be better understood is in relation to the explicit right-moving Ramond ground states in a given $\lam$-sector. With the currently presented method of Young symmetrizers in Appendix~\ref{app:S-W++}, one does not exactly obtain the physical right-moving states that fill $\psit$ Clifford quartets. This can be seen from the simple example of the $\lam=\YTnormalsize \ydiagram{1,1}$ sector with general strand lengths $k_1,k_2$. The physical state with $\mt=\mt_2=\frac12$ can be obtained from the highest-weight state of the representation, $\ket*{\dot{1}}^{[1]}_{k_1}\ket*{\dot{1}}^{[2]}_{k_2}$, by the action of the total mode $\psit^{\dot{+}\dot{2}}_0$ as $(\sqrt{k_2}\ket*{\dot{1}}^{[1]}_{k_1}\ket*{\dot{+}}^{[2]}_{k_2} - \sqrt{k_1}\ket*{\dot{+}}^{[1]}_{k_1}\ket*{\dot{1}}^{[2]}_{k_2})$. By comparison, the state obtained from the appropriate Young symmetrizer (given in \eqref{eq.YS_N=2_2}) is $(\ket*{\dot{1}}^{[1]}_{k_1}\ket*{\dot{+}}^{[2]}_{k_2} - \ket*{\dot{+}}^{[1]}_{k_1}\ket*{\dot{1}}^{[2]}_{k_2})$, with no factors of $\sqrt{k_{1,2}}$. At the level of state counting, which is the main goal of this paper, this difference is not relevant. We nevertheless hope to understand this detail better in the near future.

Due to the D1-D5 CFT's central role in the study of black-hole microstate physics, the REG has the possibility of providing a new handle on longstanding and actively pursued problems, including: the lifting problem \cite{Hampton:2018ygz,Benjamin:2021zkn,Guo:2019pzk,Gaberdiel:2024nge,Guo:2022ifr}, BPS chaos \cite{Chen:2024oqv}, the holographic dictionary for multi-center configurations \cite{Dabholkar:2009dq, Bena:2011zw, Denef:2007vg}, fortuity~\cite{Chang:2024lxt},
and the relation to the generalized supergravity index \cite{Hughes:2025tdy}. We hope to investigate some of these applications in the future, and conclude by considering some of these future directions.

The Schur-Weyl formalism developed here for the symmetric orbifold of $T^4$ should, with minimal modifications, be applicable to general symmetric orbifold theories -- for studies of spaces of such theories and their relevance to holography, see \textit{e.g.}~\cite{Belin:2019rba}. In particular, the formalism should be applicable to the D1-D5 CFT on $K3$. While the Hilbert space of $\mathrm{Sym}^N(K3)$, defined using the torus orbifold description of $K3$, does not completely factorize into left- and right-moving sectors (contrary to the assumption of \eqref{eq:Hn_LxR}), a slightly modified formalism should still apply. Other applications should be to the settings of AdS$_3\times S^3\times S^3\times S^1$~\cite{Murthy:2025moj,Eberhardt:2017pty} and ${\rm AdS}_3 \times ({\rm S}^3 \times {\rm S}^3 \times {\rm S}^1)/\mathbb Z_2$ holography~\cite{Eberhardt:2018sce}, which in both cases have a conjectured symmetric orbifold dual. We hope to report progress on some of these in the near future.

Recent ideas, termed ``fortuity'' \cite{Chang:2024zqi}, for a purely conformal field theoretic classification of BPS states in holographic systems, roughly separating those dual to perturbative supergravity excitations from those dual to black hole microstates, have gained a lot of traction. In this framework one asks not just whether a state is BPS (with respect to the deformed supercharges) at a particular value of $N$, but whether it remains BPS under an embedding into the Hilbert space of theories with all larger values of $N$. While these ideas have proved extremely fruitful in other holographic systems\footnote{For a discussion of analogous phenomena in non-holographic systems, see~\cite{Gorsky:2025ylb,deMelloKoch:2025cec}.} \cite{Chang:2013fba,Chang:2022mjp,Chang:2024lxt,Choi:2022caq,Choi:2023vdm,Chang:2023zqk,Budzik:2023vtr,Choi:2023znd,Kim:2025vup,Behan:2025hbx,Choi:2025pqr,Belin:2025hsg,Chen:2025sum}, their application to the D1-D5 CFT is complicated, in part, by the presence of singleton states \cite{Hughes:2025tdy}, and as such is still ongoing \cite{Chang:2025rqy,Chang:2025wgo}.

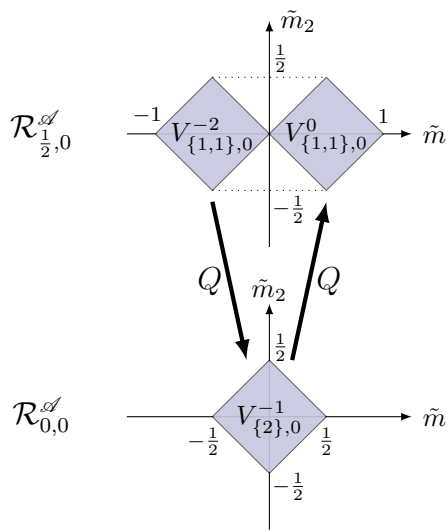
\begin{figure}[tb]
\begin{center}
\begin{tikzpicture}[scale=1.5]
\begin{scope}[shift={(0,0)}]
\node at (-2,0) {$\cR^{\scrA}_{\half,0}$};
\draw[-latex] (-1.25,0) -- (1.25,0) node [right,font=\footnotesize] {$\mt$};
\draw[-latex] (0,-1) -- (0,1) node [right,font=\footnotesize] {$\mt_2$};
\diam{-1/2}{0}
\diam{1/2}{0}
\node at (-1/2,0) [yshift=-1,font=\footnotesize] {$V^{-2}_{\{1,1\},0}$};
\node at ( 1/2,0) [yshift=-1,font=\footnotesize] {$V^{0}_{\{1,1\},0}$};
\node at (-1,0) [above,font=\scriptsize,xshift=-3] {$-1$};
\node at (1,0)  [above,font=\scriptsize] {$1$};
\draw[dotted] (-1/2,1/2) -- (1/2,1/2);
\draw[dotted] (-1/2,-1/2) -- (1/2,-1/2);
\node at (0,1/2)  [right,font=\scriptsize,yshift=7,xshift=-3] {$\frac12$};
\node at (0,-1/2)  [right,font=\scriptsize,yshift=-7,xshift=-3] {$-\frac12$};
\end{scope}
\begin{scope}[shift={(0,-2.5)}]
\node at (-2,0) {$\cR^{\scrA}_{0,0}$};
\draw[-latex] (-1.25,0) -- (1.25,0) node [right,font=\footnotesize] {$\mt$};
\draw[-latex] (0,-1) -- (0,1) node [above,yshift=-3,font=\footnotesize] {$\mt_2$};
\diam{0}{0}
\node at (0,0) [yshift=-1,font=\footnotesize] {$V^{-1}_{\{2\},0}$};
\node at (-1/2,0)  [below,font=\scriptsize,xshift=-3] {$-\frac12$};
\node at (1/2,0)  [below,font=\scriptsize] {$\frac12$};
\node at (0,1/2)  [right,font=\scriptsize,yshift=5,xshift=-3] {$\frac12$};
\node at (0,-1/2)  [right,font=\scriptsize,yshift=-5,xshift=-3] {$-\frac12$};
\end{scope}
\draw[-latex,ultra thick] (-0.5,-0.6) -- (-0.2,-2) node [midway,left,xshift=2] {$Q$};
\draw[-latex,ultra thick] (0.2,-2) -- (0.5,-0.6) node [midway,right,xshift=-2] {$Q$};
\end{tikzpicture}
\hspace*{1.5cm}
\vspace*{-0.5cm}
\end{center}
\caption{\sl The diamond diagrams involved in the longest $N=2$ cochain complex \eqref{eq.cochain_ex}.  The top diagram gives the $\scrA$-representation $\cR^{\scrA}_{\half,0}$ while the bottom diagram gives $\cR^{\scrA}_{0,0}$.  Each diamond corresponds to a $Q$-cohomology class.  These diamond diagrams are the same as the ones in Figure~\ref{fig:diam_gnt_1}, except that we are not showing the multiplicity 2 of $\cR^{\scrA}_{0,0}$ coming from $m_1=\pm \half$.
\label{fig:diam_cochain_cplx}
}
\end{figure}

The formulation of \cite{Chang:2025rqy} is in terms of the cohomology of the first-order deformed right-moving supercharge (related to one of the Gava-Narain operators $\cGt^{\dot{+}A}$ used in this paper). The classes of this ``$Q$-cohomology'', which are in one-to-one correspondence with BPS states, then decompose into long exact sequences (cochain complexes) under the action of $Q$. As discussed in \cite{Chang:2025wgo}, if $V^{2\mt}_{\{n_I\},2\mt_2}$ is the $Q$-cohomology class labelled by twist sector $\{n_I\}$, $su(2)_R$ and $\widetilde{su}(2)_2$ third-component charges $\mt$, $\mt_2$,\footnote{In the conventions of \cite{Chang:2025wgo}, these charges are of the state with minimum $su(2)_R$ charge in the Clifford quartet generated by $\psit^{\alphad\Ad}_0$ total modes. In the diamond diagram picture, this state is the left-most in any given diamond. Note that we choose to instead label $\scrA$-representations and REG sectors by the Casimirs of these $su(2)$'s after factoring out the $\psit$ Clifford structure.} then a cochain complex contains cohomology classes labelled by a fixed $\mt_2$ value. The REG sector $\cE_{N,\tj_2}$ then corresponds, at the level of BPS states, to the cochain complex labelled by $\mt_2=\tj_2$.  For example, the longest cochain complex at $N=2$ (given in~\cite[App.~B]{Chang:2025wgo} and translated into our notation in the Ramond sector)
\begin{equation} \label{eq.cochain_ex}
    0 \overset{Q}{\lto} V^{-2}_{\{1,1\},0} \overset{Q}{\lto} V^{-1}_{\{2\},0} \overset{Q}{\lto} V^{0}_{\{1,1\},0} \overset{Q}{\lto} 0
\end{equation}
corresponds to the diamond diagrams shown in Figure~\ref{fig:diam_cochain_cplx}, where one sees that each class corresponds to a diamond from an $\scrA$-representation. The cochain complex in \eqref{eq.cochain_ex} is therefore captured by the REG $\cE_{2,0}$.

\let\diamcolorsave=\diamcolor
\let\diamopacitysave=\diamopacity
%
\def\diamopacity{0.90}
\def\diamcolor{dfill3!25}
\let\diamcolor=\diamcolorsave
\let\diamopacity=\diamopacitysave

One way in which the formalism presented in this paper provides more fine-grained information about the structure of cochain complexes is with regards to the ``R-charge concentration'' conjecture of \cite{Chang:2025wgo}. Roughly speaking, this conjecture states that fortuitous states (classes) exist only in the middle of a cochain complex, which generically contains a direct sum of many cohomology classes. This phenomenon, also observed in SYK models~\cite{Chang:2024lxt}, is thought to be related to the claims of~\cite{Sen:2009vz, Dabholkar:2010rm, Chowdhury:2015gbk}, that for 4-dimensional single-centred black holes the microstates all have vanishing angular momentum. In the AdS$_3\times S^3$ setting considered here, this angular momentum corresponds to $su(2)_R$~\cite{Gaiotto:2005gf}, and so we expect that R-charge concentration can be equally stated as all fortuitous states being in $\scrA$-representations labelled by $\tj=0$. In other words, being in a sector whose diamond diagram is a single column of diamonds centred on $\tj=0$. From \eqref{eq.gl22_to_A_Def} we can translate this into a condition on the relevant $\lam$-sectors: for the example of the longest cochain complex at general $N$, one finds that $\tilde{S}_{\lam}\supset \tilde{\chi}^{\scrA}_{0,0}$ (in the sense of \eqref{eq.gl22_to_A_Def}) when
\begin{equation}
    \lam = \{1\},\ \{2,1\},\ \{a,a,2,\dots,2,1\},\ \{a+1,a,2,\dots,2\} \quad,\quad a\geq2 \ .
\end{equation}
Or, in terms of Young diagrams,
\begin{align}
\lambda
~=~
\vcenter{\hbox{\begin{tikzpicture}[scale=0.3]
\draw (0,0) rectangle +(1,1);
\end{tikzpicture}}}
~,\hspace{3ex}
\vcenter{\hbox{\begin{tikzpicture}[scale=0.3]
\draw (0,0) rectangle +(1,1);
\draw (1,0) rectangle +(1,1);
\draw (0,-1) rectangle +(1,1);
\end{tikzpicture}}}
~,\hspace{3ex}
\vcenter{\hbox{\begin{tikzpicture}[scale=0.3]
\draw[color=white!70!black] (2,0) rectangle +(1,1);
\draw[color=white!70!black] (3,0) rectangle +(1,1);
\draw[color=white!70!black] (6,0) rectangle +(1,1);
\node[color=white!70!black,font=\scriptsize] at (5.1,0) {$\cdots$};
\draw[color=white!70!black] (2,-1) rectangle +(1,1);
\draw[color=white!70!black] (3,-1) rectangle +(1,1);
\draw[color=white!70!black] (6,-1) rectangle +(1,1);
\draw[color=white!70!black] (0,-2) rectangle +(1,1);
\draw[color=white!70!black] (1,-2) rectangle +(1,1);
\draw[color=white!70!black] (0,-3) rectangle +(1,1);
\draw[color=white!70!black] (1,-3) rectangle +(1,1);
\node[color=white!70!black,font=\scriptsize] at (1,-3.7) {$\vdots$};
\draw[color=white!70!black] (0,-6) rectangle +(1,1);
\draw[color=white!70!black] (1,-6) rectangle +(1,1);
\draw (0,0) rectangle +(1,1);
\draw (1,0) rectangle +(1,1);
\draw (0,-7) rectangle +(1,1);
\draw (0,-1) rectangle +(1,1);
\draw (1,-1) rectangle +(1,1);
\draw [decorate,decoration={brace,amplitude=5pt,raise=2pt}] (2,1) -- (7,1);
\draw [decorate,decoration={brace,amplitude=5pt,raise=2pt,mirror}] (0,-1) -- (0,-6);
\end{tikzpicture}}}
 ~~,\hspace{3ex}
{\vcenter{\hbox{\begin{tikzpicture}[scale=0.3]
\draw[color=white] (0,-7) rectangle +(1,1); 
\draw[color=white!70!black] (2,0) rectangle +(1,1);
\draw[color=white!70!black] (3,0) rectangle +(1,1);
\draw[color=white!70!black] (6,0) rectangle +(1,1);
\node[color=white!70!black,font=\scriptsize] at (5.1,0) {$\cdots$};
\draw[color=white!70!black] (2,-1) rectangle +(1,1);
\draw[color=white!70!black] (3,-1) rectangle +(1,1);
\draw[color=white!70!black] (6,-1) rectangle +(1,1);
\draw[color=white!70!black] (0,-2) rectangle +(1,1);
\draw[color=white!70!black] (1,-2) rectangle +(1,1);
\draw[color=white!70!black] (0,-3) rectangle +(1,1);
\draw[color=white!70!black] (1,-3) rectangle +(1,1);
\node[color=white!70!black,font=\scriptsize] at (1,-3.7) {$\vdots$};
\draw[color=white!70!black] (0,-6) rectangle +(1,1);
\draw[color=white!70!black] (1,-6) rectangle +(1,1);
\draw (0,0) rectangle +(1,1);
\draw (1,0) rectangle +(1,1);
\draw (7,0) rectangle +(1,1);
\draw (0,-1) rectangle +(1,1);
\draw (1,-1) rectangle +(1,1);
\draw [decorate,decoration={brace,amplitude=5pt,raise=2pt}] (2,1) -- (7,1);
\draw [decorate,decoration={brace,amplitude=5pt,raise=2pt,mirror}] (0,-1) -- (0,-6);
\end{tikzpicture}}}}
~~\supset~~\vtikz{\diam[0.2]{0}{0}}~~
~~,
\end{align}
where the length of the gray, braced parts can be arbitrary.
For fixed $N$, the contributing $\lam$-sectors will be those for which $\lam\vdash n\leq N$. While there will generally be other cohomology classes in the middle of this cochain, these correspond to central diamonds within an $\scrA$-representation with $\tj\neq0$, and so are only relevant for monotone states (under the R-charge concentration conjecture).

The REG $\cE_{\jt_2}(p,q,y)$ was defined in \eqref{eq.REG1} by restricting the $\scrA$-representation sum expression of the MEG \eqref{MEGdecompA} to contributions from $\scrA$-representations carrying a fixed $\widetilde{su}(2)_2$ label $\jt_2$.  This definition can be further ``resolved'' by restricting the sum to contributions from $\scrA$-representations with fixed $su(2)_R$ and $\widetilde{su}(2)_2$ labels, $(\jt,\jt_2)$.   The resulting object, which we denote by $\cE_{\jt,\jt_2}(p,q,y)$, may be called a  ``$(\jt,\jt_2)$-resolved REG''\@.  Note that this is \emph{not} a protected quantity because long multiplets are formed by combining $\scrA$-representations with different values of $\jt$.  Nevertheless, it may still be useful for studying the R-charge concentration, because it allows one to isolate the contribution from $\jt=0$.   In Appendix \ref{ss:gen_fun_(jt,jt2)-REG}, we derive a closed-form expression for the generating function for $\cE_{\jt,\jt_2}(p,q,y)$.

The REG differs from the refined indices obtained by inserting fugacities for conserved charges into the trace defining the partition function. The REG counts $\scrA$-representations, rather than individual states.  When additional conserved charges are present, the REG can itself be further refined by them.  For example, at the free point, the left-moving $su(2)_2$ is a symmetry, and one can therefore insert $\eta_2^{2K^3_{2,0}}$ in the trace to define a refined REG, $\cE_{N,\jt_2}(q,y,\eta_2)$.  Such quantities should be useful for explicit computations of lifted states.

One obvious question is whether the REG, in particular its generating function form \eqref{genFuncREG}, admits a gravity dual.\footnote{For recent developments on the bulk computations of supersymmetry indices, often referred to as gravitational indices, see {\it e.g.}\ \cite{Cabo-Bizet:2018ehj,Turetta:2026mgq}}  Such a construction would not be straightforward, because the parameter $\etatb$ is not a simple fugacity for the $\widetilde{su}(2)_2$ charge associated with the (broken) rotational symmetry of the bulk $T^4$, but rather counts entire $\scrA$-multiplets.  In addition, the definition of the REG is given only in the neighborhood of the free orbifold point in moduli space, far from the supergravity point.  Nevertheless, one may still hope to construct a bulk version of the REG in the neighborhood of the special point where all hypermultiplet scalars vanish.

\section*{Acknowledgements}

We would like to thank Stefano Giusto, Bin Guo, Hiroaki Kanno, Samir Mathur, Rodolfo Russo and David Turton for fruitful discussions and useful feedback during the latter stages of this work and to Rafi Rizqy Firdaus for pointing out typos in version 1. We thank Stefano Giusto, Rodolfo Russo and James Inglis for discussions and collaboration on related work.
MRRH gratefully acknowledges the hospitality of the Simons Center for Geometry and Physics, Stony Brook University, during the 50 years of the black hole information paradox program at which some of the research for this paper was presented and refined.
This work was supported in part by MEXT
KAKENHI Grant Numbers 21H05184 and 24K00626.

\appendix

\section{Notations and conventions}
\label{app:conventions}

In this Appendix we give our conventions for the symmetry algebras, modes and spectral flow used in the main body.

\subsection{Symmetry algebra}

The small $d=2$, $\mathcal{N}=4$ superconformal algebra with central charge $c$ is given by
\begin{subequations} \label{eq.commutators1}
    \begin{align}
        \big[L_m,L_n\big] &= (m-n)L_{m+n} + \frac{c}{12}m(m^2-1)\delta_{m+n,0} \ , \label{eq.[L,L]}\\
        \big[J^a_{m},J^b_{n}\big] &= \frac{c}{12}m\,\delta^{ab}\delta_{m+n,0} +  i\epsilon^{ab}_{\ \ c}\,J^c_{m+n} \ ,\label{eq.[J,J]}\\
        \big\{ G^{\alpha A}_{r} , G^{\beta B}_{s} \big\} &= \epsilon^{AB} \bigg[\epsilon^{\alpha\beta}\frac{c}{6}\Big(\frac14-r^2\Big)\delta_{r+s,0} + \big(\sigma^{a}\big)^{\alpha}{}_{\gamma}\:\epsilon^{\beta\gamma}(r-s)J^a_{r+s} - \epsilon^{\alpha\beta}L_{r+s} \bigg] \ ,\label{eq.[G,G]}\\
        \big[J^a_{m},G^{\alpha A}_{r}\big] &= \frac12 \big(\sigma^{a}\big)^{\!\alpha}{}_{\beta}\, G^{\beta A}_{m+r} \ ,\label{eq.[J,G]}\\
        \big[L_{m},G^{\alpha A}_{r}\big] &= \Big(\frac{m}{2}  -r\Big)G^{\alpha A}_{m+r} \ ,\label{eq.[L,G]}\\
        \big[L_{m},J^a_n\big] &= -nJ^a_{m+n} \ . \label{eq.[L,J]}
    \end{align}
\end{subequations}
We define
\begin{align}
    (\sigma^a)_\alpha{}^{\beta}=\tau^a,\qquad
\end{align}
where $\tau^a$ denotes the Pauli matrices, independent of the index position.  The $\alpha,A$ indices are raised and lowered by the antisymmetric symbols $\epsilon_{\alpha\beta},\epsilon_{AB}$ and $\epsilon^{\alpha\beta},\epsilon^{AB}$ with
\begin{align}
    \epsilon^{+-}=\epsilon^{12}=-1,\qquad
    \epsilon_{+-}=\epsilon_{12}=1.
\end{align}
$\epsilon^{\Ad\Bd},\epsilon_{\Ad\Bd}$  are defined similarly.
For example,
\begin{align}
    (\sigma^a)^\alpha{}_\beta=\epsilon^{\alpha\gamma}\epsilon_{\beta\delta}(\sigma^{a})_{\gamma}{}^{\delta}=(\tau^a)^T=(\sigma^a)_\beta{}^{\alpha}.
\end{align}
Note that $(\sigma^a)_{\alpha\beta}$, $a=1,2,3$ give a
basis of symmetric matrices:
\begin{align}
 (\sigma^1)_{\alpha\beta}=\tau^3,\qquad
 (\sigma^2)_{\alpha\beta}=-i{\bf 1}_2 ,\qquad
 (\sigma^3)_{\alpha\beta}=-\tau^1.
\end{align}

The mode algebra \eqref{eq.commutators1} is enhanced to the contracted large $d=2$, $\cN=4$ superconformal algebra, again with central charge $c$, including the additional relations
\begin{subequations} \label{eq.commcurrents2}
    \begin{align}
        \big[ \alpha^{\dot{A}A}_{n} ,  \alpha^{\dot{B}B}_{m}\big] &= n \frac{c}{6} \ep^{AB}\ep^{\dot{A}\dot{B}} \delta_{n+m,0} \ , \label{eq.[al,al]}\\
        \big\{ \psi^{\alpha \dot{A}}_{r}, \psi^{\beta \dot{B}}_{s}\big\} &= -\frac{c}{6}\ep^{\alpha\beta}\ep^{\dot{A}\dot{B}} \delta_{r+s,0} \ , \label{eq.[psi,psi]}\\
        \big[ L_n, \alpha^{\dot{A}A}_{m} \big] &= -m\, \alpha^{\dot{A}A}_{n+m} \ ,\label{eq.[L,al]} \\
        \big[ L_n, \psi^{\alpha\dot{A}}_{r} \big] &= -\big(\tfrac{n}{2} + r\big) \psi^{\alpha\dot{A}}_{n+r} \ , \label{eq.[L,psi]}\\
        \big[ J^a_n, \alpha^{\dot{A}A}_{m} \big] &= 0 \ , \label{eq.[J,al]}\\
        \big[ J^a_n, \psi^{\alpha\dot{A}}_{r} \big] &= \tfrac12 \big(\sigma^{a}\big)^{\alpha}{}_{\beta}\, \psi^{\beta\dot{A}}_{n+r} \ , \label{eq.[J,psi]}\\
        \big[G^{\alpha A}_{r} , \alpha^{\dot{B}B}_{m}\big] &= -m \ep^{AB}\psi^{\alpha\dot{B}}_{r+m} \ , \label{eq.[G,al]}\\
        \big\{G^{\alpha A}_{r} , \psi^{\beta\dot{A}}_{s} \big\} &= \ep^{\alpha\beta} \alpha^{\dot{A}A}_{r+s} \ .\label{eq.[G,psi]}
    \end{align}
\end{subequations}

A different basis for the $SU(2)_L$ currents is often used, where instead of the $J^a_n$ we have $J^{\pm,3}_n$ defined as
\begin{equation}
    J^{\pm}_n \equiv J^1_n \pm i J^2_n \ .
\end{equation}
The non-zero commutation relations involving the $J^{\pm}_{n}$ are
\begin{subequations} \label{eq.commcurrentsJpm}
    \begin{align}
        \big[J^+_n, J^-_m \big] &= n\frac{c}{6}\delta_{n+m,0} + 2J^3_{n+m} \ , \label{eq.[Jp,Jm]}\\
        \big[J^3_n,J^{\pm}_m\big] &= \pm J^{\pm}_{n+m} \ , \label{eq.[J3,Jpm]}\\
        \big[L_n,J^{\pm}_m\big] &= -m J^{\pm}_{n+m} \ , \label{eq.[L,Jpm]}\\
        \big[J^{\pm}_{n}, G^{\mp A}_{r}] &= G^{\pm A}_{n+r} \ , \label{eq.[Jpm,G]}\\
        \big[J^{\pm}_{n}, \psi^{\mp \Ad}_{r}] &= \psi^{\pm \Ad}_{n+r} \ .\label{eq.[Jpm,psi]}
    \end{align}
\end{subequations}
The hermiticity of $G,\alpha,\psi$ is given by
\begin{subequations}
\label{eq:hermiticity}
\begin{align}
    (G^{\alpha A})^\dagger&=-\epsilon_{\alpha\beta}\epsilon_{AB}G^{\beta B}
    =-G_{\alpha A}
    \ ,\\
    (\alpha^{\Ad A})^\dagger&=\epsilon_{\Ad\Bd}\epsilon_{AB}\alpha^{\Bd B}
    =\alpha_{\Ad A}
    \ ,\\
    (\psi^{\alpha \Ad})^\dagger&=-\epsilon_{\alpha\beta}\epsilon_{\Ad\Bd}\psi^{\beta\Bd}
    =-\psi_{\alpha \Ad}\ .
\end{align}
\end{subequations}

In the context of the symmetric orbifold theory $\mathrm{Sym}^N(T^4)$, the modes on the $i$th copy of the $T^4$ theory ($i=1,\dots,N$) are defined on the Euclidean cylinder $w=\tau + i\sigma$ as
\begin{equation} \label{eq.copy_mode_def}
    \cO_m^{(i)} \equiv \frac1{2\pi i} \int_{\sigma=0}^{2\pi } dw\ e^{mw}\, \cO^{(i)}(w) \ ,
\end{equation}
where $\cO_m^{(i)}$ represents one of the modes in the above contracted large algebra and $\cO^{(i)}(w)$ the associated current. For these individual copy modes, the central charge appearing in the algebra is that of the seed theory, \textit{i.e.}\ $c=c_0=6$.

On the $I$th strand, involving $k_I$ copies twisted together, the modes in our conventions are similarly defined as 
\begin{equation} \label{eq.strand_mode_def}
    \cO_m^{[I]} \equiv \frac1{2\pi i} \int_{\sigma=0}^{2\pi k_I}\!dw\ e^{mw}\, \cO(w) \ ,
\end{equation}
and which satisfy the contracted large algebra with a central charge of $c=6k_I$.

The symmetry mode algebra of $\mathrm{Sym}^N(T^4)$ is the diagonal contracted large $\cN=4$ algebra, for which the modes are the ``total'' modes
\begin{equation} \label{eq.total_mode_def}
    \cO_m^{({\rm T})} \equiv \sum_{i=1}^{N} \cO_m^{(i)} = \sum_{I=1}^{n} \cO_m^{[I]} \ ,
\end{equation}
where $n$ is the number of strands in a particular sector of the theory, satisfying $\sum_{I=1}^n k_I = N$. These total modes satisfy the algebra relations \eqref{eq.commutators1}, \eqref{eq.commcurrents2} with central charge $c=6N$. Throughout the rest of this paper we have dropped the total mode label, writing $\cO^{({\rm T})}_m$ as simply $\cO_m$, for the sake of clarity.

The 2-dimensional $\cN=4$ superconformal algebra has an $SU(2)$ inner automorphism group which leads equivalent algebras under the spectral flow transformations
\begin{equation} \label{eq.symmModeDeform}
\begin{aligned}
    L_n(\eta) &= L_n + \eta J^3_n + \eta^2\frac{c}{24} \delta_{n,0} \quad , \quad J^3_n(\eta) = J^3_n + \eta \frac{c}{12} \delta_{n,0} \ , \\
    J^{\pm}_{n}(\eta) &= J^{\pm}_{n\mp\eta} \quad ,\quad G^{\pm A}_{r}(\eta) = G^{\pm A}_{r\mp \frac{\eta}{2}} \quad ,\quad \psi^{\pm \Ad}_{r}(\eta) = \psi^{\pm \Ad}_{r\mp \frac{\eta}{2}} \ ,
\end{aligned}
\end{equation}
where $\eta$ is the flow parameter. In particular $\eta\in\mathbb{Z}$ interpolates between the NS and R sectors. In this paper we mostly work in the Ramond sector, and when required, spectral flow to the NS sector using $\eta=+1$. The maps \eqref{eq.symmModeDeform} lead to the maps of conformal dimension $h$ and R-charge $m$ of the form
\begin{equation} \label{eq.SF_hm}
    h(\eta) = h + m\eta + \frac{c}{24}\eta^2 \quad,\quad m(\eta) = m + \frac{c}{12}\eta \ . 
\end{equation}
Spectral flow of the right-moving algebra is defined analogously.

\subsection{$gl(2|2)$ and $\scrA$-algebras}
\label{ss:gl(2|2)_and_A-alg}

Here we study how the generators of the $\scrA$-algebra, $\Jt^a,\Kt^a_2,\psit^{\alphad\Ad}$, are embedded in $gl(2|2)$, and what $gl(2|2)$ generators are missing in the $\scrA$-algebra. We are particularly interested in odd (fermionic, or off-block-diagonal) $gl(2|2)$ generators that are not part of the $\scrA$-algebra.
We suppress the zero-mode subscript ``0'' on the generators.

The action of the fermionic zero mode $\psit^{\alphad \Ad}$ on the Ramond ground states~\eqref{eq.R_GS_rels} in $\Vt$ is
\begin{subequations} 
\label{gicb28Sep25}
 \begin{align}\label{eggh21Mar25}
 \psit^{\alphad\Ad}\ket*{\betad}_k
 =-\sqrt{k}\,\epsilon^{\alphad\betad}\ket*{\Ad}_k,\qquad
 \psit^{\alphad \Ad}\ket*{\Bd}_k=
 -\sqrt{k}\,\epsilon^{\Ad\Bd}\ket{\alphad}_k\ .
 \end{align}
\end{subequations}
Thus the operators
$\psit^{\alphad\Ad }$ can be represented on $\Vt$ by the odd $gl(2|2)$ matrices\footnote{%
We define the matrix elements of an operator $\cO$ so that, when acting on a state $\ket{v}=\sum_I v_I \ket{I}$ where $\ket{I}$ is a basis of vectors, its action $\ket{v'}=\cO\ket{v}$ is written as $v'_I = \cO_I{}^J v_J$.}
\begin{align}
 \psit^{\alphad\Ad}
 &=
 \mqty(
 (\psit^{\alphad\Ad})_{\betad}{}^{\gammad}&(\psit^{\alphad\Ad})_{\betad}{}^{\Cd}\\
 (\psit^{\alphad\Ad})_{\Bd}{}^{\gammad}&(\psit^{\alphad\Ad})_{\Bd}{}^{\Cd}\\
 )
=
 \sqrt{k}\mqty(
    0&-\delta^{\alphad}_{\betad}\epsilon^{\Ad\Cd}\\
    -\delta^{\Ad}_{\Bd}\epsilon^{\alphad\gammad}&0
 )\ .
\label{psit_4x4mat}
\end{align}
On the other hand, the $su(2)_R\oplus \widetilde{su}(2)_2$ generators in this representation are
\begin{align}
    \Jt^a=\mqty(\tau^a/2&0\\ 0&0)\ ,\qquad
    \Kt^a_2=\mqty(0&0\\ 0&\tau^a/2)\ ,
\end{align}
where $\tau^a$ are the Pauli matrices.

The above $\psit^{\alphad\Ad}$ in \eqref{psit_4x4mat} give only half of the eight odd $gl(2|2)$ generators.  One way to construct all eight odd generators is to use as projectors the Casimirs of the two $su(2)$'s, namely,
\begin{align}
 \tfrac43 \Jt^2=\mqty(1&0\\0&0)\equiv P\ ,\qquad
 \tfrac43 \Kt_2^2=\mqty(0&0\\0&1)\equiv P_2 \ .
\end{align}
Then we can define $Q$ and $S$ by
\begin{align}
 Q^{\alphad \Ad} \equiv P \psit^{\alphad\Ad} = \psit^{\alphad\Ad} P_2= P\psit^{\alphad\Ad} P_2\ ,\qquad
 S^{\alphad\Ad} \equiv \psit^{\alphad\Ad} P = P_2 \psit^{\alphad\Ad}=P_2 \psit^{\alphad\Ad} P\ ,
\end{align}
which satisfy $\psit=Q+S$ and
whose matrix form is, in the basis of \eqref{psit_4x4mat},
\begin{align}
 Q^{\alphad \Ad}=
 \sqrt{k}\mqty(0& - \delta^{\alphad}_{\betad}\epsilon^{\Ad\Cd}\\ 0&0)\ ,\qquad
 S^{\alphad \Ad}=
 \sqrt{k}\mqty(0&0\\ -\delta^{\Ad}_{\Bd}\epsilon^{\alphad\gammad}&0)\ .
\end{align}
These satisfy
\begin{align}
 \{Q,Q\}&=\{S,S\}=0\ ,\quad
 \{Q^{\alphad}_{\Ad},S^{\Bd}_\beta\}_{\gammad}{}^{\dot\delta }
 =k\delta_{\Ad}^{\Bd} \delta ^{\alphad} _{\gammad} \delta^{\dot{\delta}} _{\betad}\ ,
 \quad
 \{Q^{\alphad}_{\Ad},S^{\Bd}_{\betad}\}_{\Cd}{}^{\dot{D}}
 =k\delta^{\alphad}_{\betad} \delta^{\Bd}_{\Cd }\delta^{\dot{D}}_{\Ad}\ ,
 \label{eq:Q,S_anticomm}
\end{align}
and $Q^{\alphad \Ad}$ and $S^{\alphad \Ad}$
give eight independent odd $gl(2|2)$ generators.
Therefore, in the case of a single strand, we can construct all odd $gl(2|2)$ generators using the modes $\psit^{\alphad \Ad}$, 
$\Jt^{a}$ and $\Kt^{a}_2$, which are total modes in this case.

This is not the case in the multiple-strand setting.  Namely, missing odd $gl(2|2)$ generators cannot be constructed using other total generators of the contracted large $\cN=4$ algebra.
At the free point, the eight total odd $gl(2|2)$ generators can be written as the sum of strand-wise ones as
\begin{align}
 Q^{\alphad\Ad} 
 &=\sum_I Q^{\alphad\Ad[I]}
 =\sum_I P^{[I]} \psit^{\alphad \Ad[I]}
 =\sum_I \psit^{\alphad \Ad[I]} P_2^{[I]}\ ,
 \notag\\
 S^{\alphad \Ad} 
 &=\sum_I S^{\alphad \Ad [I]}
 =\sum_I P_2^{[I]}\psit^{\alphad \Ad[I]}
 =\sum_I \psit^{\alphad \Ad[I]} P^{[I]}\ ,
\end{align}
which satisfy \eqref{eq:Q,S_anticomm} (with the right hand side multiplied by $N$).
However, they cannot be constructed from other total modes, such as $\sum_I\psit^{\alphad\Ad[I]}$, $\sum_I \Jt^{a[I]}$ and  $\sum_I \Kt^{a[I]}_2$, that exist everywhere on moduli space.

\subsection{Representations of the $\scrA$-algebra: simple cases}
\label{app:A-alg_rep_simple_cases}

In Section \ref{sssec:diamonds}, we discussed the representation theory of the $\scrA$-algebra \eqref{A-alg_comm} in general. Here we study concrete $\scrA$-representations  realized on the right-moving Hilbert space of a small number of strands.

\subsubsection*{Single strand}

Let us start with the case of a single strand of length one, for which the right-moving Hilbert space is $\Vt={\rm span}\{\ket*{\alphad}_1,\ket*{\Ad}_1\}$. In this case, the $cl_4^\psit$ relation \eqref{A-alg_comm_psit} is
\begin{align}
 \big\{\psit^{\alphad\Ad},\psit^{\betad\Bd}\big\}=-\epsilon^{\alphad\betad}\epsilon^{\Ad\Bd}\ .
 \label{cl_4_single}
\end{align}
If we want, we can rewrite this in the more familiar, canonical form
\begin{align}
 \{\Gamma^\mu,\Gamma^\nu\}=2\delta^{\mu\nu},\qquad \mu,\nu=1,2,3,4,
 \label{cl_4_single_Gamma}
\end{align}
with hermitian $\Gamma^{\mu}$ by a change of basis.\footnote{One can define $\Gamma^\mu=\sigma^\mu_{\alphad \Ad}\psi^{\alphad \Ad}$, for $\mu=(a,4)$, with $(\sigma^a)^{\alphad \Ad}=\tau^a,(\sigma^4)^{\alphad \Ad}=i{\bf 1}$, where $\tau^a$ are the Pauli matrices.}
The Dirac representation of the $cl_4^\psi$ algebra \eqref{cl_4_single} (or \eqref{cl_4_single_Gamma}) is four dimensional, which is nothing but $\Vt$.  

Let us turn to the remaining $su(2)_R\oplus \widetilde{su}(2)_2\cong so(4)$ part of the $\scrA$-algebra, \eqref{A-alg_comm_Jt} and \eqref{A-alg_comm_Kt2}. In fact, in this case, the $so(4)$ generators can be constructed solely out of the total modes $\Gamma^\mu$ by the usual formula, $\Sigma^{\mu\nu}\equiv \frac{i}{4}[\Gamma^\mu,\Gamma^\nu]$.   These decompose under $so(4)\cong su(2)_R\oplus \widetilde{su}(2)$ into the following generators:
\begin{align}
 \Jt^a
 = -(\sigma^a)_\alphad{}^\betad \psit^{\alphad\Ad}\psit_{\betad\Ad} \quad ,
\quad
\Kt_2^a
=
  -(\sigma^a)_{\Ad}{}^{\Bd} \psit^{\alphad\Ad}\psit_{\alphad\Bd}\ .
  \label{eq:Jt,Kt2_ito_psit}
\end{align}
One can show that these satisfy the relations \eqref{A-alg_comm_Jt} and \eqref{A-alg_comm_Kt2}.
With respect to the 
$SO(4)\cong SU(2)_R\times \widetilde{SU}(2)_2$ rotations generated by $\Sigma^{\mu\nu}$, the Dirac spinor representation $\Vt$ decomposes into $(\half,0)\oplus (0,\half)$, which is nothing but the states $\ket*{\alphad}_1$ and $\ket*{\Ad}_1$.

So, in the single-string case, $\Vt$ gives an irreducible representation of the $\scrA$-algebra.  We can display the quantum numbers $(\mt,\mt_2)$ (the eigenvalues of $(\Jt^3,\Kt^3_2)$; see Table \ref{tbl:su(2)s}) of the states in this representation as below:
\begin{align}
\vcenter{\hbox{\begin{tikzpicture}[scale=1.25]
\draw[-latex] (-1.3,0) -- (1.3,0) node [right] {$\mt$};
\draw[-latex] (0,-1.3) -- (0,1.3) node [above] {$\mt_2$};
\diam{0}{0}
\draw[fill=black] (-0.5,0) circle (0.03) node [above left,xshift=4] {$-\half$};
\draw[fill=black] ( 0.5,0) circle (0.03) node [above right,xshift=-4] {$\half$};
\draw[fill=black] (0,-0.5) circle (0.03) node [below left ,yshift=2,xshift=2] {$-\half$};
\draw[fill=black] (0, 0.5) circle (0.03) node [above left ,yshift=-2,xshift=2] {$\half$};
\end{tikzpicture}}}
\label{eq:1-strand_A-alg_reps}
\end{align}
where each dot corresponds to a state, and the bluish ``diamond'' represents the Dirac quartet of the mode $\Gamma^\mu$ (or $\psit^{\alphad \Ad}$).

In the main text, we call this representation $\cR^{\scrA}_{0,0}$.

\subsubsection*{Two strands}

Let us now consider the case of two strands of length one,
where the right-moving Hilbert space is $\Vt^{[1]}\otimes \Vt^{[2]}$ where $[I]$ labels the strand number.
The total fermion zero modes $\psit^{\alphad \Ad}$ now satisfy the $cl_4^\psit$ relation \eqref{A-alg_comm_psit} with $N=2$.

In the free theory, the total mode $\psit^{\alphad \Ad}$ can be written as a sum of strand-wise generators $\psit^{\alphad \Ad[I]}$ as 
\begin{align}
\psit^{\alphad \Ad}=\psit^{\alphad \Ad[1]}+\psit^{\alphad \Ad[2]}  ,
\label{eq:total_psit_2-strands}
\end{align}
and they each satisfy the Clifford algebra relation:
\begin{align}
 \big\{\psit^{\alphad\Ad[I]},\psit^{\betad\Bd[J]}\big\}=-\epsilon^{\alphad\betad}\epsilon^{\Ad\Bd}\delta^{IJ}\ ,\qquad  
 I,J=1,2.
 \label{cl_4_two}
\end{align}
Or equivalently, by a basis change, we can write this in terms of $\Gamma^{\mu[I]}$ as
\begin{align}
    \big\{\Gamma^{\mu[I]},\Gamma^{\nu[J]}\big\}=2\delta^{\mu\nu}\delta^{IJ}.
\end{align}
To discuss the $\scrA$-representation of the total modes that we are interested in, let us define
\begin{align}
    \Gamma^{\mu}\equiv \Gamma^{\mu[1]}+\Gamma^{\mu[2]}\ ,
    \qquad
    \Gamma'^{\mu}\equiv \Gamma^{\mu[1]}-\Gamma^{\mu[2]} \ ,
\end{align}
where $\Gamma^{\mu}$ are the total modes, while $\Gamma'^{\mu}$ are ``relative'' modes.  These combinations satisfy separate Clifford algebras:
\begin{align}
  \{\Gamma^{\mu},\Gamma^{\nu}\}=4\delta^{\mu\nu},\qquad
  \{\Gamma'^{\mu},\Gamma'^{\nu}\}=4\delta^{\mu\nu},\qquad
  \{\Gamma^{\mu},\Gamma'^{\nu}\}=0\ .
\end{align}
This means that the Hilbert space $\Vt^{[1]}\otimes \Vt^{[2]}$ can be factorized as $\Vt\otimes \Vt'$, where $\Vt$ and $\Vt'$ are the Dirac spinor spaces associated with $\Gamma^{\mu}$ and $\Gamma'^{\mu}$, respectively. As far as the representation of the total modes $\Gamma^{\mu}$ (or equivalently $\psit^{\alphad \Ad}$) is concerned, the second factor $\Vt'$ just gives a multiplicity of 4.

Let us next turn to the $so(4)\cong su(2)_R\oplus \widetilde{su}(2)_2$ part of the $\scrA$-algebra, \eqref{A-alg_comm_Jt} and \eqref{A-alg_comm_Kt2}.  The total $so(4)$ generators are the sum of generators on each strand,
\begin{align}
\Sigma^{\mu\nu}
=\Sigma^{\mu\nu[1]}+\Sigma^{\mu\nu[2]},
\qquad
    \Sigma^{\mu\nu[I]}
    =\tfrac{i}{4}
    \left[\Gamma^{\mu[I]},\Gamma^{\nu[I]}\right].
\end{align}
Expressed in terms of the total $\Gamma^\mu$ and the relative $\Gamma'^{\mu}$, this becomes:
\begin{align}
\Sigma^{\mu\nu}
=\tfrac{i}{8}[\Gamma^\mu,\Gamma^\nu]
+\tfrac{i}{8}[\Gamma'^\mu,\Gamma'^\nu]
\equiv \Sigma^{\mu\nu}_{(\Gamma)}+\Sigma'^{\mu\nu}.
\label{eq:Sigma=tot+rel}
\end{align}
So, the total $\Sigma^{\mu\nu}$ is the sum of the contribution $\Sigma^{\mu\nu}_{(\Gamma)}$ from the total mode $\Gamma^\mu$ and the contribution $\Sigma'^{\mu\nu}$ from the relative mode $\Gamma'^\mu$.
If we interpret $\Sigma^{\mu\nu}$ as spin, this is simply the addition of the spins of two particles: a ``total'' particle with spin wavefunction in $\Vt$, and a ``relative'' particle with spin wavefunction in $\Vt'$.  Because $\Vt'$ furnishes the Dirac spinor representation of $\Gamma'^\mu$, the relative particle's wavefunction decomposes into two irreducible representations of $su(2)_R\oplus \widetilde{su}(2)_2$, namely $(\half,0)$ and $(0,\half)$.  The corresponding quantum numbers $(\mt,\mt_2)$ of the relative particle can be depicted as below:
\begin{align}
\vcenter{\hbox{\begin{tikzpicture}
\draw[-latex] (-1.2,0) -- (1.2,0) node [right] {$\mt$};
\draw[-latex] (0,-1.2) -- (0,1.2) node [above] {$\mt_2$};
 \draw[fill=black] ( 0.5,0) circle (0.05) node [above] {$\half$};
 \draw[fill=black] (-0.5,0) circle (0.05) node [above] {$-\half\phantom{-}$};
\end{tikzpicture}}}
\qquad \raisebox{-1.7ex}{$\oplus$} \qquad
\vcenter{\hbox{\begin{tikzpicture}
\draw[-latex] (-1.2,0) -- (1.2,0) node [right] {$\mt$};
\draw[-latex] (0,-1.2) -- (0,1.2) node [above] {$\mt_2$};
 \draw[fill=black] (0, 0.5) circle (0.05) node [left] {$\half$};
 \draw[fill=black] (0,-0.5) circle (0.05) node [left] {$-\half$};
\end{tikzpicture}}}\label{ewbg15Jan26}
\end{align}
where each dot corresponds to a state.  The total particle likewise has spin content $(\half,0)\oplus (0,\half)$, with quantum numbers $(\mt,\mt_2)=(\pm\half,0),(0,\pm\half)$. The $\scrA$-algebra representations are obtained by adding this to the spin of the relative particle, and can be depicted as below:
\begin{align}
\vcenter{\hbox{\begin{tikzpicture}[scale=1.25]
\draw[-latex] (-1.3,0) -- (1.3,0) node [right] {$\mt$};
\draw[-latex] (0,-1.3) -- (0,1.3) node [above] {$\mt_2$};
\diam{-1/2}{0}
\diam{1/2}{0}
\draw[fill=black] (-1,0) circle (0.03);
\draw[fill=black] (-0.03,0) circle (0.03);
\draw[fill=black] (0.03,0) circle (0.03);
\draw[fill=black] (1,0) circle (0.03);
\draw[fill=black] (-1/2,1/2) circle (0.03);
\draw[fill=black] (1/2,1/2) circle (0.03);
\draw[fill=black] (-1/2,-1/2) circle (0.03);
\draw[fill=black] (1/2,-1/2) circle (0.03);
\end{tikzpicture}}}
\qquad \raisebox{-1.7ex}{$\oplus$} \qquad
\vcenter{\hbox{\begin{tikzpicture}[scale=1.25]
\draw[-latex] (-1.3,0) -- (1.3,0) node [right] {$\mt$};
\draw[-latex] (0,-1.3) -- (0,1.3) node [above] {$\mt_2$};
\diam{0}{-1/2}
\diam{0}{1/2}
\draw[fill=black] (0,-1) circle (0.03);
\draw[fill=black] (0,-0.03) circle (0.03);
\draw[fill=black] (0,0.03) circle (0.03);
\draw[fill=black] (0,1) circle (0.03);
\draw[fill=black] (-1/2,1/2) circle (0.03);
\draw[fill=black] (1/2,1/2) circle (0.03);
\draw[fill=black] (-1/2,-1/2) circle (0.03);
\draw[fill=black] (1/2,-1/2) circle (0.03);
\end{tikzpicture}}}
\label{eq:2-strand_A-alg_reps}
\end{align}
where each dot corresponds to a state, and a ``diamond'' represents the Dirac quartet of the total mode $\Gamma^\mu$ (or $\psit^{\alphad \Ad}$).
So, the 2-strand Hilbert space $\Vt^{[1]}\otimes \Vt^{[2]}\cong \Vt\otimes \Vt'$ contains two $\scrA$-algebra representations, corresponding to the two diagrams in \eqref{eq:2-strand_A-alg_reps}.
In the main text, we call these representations $\cR^{\scrA}_{\half,0}$ and  $\cR^{\scrA}_{0,\half}$, respectively.

We can also state the above ``$\text{total}+\text{relative}$'' decomposition \eqref{eq:Sigma=tot+rel} in terms of $su(2)_R$ and $\widetilde{su}(2)_2$ generators $\Jt^a$ and $\Kt^a_2$, instead of the $so(4)$ generators $\Gamma^\mu$. 
The total $su(2)_R$ and $\widetilde{su}(2)_2$ generators are obtained simply by summing \eqref{eq:Jt,Kt2_ito_psit} over strands:
\begin{align}
 \Jt^a
 = -\sum_{I=1,2}(\sigma^a)_\alphad{}^\betad \psit^{\alphad\Ad[I]}\psit_{\betad\Ad}^{[I]} \quad ,
\quad
\Kt_2^a
=
  -\sum_{I=1,2}(\sigma^a)_{\Ad}{}^{\Bd} \psit^{\alphad\Ad[I]}\psit_{\alphad\Bd}^{[I]}\ .
\end{align}
It is easy to see that these generators decompose as
\begin{align}
 \Jt^a&= \Jt^a_{(\psit)} + \Jt'^a,\qquad
  \Kt_2^a= \Kt^a_{2(\psit)} + \Kt'^a_2\ ,
\end{align}
where
\begin{align}
 \Jt^a_{(\psit)} &\equiv -\tfrac12(\sigma^a)_\alphad{}^\betad \psit^{\alphad\Ad}\psit_{\betad\Ad},\qquad
 \Kt^a_{2(\psit)} \equiv -\tfrac12(\sigma^a)_{\Ad}{}^{\Bd}
 \psit^{\alphad\Ad}\psit_{\betad\Ad}\ ,
 \label{def_total_Jt,Kt2_2strands}
\end{align}
are contributions from the total modes $\psit^{\alphad \Ad}$, while
\begin{align}
\Jt'^a &\equiv -\tfrac12(\sigma^a)_\alphad{}^\betad \psit'^{\alphad\Ad}\psit'_{\betad\Ad}, \qquad
\Kt'^a_2 \equiv -\tfrac12(\sigma^a)_{\Ad}{}^{\Bd}
\psit'^{\alphad\Ad}\psit'_{\betad\Ad} \ ,
\end{align}
are contributions from the relative modes
\begin{align}
\psit'^{\alphad \Ad}=\psit^{\alphad \Ad[1]}-\psit^{\alphad \Ad[2]}\ .
\end{align}
It is also easy to show that $\Jt'^a_{(\psit)}$, $\Jt'^a$, $\Kt^a_{2(\psit)}$ and $\Kt'^a_2$ separately satisfy $su(2)$ commutation relations.

\subsubsection*{$n$ strands}

If there are $n$ strands of length one, the Hilbert space is $\Vt^{[1]}\otimes\cdots\otimes \Vt^{[n]}$.  Just as in the $n=2$ case, this can be factorized as $\Vt\otimes \Vt'^{[1]}\otimes\cdots\otimes \Vt'^{[n-1]}$, where $\Vt$ is the Dirac spinor space for the total mode $\psit=\sum_{I=1}^n \psit^{[I]}$, while $\Vt'^{[I']}$, $I'=1,\dots,n-1$, are Dirac spinor spaces for appropriately defined relative modes.  The latter gives the $su(2)'_R\oplus \widetilde{su}(2)'_2$ representation $((\half,0)\oplus(0,\half))^{\otimes (n-1)}$, which can be expanded into a direct sum of multiple representations.  Multiplying this by the $\psit$-quartet, we get the $\scrA$-representations.  

For example, for $n=3$, the relative Dirac spinor spaces give the  $su(2)'_R\oplus \widetilde{su}(2)'_2$ representation
\begin{math}
((\thalf,0)\oplus(0,\thalf))^{\otimes 2}
=(1,0)\oplus (0,1)\oplus 2(\tfrac12,\tfrac12)\oplus 2(0,0).
\end{math}
When multiplied by the $\psit$-quartet, these lead to  $\scrA$-representations which we can denote using the notation in the main text by
\begin{math}
\cR^{\scrA}_{(1,0)} \oplus \cR^{\scrA}_{(0,1)}\oplus
2\cR^{\scrA}_{({1\over 2},{1\over 2})}\oplus
2\cR^{\scrA}_{(0,0)}.
\end{math}

\subsection{Characters and special functions}

Here we list some definitions and formulae for special functions and characters for various algebras used in the main body of the paper.

\subsubsection*{Special functions}

In the partition function \eqref{eq:seed_z(q,y)} of the seed theory used to construct the symmetric orbifold $\mathrm{Sym}^N(T^4)$, we require the first Jacobi theta function and the Dedekind eta function. These are defined for our purposes as the products
\begin{subequations} \label{eq.theta1etaDef}
\begin{align}
 \vartheta_{1}(\nu,\tau) &\equiv -iq^{\frac{1}{8}}\big(y^{\frac12}-y^{-\frac12}\big) \prod_{m=1}^{\infty} (1-q^m)(1-zq^{m})(1-z^{-1}q^{m})\ ,\label{eq.theta1_def}\\
 \eta(\tau) &\equiv q^{\frac{1}{24}}\prod_{m=1}^{\infty}(1-q^m)\ ,\label{eq.eta_def}
\end{align}
\end{subequations}
where $q=e^{2\pi i\tau}$, $y=e^{2\pi i \nu}$.

\subsubsection*{$SU(2)$}

The character of a spin $j$ representation of $SU(2)$, with $j=0,\half,1,\dots$, is given by the standard form
\begin{align} \label{eq:su(2)_char_theta}
 \chi_j(\theta)=\sum_{m=-j}^{j}e^{i\theta}={\sin(j+\half)\theta\over \sin{\theta\over 2}}\ ,
\end{align}
which can equally be found from the Weyl character formula. These characters satisfy the orthogonality relation
\begin{align} \label{eq:su(2)_char_orth_theta}
 {1\over 2\pi}\int_0^{4\pi} d\theta\,\sin^2\!\frac{\theta}{2}\,\,\chi_j(\theta)\chi_{j'}(\theta)=\delta_{jj'}\ .
\end{align}
Setting
\begin{align} \label{eq:theta_to_y}
 e^{i\theta/2}=y\ ,
\end{align}
allows for connection to the partition functions and indices discussed in the main body, giving
\begin{equation} \label{eq:su(2)_char_y}
 \chi_j(y)=\sum_{m=-j}^{j}y^{2m}={y^{2j+1}-y^{-2j-1}\over y-y^{-1}}\ ,
\end{equation}
and which satisfy the orthogonality relation
\begin{equation} \label{eq:su(2)_char_orth_y}
     {i\over 4\pi}\oint {dy\over y}(y-y^{-1})^2\,\chi_j(y)\chi_{j'}(y)=\delta_{jj'}\ .
\end{equation}

\subsubsection*{$SU(1,1\,|\,2)$}

The Kaluza-Klein spectrum on AdS$_3\times S^3$, and so the spectrum of supergraviton states in the D1-D5 CFT, fall into short representations of global $SU(1,1\,|\,2)$ both for the left- and right-moving sectors.
A short representation is generated by the action of global generators $L_{-1},G^{-A}_{-\half},J^-_0$ on a chiral primary with $h=j=0,\frac12,1,\dots$, and its character is given by
\begin{equation} \label{eq.su112short}
    \phi^{(s)}_{j}(q,y) = \frac{q^j}{1-q}\frac{y^{2j}(y-2\sqrt{q}+y^{-1}q)-y^{-2j}(y^{-1}-2\sqrt{q}+yq)}{y-y^{-1}} \ ,
\end{equation}
where for our purposes the fugacities $q$ and $y$ are fugacities for eigenvalues of $L_0$ and $2J^3_0$ respectively.

\section{The Gava-Narain operator}
\label{app:GN_op}

As discussed in Section \ref{sss:struct_lifting}, the Gava-Narain operator can be defined by \cite{Gava:2002xb,Guo:2019pzk}
\begin{equation} 
    \tilde{\cG}^{\alphad A} = \pi \cP \big(G^{+A}_{-\frac12} \sigma_2^{-\alphad}\big)(w_0)\,\cP = -\pi \cP \big(G^{-A}_{-\frac12} \sigma_2^{+\alphad}\big)(w_0)\,\cP\ ,
    \label{eq:GN_def_app}
\end{equation}
where $w_0$ is the insertion point on the cylinder of the twist operator (where $\sigma^{\alpha\alphad}_2$ with $\alpha=\pm$, $\alphad=\dot{\pm}$ is the twist operator in the R sector with the displayed R-charges) and $G^{+A}_{-\frac12}$ is defined on a contour around $w_0$. The projection operator $\cP$ is onto the space of states with free conformal dimensions $(n+{N\over 4},{N\over 4})$ with some fixed $n\in\bbZ_{\ge 0}$.  Therefore, the Gava-Narain operator is defined only 
in the space of free-theory BPS states. The projection operator $\cP$ also causes the action of the Gava-Narain operator to be independent of the insertion point $w_0$, hence why it is treated as an operator (zero) mode.

The hermiticity of the Gava-Narain operator is given by
\begin{align}
\big(\tilde{\cG}^{\alphad A}\big)^{\dagger}=-\epsilon_{\alphad\betad}\epsilon_{AB}\tilde{\cG}^{\betad B}.
\label{eq:GN_hermiticity}
\end{align}

The power of the Gava-Narain operator is that it captures both the action of 
the first-order deformed supercharge on the free-theory BPS states
and 
the free-theory right-moving supercharge on the first-order deformed BPS states:
\begin{equation} \label{eq.GN_psi_def}
    \tilde{\mathcal{G}}^{\dot{\alpha}A}\ket*{\psi^{(0)}} = \tilde{G}^{\dot{\alpha}A(1)}_{0}\ket*{\psi^{(0)}} + \tilde{G}^{\dot{\alpha}A(0)}_{0}\ket*{\psi^{(1)}} \ .
\end{equation}

We show in \ref{app:GN_alg} that the Gava-Narain operator (anti)commutes with all free-theory left-moving generators
\begin{align}
\big[\cGt^{\alphad A},\cA\big]_\pm=0\quad,\qquad
\cA=L_n,J^a_n,G^{\alpha A}_n,\alpha^{\Ad A}_n,\psi^{\alpha \Ad}_n.
\end{align}
as well as the zero modes of the right-moving free-theory fermions
\begin{align}
\big\{\cGt^{\alphad A},\psit^{\betad \Ad}_0\big\}=0\ .
\end{align}
We can define the lifting operator $\hat{E}^{(2)}$ by
\begin{equation}
    2\big\{ \tilde{\cG}^{\alphad A}, \tilde{\cG}^{\betad B} \big\}
    = -\epsilon^{\alphad\betad}\epsilon^{AB} \hat{E}^{(2)} \ ,
    \label{eq:anom_dim1}
\end{equation}
or equivalently by
\begin{equation}
    2\big\{ \tilde{\cG}^{\alphad A}, (\tilde{\cG}^{\betad B})^\dagger \big\}
    = \delta^{\alphad}_{\betad}\delta^A_B \hat{E}^{(2)} \ .
    \label{eq:anom_dim2}
\end{equation}
The fact that the anti-commutator on the left-hand side of \eqref{eq:anom_dim1} has the particular structure with respect to the indices $\alphad,\betad,A,B$ as dictated by the right-hand side is a nontrivial statement shown in \cite{Guo:2019pzk, Guo:private}  In particular, 
\begin{equation}
\hat{E}^{(2)}=    2\big\{ \tilde{\cG}^{\dot{+}1}, (\tilde{\cG}^{\dot{+}1})^\dagger \big\}
    =    2\big\{ \tilde{\cG}^{\dot{+}2}, (\tilde{\cG}^{\dot{+}2})^\dagger \big\}\ .
\end{equation}
One can show that \cite{Guo:2019pzk}
\begin{equation}
    \big[\cGt^{\alphad\Ad},\hat{E}^{(2)}\big]=0\ ,\label{eq:[cGt,E(2)]=0}
\end{equation}
which, given the structure \eqref{eq:anom_dim1},  implies that
\begin{equation}
    \big[\cGt^{\alphad A},\{ \tilde{\cG}^{\betad B }, \tilde{\cG}^{\gammad C} \}\big]=0\ .
\end{equation}

\subsection{The Gava-Narain operator and the symmetry algebra} \label{app:GN_alg}

Here we first show that the Gava-Narain operator (anti-)commutes with the generators of the free-theory \emph{left}-moving total-mode symmetry algebra, which here is the contracted large $\cN=4$ superconformal algebra (as given in Appendix~\ref{app:conventions}\@). This extends a similar result~\cite{Guo:2019pzk} for some of the contracted large $\cN=4$ generators.

Given a mode $\cA$ of the left-moving affine symmetry algebra, its (anti-)commutator with the right-moving supercharges $\tilde{G}^{\dot{\alpha}A}_{0}$ vanishes at any point on the moduli space. Let us expand these generators perturbatively around the free theory as $\cA = \cA^{(0)} + \cA^{(1)} + \cdots$ and $\tilde{G}^{\alphad A}=\tilde{G}^{\dot{\alpha}A(0)}_{0} +   \tilde{G}^{\dot{\alpha}A(1)}_{0} + \cdots$, where the superscript $(i)$ means the order of perturbative expansion (and not the copy index as in other parts of this paper).  Then we have
\begin{equation}
    \big[\cA^{(0)} +   \cA^{(1)} + \cdots , \tilde{G}^{\dot{\alpha}A(0)}_{0} +   \tilde{G}^{\dot{\alpha}A(1)}_{0} + \cdots\big]_{\pm} = 0 \ ,
\end{equation}
which vanishes at each order separately. By acting on a state $\ket*{\psi} = \ket*{\psi^{(0)}} + \ket*{\psi^{(1)}} + \cdots$ whose free-theory part is BPS, \textit{i.e.}\  $\tilde{G}^{\dot{\alpha}A(0)}_{0}\ket*{\psi^{(0)}}=0$, we find up to first order
\begin{align}
    0 &= \big[\cA^{(0)} +   \cA^{(1)}, \tilde{G}^{\dot{\alpha}A(0)}_{0} +  \tilde{G}^{\dot{\alpha}A(1)}_{0}\big]_{\pm}\big( \ket*{\psi^{(0)}} +   \ket*{\psi^{(1)}}\big) \nonumber\\
    &= \big[\cA^{(0)},\tilde{G}^{\dot{\alpha}A(0)}_{0}\big]_{\pm}\ket*{\psi^{(0)}} + \Big( \big[\cA^{(1)},\tilde{G}^{\dot{\alpha}A(0)}_{0}\big]_{\pm}\ket*{\psi^{(0)}} + \big[\cA^{(0)},\tilde{G}^{\dot{\alpha}A(1)}_{0}\big]_{\pm}\ket*{\psi^{(0)}} \nonumber\\
    &\hspace{8cm}+ \big[\cA^{(0)},\tilde{G}^{\dot{\alpha}A(0)}_{0}\big]_{\pm}\ket*{\psi^{(1)}}\Big) + \cdots \ .
\end{align}
The zeroth order condition $\big[\cA^{(0)},\tilde{G}^{\dot{\alpha}A(0)}_{0}\big]_{\pm}\ket*{\psi^{(0)}} =0$ follows trivially from the superconformal algebra, and at first order we get
\begin{align}
    0 &=\big[\cA^{(1)},\tilde{G}^{\dot{\alpha}A(0)}_{0}\big]_{\pm}\ket*{\psi^{(0)}} + \big[\cA^{(0)},\tilde{G}^{\dot{\alpha}A(1)}_{0}\big]_{\pm}\ket*{\psi^{(0)}} + \big[\cA^{(0)},\tilde{G}^{\dot{\alpha}A(0)}_{0}\big]_{\pm}\ket*{\psi^{(1)}}  \nonumber\\
    &= \big( \cA^{(0)}\tilde{G}^{\dot{\alpha}A(1)}_{0} \pm \tilde{G}^{\dot{\alpha}A(1)}_{0}\cA^{(0)} \pm \tilde{G}^{\dot{\alpha}A(0)}_{0}\cA^{(1)}\big)\ket*{\psi^{(0)}} + \big(\cA^{(0)}\tilde{G}^{\dot{\alpha}A(0)}_{0} \pm \tilde{G}^{\dot{\alpha}A(0)}_{0}\cA^{(0)}\big)\ket*{\psi^{(1)}} \nonumber\\
    &= \cA^{(0)} \big( \tilde{G}^{\dot{\alpha}A(1)}_{0}\ket*{\psi^{(0)}} + \tilde{G}^{\dot{\alpha}A(0)}_{0}\ket*{\psi^{(1)}}\big) \notag\\
        &\qquad\qquad
    \pm \tilde{G}^{\dot{\alpha}A(0)}_{0}\big( \cA^{(0)}\ket*{\psi^{(1)}} + \cA^{(1)}\ket*{\psi^{(0)}} \big) 
    \pm \tilde{G}^{\dot{\alpha}A(1)}_{0} \cA^{(0)}\ket*{\psi^{(0)}} \nonumber\\
    &= \cA^{(0)} \tilde{\mathcal{G}}^{\dot{\alpha}A}\ket*{\psi^{(0)}} \pm 
    \tilde{G}^{\dot{\alpha}A(0)}_{0} \ket*{\phi^{(1)}}
    \pm 
    \tilde{G}^{\dot{\alpha}A(1)}_{0} \ket*{\phi^{(0)}} 
    \nonumber\\
    &= \cA^{(0)} \tilde{\mathcal{G}}^{\dot{\alpha}A}\ket*{\psi^{(0)}} \pm \tilde{\mathcal{G}}^{\dot{\alpha}A}\ket*{\phi^{(0)}} \nonumber\\
    &= \big[\cA^{(0)},\tilde{\mathcal{G}}^{\dot{\alpha}A}\big]_{\pm}\ket*{\psi^{(0)}} \ ,
\end{align}
where we used the Gava-Narain operator's action on free-theory BPS states given in \eqref{eq.GN_psi_def}, and we defined
$\ket*{\phi}\equiv\cA\ket*{\psi}$ whose expansion gives $\ket*{\phi^{(0)}}=\cA^{(0)}\ket*{\psi^{(0)}}$
and $\ket*{\phi^{(1)}}=\cA^{(1)}\ket*{\psi^{(0)}}+\cA^{(0)}\ket*{\psi^{(1)}}$.

The above shows that $[\cA^{(0)},\tilde{\mathcal{G}}^{\dot{\alpha}A}]_{\pm}$ vanishes on free-theory BPS states.  It is easy to see that $[\cA^{(0)},\tilde{\mathcal{G}}^{\dot{\alpha}A}]_{\pm}$ vanishes also on non-BPS states, because of the projection $\cP$ onto the BPS subspace in the definition of $\cGt^{\alphad A}$ and because the left-moving generator $\cA^{(0)}$ maps a non-BPS state into a non-BPS state.  This proves $[\cA^{(0)},\tilde{\mathcal{G}}^{\dot{\alpha}A}]_{\pm}=0$ as an operator equation.

Exactly along the same line, one can show that the (anti)commutation relation between $\cGt^{\alphad A}$ and free-theory \emph{right}-moving zero-mode generators takes the same form as the (anti)commutation relation between $\tilde{G}_0^{\alphad A}$ and the corresponding right-moving zero-mode generators in the interacting theory. For example,
\begin{subequations}\label{eq:[At,cGt]}
\begin{align}
    \big[\Jt^{a(0)}_0,\cGt^{\alphad A}\big]&={1\over 2}(\sigma^a)^\alphad{}_{\betad}\, \cGt^{\betad A}\ ,\label{eq:[Jt,cGt]}\\
    \big[\tilde{L}_0^{(0)},\cGt^{\alphad A}\big]&=0\ ,\label{eq:[Lt,cGt]}\\
    \big\{\tilde{\psi}^{\alphad \Ad(0)}_0,\cGt^{\betad A}\big\}&=0\ ,\label{eq:[psit,cGt]}
\end{align}
\end{subequations}
which correspond to \eqref{eq.[J,G]},  \eqref{eq.[L,G]}, and \eqref{eq.[G,psi]}, respectively.
To show \eqref{eq:[psit,cGt]}, we assumed that the bosonic zero mode $\tilde{\alpha}_0^{\Ad A}$ vanishes in the interacting theory.
Some of these relations were shown in \cite{Guo:2019ady}.

\section{Symmetric functions}
\label{app:symmPoly}

The Schur-Weyl formalism for decomposing the Hilbert space of symmetric orbifold CFTs introduced in Section~\ref{sec:S-W} heavily relies on results from the study of symmetric polynomials. In this Appendix we list basic facts about these, which we draw from throughout the main part of this paper.

\begin{figure}[tb]
    \centering
    \begin{tikzpicture}[scale=1.25]
 \draw (0  ,0) rectangle +(1.6ex,1.6ex);
 \draw (1.6ex,0) rectangle +(1.6ex,1.6ex);
 \draw (3.2ex,0) rectangle +(1.6ex,1.6ex);
 \draw (4.8ex,0) rectangle +(1.6ex,1.6ex);
 \draw (6.4ex,0) rectangle +(1.6ex,1.6ex);
 \draw (8ex,0) rectangle +(1.6ex,1.6ex);
 \draw (0  ,-1.6ex) rectangle +(1.6ex,1.6ex);
 \draw (1.6ex,-1.6ex) rectangle +(1.6ex,1.6ex);
 \draw (3.2ex,-1.6ex) rectangle +(1.6ex,1.6ex);
 \draw (4.8ex,-1.6ex) rectangle +(1.6ex,1.6ex);
 \draw (6.4ex,-1.6ex) rectangle +(1.6ex,1.6ex);
 \draw (0,-3.2ex) rectangle +(1.6ex,1.6ex);
 \draw (1.6ex,-3.2ex) rectangle +(1.6ex,1.6ex);
 \draw (3.2ex,-3.2ex) rectangle +(1.6ex,1.6ex);
 \draw (0,-4.8ex) rectangle +(1.6ex,1.6ex);
 \draw (1.6ex,-4.8ex) rectangle +(1.6ex,1.6ex);
 \draw (0,-6.4ex) rectangle +(1.6ex,1.6ex)  node  [below,xshift=-1.2ex,yshift=-2ex] {$\vdots$};
 \draw[latex-latex] (0, 0.8ex) -- +(9.6ex,0) node  [font=\footnotesize,right,xshift=-0.3ex,yshift=-0.2ex] {$\lambda_1$};
 \draw[latex-latex] (0,-0.8ex) -- +(8.0ex,0) node [font=\footnotesize,right,xshift=-0.3ex,yshift=-0.2ex] {$\lambda_2$};
 \draw[latex-latex] (0,-2.4ex) -- +(4.7ex,0) node [font=\footnotesize,right,xshift=-0.3ex,yshift=-0.2ex] {$\lambda_3$};
 \draw[latex-latex] (0,-4.0ex) -- +(3.2ex,0) node [font=\footnotesize,right,xshift=-0.3ex,yshift=-0.2ex] {$\lambda_4$};
 \draw[latex-latex] (0,-5.6ex) -- +(1.6ex,0) node [font=\footnotesize,right,xshift=-0.3ex,yshift=-0.2ex] {$\lambda_5$};
\end{tikzpicture}
    \caption{\sl A Young diagram labeled by $\{\lambda_i\}$, with $\lam_1\geq\lam_2\geq\cdots$.}
    \label{fig:Y_diag}
\end{figure}
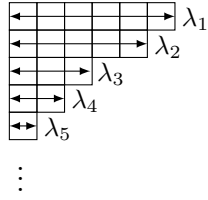

Homogeneous symmetric polynomials $Q_n(x)$ of degree $n$ in $b$ variables $x = (x_1,\dots,x_b)$ form a vector space and satisfy $Q_n(\sigma{x}) = Q_n({x})$, where $\sigma\in S_b$ permutes the variables via $\sigma{x} = (x_{\sigma(1)}, \dots, x_{\sigma(b)})$. While there are many bases on this space, we focus on two in particular: the Schur polynomials and the power sum polynomials. Firstly, Schur polynomials $S_{\lam}({x})$ are indexed by partitions $\lam\vdash n$ into at most $b$ parts, which we denote by 
\begin{equation}
    \lam = \{\lam_1,\dots,\lam_b\} \ ,\qquad \lam_1\geq \lam_2 \geq \cdots\geq \lam_b\ge 0 \ ,\qquad \sum_{j=1}^{b} \lam_{j} = n \ .
\end{equation}
Equivalently, Schur polynomials can be labelled by Young diagrams with $n$ boxes and at most $b$ rows; see Figure \ref{fig:Y_diag}.

While Schur polynomials have various determinantal formulae \cite{Fulton:1991} from which they can be defined, for the purposes of this paper we will only need their relation to the power sum polynomials.

Forming another basis on the space of homogeneous symmetric polynomials of degree $n$ and $b$ variables, the power sum polynomials are defined simply as
\begin{align} \label{eq:def_power_sum}
    P^{({\bf i})}({x}) \equiv P_1({x})^{i_1}\,\cdots\, P_n({x})^{i_n}\ ,\qquad P_{\alpha}({x}) \equiv x_1^{\alpha} + \cdots + x_b^{\alpha} \ ,
\end{align}
where ${\bf i}=(i_1,\dots,i_n)$, with $\sum_{\alpha=1}^{n} \alpha i_{\alpha} = n$, is a partition of $n$. The relation between these two bases is given by
\begin{equation} \label{eq:Schur_ito_power_sum}
    S_\lambda({x}) = \sum_{{\bf i}\,\vdash n} \frac{\omega_\lambda({\bf i})}{z({\bf i})} P^{({\bf i})}({x})\ ,\qquad z({\bf i}) \equiv \prod_{\alpha=1}^n i_\alpha!\,\alpha^{i_\alpha}\ ,
\end{equation}
where $\omega_{\lam}$ is defined in terms of power sum polynomials and the discriminant $\Delta({x})$ by
\begin{equation} \label{eq:def_omega_lambda}
    \omega_\lambda({\bf i}) = \big[\Delta({x})\, P^{({\bf i})}({x})\big]_{\ell}\ ,\quad \Delta({x}) = \prod_{1\leq j<k}^{b}(x_j-x_k)\ ,\quad \ell\equiv (\lambda_1+b-1,\lambda_2+b-2,\dots,\lambda_b)\ ,
\end{equation}
and $[Q({x})]_\lambda \equiv (\text{coefficient of $x_1^{\lambda_1}\cdots\, x_n^{\lambda_n}$})$ for a symmetric polynomial $Q({x})$. 

The Schur polynomials for some simple Young diagrams are:
\begin{align}
\label{Schur_poly_example}
\begin{split}
\YTscriptsize
    S_{\ydiagram{1}}({x})&=P_1=\sum_i x_i\ ,\\
    S_{\ydiagram{2}}({x})
    &=\frac12 P_1^2 +\frac12 P_2
    =\sum_{i<j} x_i x_j + \sum_i x_i^2\ ,
    \\
    S_{\ydiagram{1,1}}({x})
    &=\frac12 P_1^2-\frac12 P_2
    = \sum_{i<j} x_i x_j\ ,\\
    S_{\ydiagram{3}}({x})    &=\frac16 P_1^3+\frac12 P_1 P_2 + \frac13 P_3
    = \sum_{i<j<k} x_i x_j x_k  + \sum_{i\neq j} x_i^2 x_j +\sum_i x_i^3\ ,\\
    S_{\ydiagram{2,1}}({x})    &=\frac13 P_1^3- \frac13 P_3
    = 2\sum_{i<j<k} x_i x_j x_k  + \sum_{i\neq j} x_i^2 x_j\ ,\\
    S_{\ydiagram{1,1,1}}({x})   &=\frac16 P_1^3-\frac12 P_1 P_2 + \frac13 P_3
    = \sum_{i<j<k} x_i x_j x_k  \ .
\end{split}
\end{align}

Schur polynomials satisfy the Cauchy identity
\begin{equation} \label{eq:Cauchy_id}
    \sum_{\lam} S_{\lam}(x) S_{\lam}(\xt) = 
    \prod_{j=1}^{b} \prod_{k=1}^{\tilde b}\frac1{1-x_j \xt_k} \ ,
\end{equation}
where $\xt=(\xt_1,\dots,\xt_{\tilde b})$ is a set of $\tilde{b}$ variables.
The sum is over all Young diagrams with any number of boxes, although $S_\lambda(x)$ vanishes if $\lambda$ has more than $b$ rows and $S_\lambda(\xt)$ vanishes if $\lambda$ has more than $\tilde{b}$ rows.

For application to the symmetric orbifold of $T^4$, a theory containing both bosonic and fermionic states, super Schur functions \cite{Macdonald1992} (also sometimes called ``hook Schur functions'') are required. Super Schur functions $S_\lambda(x|x')$ are functions of two sets of variables $x=(x_1,\dots,x_b)$ and $x'=(x'_1,\dots,x'_f)$ and, for our purposes, we can define them analogously to~\eqref{eq:Schur_ito_power_sum}, \textit{i.e.}, as
\begin{equation} \label{eq:Super_Schur_power_sum}
    S_\lambda(x|x') = \sum_{{\bf i}\,\vdash n} \frac{\omega_\lambda({\bf i})}{z({\bf i})} P^{({\bf i})}({x}|x')\ ,
\end{equation}
with $\mathbf{i}$ again a partition of $n$ and $\omega_{\lam}$ and $z({\bf i})$ are still given by the bosonic formulae given in \eqref{eq:Schur_ito_power_sum} and \eqref{eq:def_omega_lambda}. The $P^{({\bf i})}({x}|x')$ in \eqref{eq:Super_Schur_power_sum} are now the super power sum polynomials
\begin{align}
    P^{({\bf i})}(x|x') \equiv P_1(x|x')^{i_1}\,\cdots\, P_n(x|x')^{i_n} \ ,\quad P_{\alpha}(x|x') \equiv \sum_{i=1}^b (x_i)^{\alpha} - \sum_{i'=1}^{f} (x'_{i'})^{\alpha} \ .
\label{eq:def_super_power_sum}
\end{align}
The super Schur function $S_\lambda$ vanishes unless $\lambda$ satisfies the $(b|f)$-hook condition.

Physically, if $\cH$ is a Hilbert space of $b$ bosonic and $f$ fermionic dimensions, $P_1(x|\tilde{x})$ can be regarded as the trace with $(-1)^F$ in $\cH$ of some operator $g$ with eigenvalues $x_1,\dots,x_b$ in the bosonic subspace and $x'_1,\dots,x'_f$ in the fermionic subspace; namely, $P_1(x|x')={\tr}[(-1)^F g]$. More generally, $P_{\alpha}(x|x')={\tr}[(-1)^Fg^{\alpha}]$.   If $g=p^{\hat{k}}q^{L_0}y^{2J^3_0}$ as used in Section~\ref{ssec:SW_T4} (for example in \eqref{eq:1-strand_z(p,q,y)}), and if we write $P_1={\tr}[(-1)^Fg]=z_1(p,q,y)$, then $P_{\alpha}={\tr}[(-1)^F g^{\alpha}]=z_1(p^{\alpha},q^{\alpha},y^{\alpha})$.  This is used to evaluate the various partition functions and indices of Sections~\ref{ssec:SW_T4}, \ref{ssec:T4_MEG} and \ref{sec:REG}.

The super Schur functions satisfy the Cauchy identity
\begin{equation}
    \sum_\lambda S_\lambda(x|x') S_\lambda(\xt|\xt')
    =
    \frac{\prod_{i,j'}(1-x_i \xt'_{j'})
    \prod_{i',j}(1-x'_{i'} \xt_{j})
    }{
    \prod_{i,j}(1-x_i \xt_{j})
    \prod_{i',j'}(1-x'_{i'}\xt'_{j'})
    }\ ,
    \label{eq:Cauchy_id_general_app}
\end{equation}
where $\xt=(\xt_1,\dots,\xt_{\tilde{b}})$ and $\xt'=(\xt'_1,\dots,\xt'_{\tilde{f}})$ are another set of bosonic and fermionic variables. The sum is over all Young diagrams $\lam$ with any number of boxes, although $S_\lambda(x|x')$ vanishes unless $\lambda$ is a $(b|f)$-hook and $S_\lambda(\xt|\xt')$ vanishes unless $\lambda$ is a $(\tilde{b}|\tilde{f})$-hook.

\section{More on the Schur-Weyl duality}
\label{app:S-W++}


In Section \ref{ssec:bosonEx} we discussed how the left- and right-moving multi-strand Hilbert spaces can be decomposed into irreducible representations (irreps) of the general linear and symmetric groups according to the Schur-Weyl duality \eqref{SchurWeyl_bosonic}. Here we give some additional details about the Schur-Weyl duality and its application to the situation of interest for this paper. We will try to make our presentation as physical and practical as possible; more mathematics-oriented readers are referred to textbooks such as \cite{Fulton:1991,SymmRepTheory}.

Given a vector space $V\cong \mathbb{C}^d$, we can consider the simultaneous commuting action of $GL(V)\cong GL(d,\mathbb{C})$ and  $S_n$ on the $n$th tensor product space $V^{\otimes n}$ as follows:  $g\in GL(V)$ acts on $v_1\otimes \cdots\otimes v_n\in V^{\otimes n}$ as standard tensor transformation as
\begin{align}
        g(v_1\otimes \cdots \otimes v_n ) &= 
        gv_1\otimes \cdots \otimes gv_n  \ ,
        \label{eq:GL_Vn_action}
\end{align}
while $\sigma\in S_n$ acts as permutation of tensor factors as\footnote{Here we really mean that a representation of the element $\sigma\in S_n$ acts by permuting tensor factors; below this will be specified more carefully. }
\begin{align}
       \sigma   (v_1\otimes \cdots \otimes v_n ) &= 
        v_{\sigma^{-1}(1)}\otimes\cdots \otimes v_{\sigma^{-1}(n)}\ .\label{eq:Sn_Vn_action}
\end{align}
In the context of symmetric orbifold CFTs, we can think of \eqref{eq:Sn_Vn_action} as  permutation $\sigma$ acting on the strand labels $[I]$ as\footnote{When there are fermionic strands, in the second equality of \eqref{eq:Sn_Vn_action_strand}, we must take into account the possible minus sign coming from exchanging fermionic strands.  The same sign must be present in \eqref{eq:Sn_Vn_action}. Furthermore, $S_n$-irreducible spaces  $V_\lambda$ such as \eqref{eq:V2,V11} must be suitably generalized to take into account such signs.
}
\begin{align}
       \sigma \!  \left(\ket{v_1}^{[1]}\cdots \ket{v_n}^{[n]} \right) 
       &= 
       \ket{v_1}^{[\sigma(1)]}\cdots \ket{v_n}^{[\sigma(n)]} \notag\\
       &= 
       \ket*{v_{\sigma^{-1}(1)}}^{[1]}\cdots \ket*{v_{\sigma^{-1}(n)}}^{[n]}  .
       \label{eq:Sn_Vn_action_strand}
\end{align}

In the case of $n=2$, the tensor product space $V^{\otimes 2}$ decomposes into the $GL(V)$ irrep spaces (or more generally, irreducible $GL(V)$-modules) $V_{\YTscriptsize\ydiagram{2}}$ and $V_{\YTscriptsize\ydiagram{1,1}}$, corresponding to elements of $V^{\otimes 2}$ which transform symmetrically and antisymmetrically with respect to $S_2$. These representation spaces $V_{\lam}$ are simply the symmetric and alternating products\footnote{The representation spaces $V_{\lam}$ can also be described by the action of the so-called Schur functors $\mathbb{S}_{\lam}$ on $V$, which in the case of $n=2$ are $\mathbb{S}_{\,\YTscriptsize\ydiagram{2}}: V \mapsto V_{\YTscriptsize\ydiagram{2}}$ and $\mathbb{S}_{\,\YTscriptsize\ydiagram{1,1}}:V \mapsto V_{\YTscriptsize\ydiagram{1,1}}$.}
\begin{equation}
    V_{\YTscriptsize\ydiagram{2}}\cong\mathrm{Sym}^2(V) \quad,\quad V_{\YTscriptsize\ydiagram{1,1}}\cong V \wedge V \ .
    \label{eq:V2,V11}
\end{equation}

Equally, for general $n$, the $n$th symmetric and alternating powers are also irreducible representation spaces corresponding to the Young diagrams $\lam=\{n\}$ and $\lam=\{1,\dots,1\}$ respectively, which intuitively generalise the $n=2$ case above to
\begin{equation}
    V_{\{n\}}\cong\mathrm{Sym}^n(V) \quad,\quad V_{\{1,\dots,1\}} \cong \wedge^n V \ .
\end{equation}
However, the form of $V_{\lam}$ for a general $\lam\vdash n$ is less intuitive and involves combinations of symmetric and alternating products, the elements of which can be constructed from the so-called Young symmetrizers in the group algebra\footnote{%
The group algebra $\bbC[G]$ of a group $G$ is, roughly speaking, an algebraic structure formed by taking all formal linear combinations of elements of $G$ with coefficients in $\bbC$. More precisely, to each group element $\sigma\in G$ we associate a basis element $e_{\sigma}\in \bbC[G]$ and the multiplication in the algebra is inherited from the group multiplication as $e_{\sigma}\circ e_{\sigma'} = e_{\sigma\sigma'}$.  More generally, the group ring $\mathcal{K}[G]$ is a free module over the ring $\mathcal{K}$. If $\mathcal{K}$ is commutative then $\mathcal{K}[G]$ forms an algebra.}  $\mathbb{C}[S_n]$.  So, 
before turning to the Schur-Weyl duality for $V^{\otimes n}$, we need to
recall a few facts about $\mathbb{C}[S_n]$ and
the representations of $S_n$.

\bigskip
Given a Young tableau\footnote{A Young tableau is a Young diagram $\lambda\vdash n$ whose $n$ boxes are filled with the integers $1,\dots,n$ in an arbitrary order.  Given a Young diagram $\lambda$, there are $n!$ different tableaux.  
A standard Young tableau is one in which the entries in each row and each column are strictly increasing. 
The canonical Young tableau is one whose entries increase left-to-right in each row by consecutive integers and is unique given shape $\lambda$.} $T$ of shape $\lam$, define $R_{T}$ and $C_{T}$ to respectively be the set\footnote{Actually $R_{T}$ and $C_{T}$ are subgroups of $S_n$, corresponding to the so-called Young subgroups respectively for $\lam$ and its conjugate $\lam'$, obtained by exchanging rows and columns in $\lam$.} of row- and column-preserving\footnote{By this we mean permutations that leave a tableau's labels in the same row or column.} permutations for $T$, where the permutations act on the labels of the tableau (we will present explicit examples of all this below). In the group algebra $\bbC[S_n]$ we can then define the elements
\begin{equation} \label{eq:rT_c_T_def}
    r_{T} \equiv \sum_{\sigma\in R_{T}} e_\sigma \in \mathbb{C}[S_n] \quad,\quad c_{T} \equiv \sum_{\sigma\in C_{T}} (-1)^{\deg(\sigma)} e_\sigma \in \mathbb{C}[S_n] \ ,
\end{equation}
with $\deg(\sigma)$ the degree\footnote{Decomposing a permutation $\sigma$ into a product of transpositions, $\deg(\sigma)$ is given by the number of transpositions in this decomposition. While such a decomposition is not generally unique, the value of $\sgn(\sigma)=(-1)^{\deg(\sigma)}$ is for a given element $\sigma$. Alternatively and equivalently, $\sgn(\sigma)$ is equal to the determinant of the permutation matrix representation of $\sigma$.} of the permutation $\sigma$. The Young symmetrizer $s_{T}\in \mathbb{C}[S_n]$, associated with the tableau $T$, is then defined as
\begin{equation} \label{eq:Young_symm_def}
    s_{T} \equiv r_T\circ c_T \ .
\end{equation}
Given a Young symmetrizer $s_{T}$, its image 
in the group algebra,
\begin{equation} \label{eq:Specht_def}
    \mathbb{C}[S_n]\circ s_T \equiv M_{\lam} \ ,
\end{equation}
gives an irreducible left-representation,\footnote{The representation $M_\lambda$ is sometimes called the Specht module.  The word Specht module often means the particular representation in the space of so-called Young tabloids which are obtained by identifying Young tableaux by permutations of the entries within each row.
}
which depends only on the shape $\lambda$ of the tableau $T$.\footnote{Namely, for two tableaux $T_1$, $T_2$ of the same shape $\lam$, the irreducible representations $\mathbb{C}[S_n]\circ s_{T_1}$ and $\mathbb{C}[S_n]\circ s_{T_2}$ are isomorphic.}  We can say that acting with the elements of $S_n$ on the left of $s_T$ generates the representation $M_\lambda$.
Letting $x\in\bbC[S_n]$, we can write the action of $\sigma\in S_n$ on $M_\lambda$ as
\begin{equation}
    \sigma\circ (x \circ s_T)
    =
    (\sigma\circ x) \circ s_T
    ~\in~ \bbC[S_n] \circ s_T
    =M_\lam,
\end{equation}
which confirms that $M_\lambda$ gives a left-representation of $S_n$.  

The dimension of the representation $M_{\lam}$ is known to be equal to the number of standard Young tableaux\footnote{Note that Young symmetrizers for different standard Young tableau $T$ of shape $\lam$ are not necessarily orthogonal~\cite{STEMBRIDGE2011576}.} of shape $\lam$, as computed by the hook length formula
\begin{equation} \label{eq:M_lam_dim}
    \dim(M_{\lam}) = f^\lambda \equiv \frac{n!}{\prod_{(i,j)\in \lam} h_{ij}} \ ,
\end{equation}
where $(i,j)$ are the (row,$\,$column) numbers of the box in the diagram $\lam$ and $h_{ij}$ is the hook length of the box.
Note that this does not mean that the irrep space \eqref{eq:Specht_def} is spanned by the Young symmetrizers of the standard tableaux; it is spanned by some $f^\lambda$ Young symmetrizers that depend on $T$.

This construction decomposes $\bbC[S_n]$ into a direct sum of irreps of $S_n$ as
\begin{equation} \label{eq:alg_decomp}
    \mathbb{C}[S_n] \cong \bigoplus_{\lam\vdash n} (M_{\lam})^{\oplus \dim(M_{\lam})} \ .
\end{equation}
We see that the multiplicity of the irrep $M_\lambda$ appearing in the decomposition is given by the irrep's dimension, $\dim(M_{\lam})$.\footnote{In more detail,  by acting with $\bbC[S_n]$ also on the right of \eqref{eq:Specht_def}, we generate the right-representation $M_\lambda$.  Combining the left and right actions, we find that  $\bbC[S_n]\circ s_T \circ\bbC[S_n]=M_\lambda\otimes M_\lambda$, which can be used to show that in fact $\bbC[S_n]=\bigoplus_{\lambda\vdash n} M_\lambda\otimes M_\lambda$.  If we forget about the right representation $M_\lambda$, this reduces to \eqref{eq:alg_decomp}. }

In our physical applications of Section~\ref{sec:S-W} we will need to decompose the tensor product of two $S_n$ irreducible representation spaces; this is given in general by
\begin{equation} \label{eq:M_lam_M_mu}
    M_{\lam}\otimes M_{\mu} = \bigoplus_{\nu} C_{\lam\mu\nu} M_{\nu} \ ,
\end{equation}
where $\lam,\mu,\nu$ are all partitions of $n$ and the $C_{\lam\mu\nu}$ are the so-called Kronecker coefficients. In particular, we require the condition on $\lam$ and $\mu$ for which the decomposition contains the trivial representation, labelled by the partition $\nu=\{n\}$. This is then a question of when the Kronecker coefficients $C_{\lam\mu\{n\}}\neq0$. It can be shown that $C_{\lam\mu\{n\}}=1$ if $\mu=\lam$ and 0 otherwise (see exercise 4.51 of \cite{Fulton:1991}). Equally, the Kronecker coefficients can be computed from the characters of the representations involved in \eqref{eq:M_lam_M_mu}.

\bigskip
Coming back to the tensor product space $V^{\otimes n}$,
$S_n$ has a natural action on it by permuting tensor factors as in \eqref{eq:Sn_Vn_action}. Given this group representation $\cR$, there is a corresponding algebra representation $\tilde{\cR}$ that defines the natural action of $\mathbb{C}[S_n]$ on $V^{\otimes n}$, where $\cR(\sigma) = \tilde{\cR}(e_\sigma)$. Under this action, the image of the Young symmetrizer \eqref{eq:Young_symm_def} in $V^{\otimes n}$ is
\begin{equation} \label{eq:s_lam_GL}
    \tilde{\cR}(s_{T})\, V^{\otimes n} \equiv V_\lambda \ .
\end{equation}
Because the $GL(V)$ and $S_n$ actions commute, this $V_\lambda$ gives a representation space of $GL(V)$, which turns out to be an 
irreducible  representation.
The dimension of this irrep is given by the so-called hook content formula\footnote{The hook content formula equivalently counts the number of semi-standard Young tableaux of a fixed shape and with labels in $\{1,\dots,\dim(V)\}$. These are tableaux whose labels need only weakly increase along rows and strictly increase down columns.}
\begin{equation} \label{eq:V_lam_dim}
    \dim(V_\lambda) = \prod_{(i,j)\in\lam} \frac{\dim(V) + j-i}{h_{ij}} \ ,
\end{equation}
which is equally given by setting all arguments to 1 in the Schur polynomial \eqref{eq:Schur_ito_power_sum} for the same Young diagram $\lam$; \textit{i.e.}, $\dim(V_\lam)= S_{\lam}(1)$.

Just as for the case of the group algebra $\bbC[S_n]$ (see below \eqref{eq:Specht_def}), we can generate the $S_n$-representation $M_\lambda$ by the left action of $S_n$.  Namely, if we act with $\sigma\in S_n$ on \eqref{eq:s_lam_GL}, we obtain
\begin{equation}
    \tilde{\cR}(\sigma \circ s_{T})\, V^{\otimes n},
\end{equation}
which produces different copies of $V_\lambda$, associated with Young symmetrizers $s_{T'}$ with $T'\neq T$.  By letting $\sigma$ over all $S_n$ elements, these copies of $V_\lambda$ span the representation $M_\lambda$.  In other words, we have generated a $GL(V)\otimes S_n$ representation $V_\lambda\otimes M_\lambda$.  

Considering all possible Young diagrams $\lambda$ leads to a
 $GL(V)\otimes S_n$ irrep decomposition of $V^{\otimes n}$:
\begin{equation} \label{eq:SW_app}
    V^{\otimes n} \cong \bigoplus_{\lam} V_{\lam} \otimes M_{\lam} \ ,
\end{equation}
which is the Schur-Weyl duality.
Below we discuss explicit examples of the Schur-Weyl decomposition of $V^{\otimes n}$ using concrete vector spaces $V$.


\subsection{Explicit low $n$ examples}
\label{app:SW_examples}

\subsubsection*{The case of $n=2$}

For $n=2$, the two irreducible representations of $S_2$ are labelled by the Young diagrams $\lam=\YTnormalsize\ydiagram{2},\ydiagram{1,1}$ each of which having only one standard (and thus canonical) tableau, namely
\begin{equation}
    \CYon T_{\YTscriptsize\ydiagram{2}} = \ytableausetup{boxsize=1.1em}\ytableaushort{12} \quad,\quad T_{\YTscriptsize\ydiagram{1,1}} = \ytableausetup{boxsize=1.1em}\ytableaushort{1,2} \ .
\end{equation}
The respective Young symmetrizers are given by
\begin{subequations} \label{eq.YS_N=2}
\begin{align}
    s_{T_{\YTscriptscriptsize\ydiagram{2}}} &= \big(e_{I} + e_{(12)}\big)\circ e_{I} = e_{I} + e_{(12)} \ , \label{eq.YS_N=2_1}\\
    s_{T_{\YTscriptscriptsize\ydiagram{1,1}}} &= e_{I} \circ \big(e_{I} - e_{(12)}\big)  = e_{I} - e_{(12)} \ , \label{eq.YS_N=2_2}
\end{align}
\end{subequations}
where we are using cycle notation for permutations, dropping cycles of length one, and $e_{I}$ is the basis element corresponding to the identity element $(1)(2)$ of $S_2$. By considering the image of the Young symmetrizers \eqref{eq.YS_N=2} in $\mathbb{C}[S_2]$ one finds $S_2$ representations
\begin{subequations} \label{eq:M_lam_N=2}
    \begin{align}
        \mathbb{C}[S_2]\circ s_{T_{\YTscriptscriptsize\ydiagram{2}}} &= \mathrm{span}\big\{s_{T_{\ydiagram{2}}}\big\} \cong M_{\ydiagram{2}}\ ,\\
        \mathbb{C}[S_2]\circ s_{T_{\ydiagram{1,1}}} &= \mathrm{span}\big\{s_{T_{\ydiagram{1,1}}}\big\} \cong M_{\ydiagram{1,1}}\ .
    \end{align}
\end{subequations}
Both $M_{\ydiagram{2}}$ and  $M_{\ydiagram{1,1}}$ are one-dimensional,
in agreement with the dimensions computed from \eqref{eq:M_lam_dim}.

To be concrete here we apply this to the right-moving Hilbert space of states in the bosonic toy model discussed in Section \ref{ssec:bosonEx}, namely the space
\begin{equation} \label{eq.BosonicToyModel}
    \Vt = \mathrm{span}\Big\{\ket{+}\, ,\ \ket{-}\Big\} \ .
\end{equation}
The $n=2$ tensor product space is spanned by the states
\begin{equation} \label{eq:V2_span}
    \Vt^{\otimes 2} = \mathrm{span}\Big\{\ket{++}\, ,\ \ket{+-}\, ,\ \ket{-+}\, ,\ \ket{--}\Big\} \ ,
\end{equation}
where $\ket{+-}\equiv \ket{+}^{[1]}\ket{-}^{[2]}$ and so on.
We can find how the elements of $S_2$ act on $\Vt^{\otimes 2}$ by using \eqref{eq:Sn_Vn_action_strand}.
We find that the representation subspaces in the Schur-Weyl decomposition \eqref{eq:SW_app} are
\begin{subequations}
\begin{align}
      \tilde{\cR}(s_{T_{\YTscriptscriptsize\ydiagram{2}}}) \,\Vt^{\otimes 2}&= \mathrm{span}\Big\{\ket{++}\, ,\ \ket{+-} + \ket{-+}\, ,\ \ket{--} \Big\} \cong \Vt_{\YTscriptscriptsize\ydiagram{2}}\ ,\label{eq:N=2_states_Rep_2}\\
      \tilde{\cR}(s_{T_{\YTscriptscriptsize\ydiagram{1,1}}}) \,
      \Vt^{\otimes 2}  &= \mathrm{span}\Big\{\ket{+-} - \ket{-+}\Big\} \cong \Vt_{\YTscriptscriptsize\ydiagram{1,1}}\ ,\label{eq:N=2_states_Rep_11}
\end{align}
\end{subequations}
the dimensions of which match those computed from \eqref{eq:V_lam_dim} with $d=\dim(\Vt)=2$. 
One can easily check that each line is separately a simultaneous representation of $GL(2)$ and $S_2$.
The above is equivalent to the discussion of the subspaces of symmetric and antisymmetric 2-tensors in Section~\ref{ssec:bosonEx}.

\subsubsection*{The case of $n=3$}

For $n=3$, the three irreducible representations are labelled by $\lam=\YTnormalsize\ydiagram{3},\ydiagram{2,1},\ydiagram{1,1,1}$ for which the canonical tableaux are
\begin{equation}
    \CYon T_{\YTscriptsize\ydiagram{3}} = \ytableausetup{boxsize=1.1em}\ytableaushort{123} \quad,\quad \CYon T_{\YTscriptsize\ydiagram{2,1}} = \ytableausetup{boxsize=1.1em}\ytableaushort{12,3} \quad,\quad \CYon T_{\YTscriptsize\ydiagram{1,1,1}} = \ytableausetup{boxsize=1.1em}\ytableaushort{1,2,3} \ ,
\end{equation}
and for which the corresponding Young symmetrizers are
\begin{subequations}
\begin{align}
    s_{T_{\YTscriptscriptsize\ydiagram{3}}} &= e_{I} + e_{(12)} + e_{(23)} + e_{(13)} + e_{(123)} + e_{(132)}  \ , \label{eq.YS_N=3_1}\\
    s_{T_{\YTscriptscriptsize\ydiagram{2,1}}} &= \big(e_{I} + e_{(12)}\big)\circ\big( e_{I} - e_{(13)}\big) = e_{I} + e_{(12)} - e_{(13)} - e_{(132)} \ , \label{eq.YS_N=3_2}\\
    %
    %
    s_{T_{\YTscriptscriptsize\ydiagram{1,1,1}}} &= e_{I} - e_{(12)} - e_{(23)} - e_{(13)} + e_{(123)} + e_{(132)} \ . \label{eq.YS_N=3_4}
\end{align}
\end{subequations}
The images of these Young symmetrizers in $\mathbb{C}[S_3]$ yield
\begin{subequations} \label{eq:M_lam_N=3}
    \begin{align}
        \mathbb{C}[S_3]\circ s_{T_{\YTscriptscriptsize\ydiagram{3}}} &= \mathrm{span}\big\{s_{T_{\ydiagram{3}}}\big\} \cong M_{\ydiagram{3}}\ ,\\
        \mathbb{C}[S_3]\circ s_{T_{\YTscriptscriptsize\ydiagram{2,1}}} &= \mathrm{span}\big\{s_{T^{[1]}_{\ydiagram{2,1}}} \ ,\ e_{(13)}\circ s_{T^{[1]}_{\ydiagram{2,1}}}\big\} \cong M_{\ydiagram{2,1}}\ ,\\
        %
        %
        \mathbb{C}[S_3]\circ s_{T_{\ydiagram{1,1,1}}} &= \mathrm{span}\big\{s_{T_{\ydiagram{1,1,1}}}\big\} \cong M_{\ydiagram{1,1,1}}\ ,
    \end{align}
\end{subequations}
whose dimensions again agree with the ones computed from \eqref{eq:M_lam_dim}.

Let us consider again the bosonic toy model \eqref{eq.BosonicToyModel}, where
\begin{align} \label{eq:V3_span}
    \Vt^{\otimes 3} &= \mathrm{span}\Big\{\ket{{+}{+}{+}} ,\ \ket{{-}{+}{+}} ,\ \ket{{+}{-}{+}} ,\ \ket{{+}{+}{-}} ,\nonumber\\
    &\qquad\qquad\ket{{+}{-}{-}} ,\ \ket{{-}{+}{-}} ,\ \ket{{-}{-}{+}} ,\ \ket{{-}{-}{-}}\Big\} \ 
\end{align}
with $\ket{{+}{+}{-}}\equiv \ket{+}^{[1]}\ket{+}^{[2]}\ket{-}^{[3]}$ and so on.
By using \eqref{eq:Sn_Vn_action_strand},
the states spanning the various $GL(2)$ representations are found to be
\begin{subequations}
\begin{align}
    \tilde{\cR}(s_{T_{\YTscriptscriptsize\ydiagram{3}}}) \,\Vt^{\otimes 3} &
    = \mathrm{span}\Big\{ \ket{{+}{+}{+}},\ \ket{{-}{-}{-}},\notag\\
    &\hspace{9.2ex}
    \ket{{-}{+}{+}} + \ket{{+}{-}{+}} + \ket{{+}{+}{-}} ,\ \notag\\
    &\hspace{9.2ex}
    \ket{{+}{-}{-}} + \ket{{-}{+}{-}} + \ket{{-}{-}{+}} \Big\} 
    = \Vt_{\YTscriptscriptsize\ydiagram{3}} \ ,
    \label{eq:V3}\\
    \tilde{\cR}(s_{T_{\YTscriptscriptsize\ydiagram{2,1}}})\,\Vt^{\otimes 3}  &
    = \mathrm{span}\Big\{ 
    2\ket{{+}{+}{-}} - \ket{{+}{-}{+}} - \ket{{-}{+}{+}} ,\ \notag\\
    &\hspace{9.3ex}
    2\ket{{-}{-}{+}} - \ket{{-}{+}{-}} - \ket{{+}{-}{-}} 
    \Big\} = \Vt_{\ydiagram{2,1}}^{(1)} \ ,
    \label{eq:V21(1)}
    \\
    \tilde{\cR}(e_{(13)}\circ s_{T_{\YTscriptscriptsize\ydiagram{2,1}}})\,\Vt^{\otimes 3} 
     &
     = \mathrm{span}\Big\{ 
    2\ket{{-}{+}{+}}-\ket{{+}{+}{-}}-\ket{{+}{-}{+}}\,,\nonumber\\
    &\hspace{9.3ex} 
    2\ket{{+}{-}{-}}-\ket{{-}{-}{+}}-\ket{{-}{+}{-}}
    \Big\}
    = \Vt_{\ydiagram{2,1}}^{(2)}\ ,
    \label{eq:V21(2)}
    \\
    \tilde{\cR}(s_{T_{\ydiagram{1,1,1}}})\,\Vt^{\otimes 3}  &
     =\{\varnothing\} = 
     \Vt_{\ydiagram{1,1,1}} \ .
\end{align}
\end{subequations}
Since our toy bosonic example has $\dim(\Vt)=2$ we cannot antisymmetrise 3 states, and so the representation space $\Vt_{\ydiagram{1,1,1}}$ vanishes in this case. 
One can check that these spaces furnish simultaneous representations of $GL(2)$ and $S_3$, labeled by the same $\lambda$.  First, $\Vt_{\ydiagram{3}}$ by itself furnishes the simultaneous $\YTnormalsize\ydiagram{3}$ representation: it spans a $\bf 4$ of $GL(2)$ because the four states in \eqref{eq:V3} rotate into one another under $GL(2)$, while each is invariant under $S_3$.  
Next, $\YTscriptsize\Vt_{\ydiagram{2,1}}^{(1)}$ and  $\Vt_{\ydiagram{2,1}}^{(2)}$ together furnish the simultaneous $\YTnormalsize\ydiagram{2,1}$ representation: under $GL(2)$, they form a $\bf 2$,  each of $\YTscriptsize\Vt_{\ydiagram{2,1}}^{(1)}$  and $\Vt_{\ydiagram{2,1}}^{(2)}$ rotating into itself. For $S_3$, they give a 2-dimensional representation $M_{\ydiagram{2,1}}$, because $S_3$ mixes the first states
in $\Vt_{\ydiagram{2,1}}^{(1)}$ and $\Vt_{\ydiagram{2,1}}^{(2)}$, and likewise the second states in $\Vt_{\ydiagram{2,1}}^{(1)}$ and $\Vt_{\ydiagram{2,1}}^{(2)}$.
So, we get the following decomposition of \eqref{eq:V3_span}:
\begin{equation} \label{eq:V3_decomp}
    \tilde{V}^{\otimes 3} 
    \cong \big(\Vt_{\ydiagram{3}}\otimes M_{\ydiagram{3}}\big) \oplus \big(\Vt_{\ydiagram{2,1}}\otimes M_{\ydiagram{2,1}}\big) \oplus \big(\Vt_{\ydiagram{1,1,1}}\otimes M_{\ydiagram{1,1,1}}\big)\ ,
\end{equation}
where the spaces $M_{\ydiagram{3}}$ and $M_{\ydiagram{1,1,1}}$ are one-dimensional (trivial and alternating) representation spaces of $S_3$.

Let us make a comment on the case with fermionic states.
Take as $\Vt$ the space of right-moving Ramond ground states of $\mathrm{Sym}^N(T^4)$ given in \eqref{eq:Vk_Vt_def}. This is a $\mathbb{Z}_2$-graded space with two states being bosonic and two being fermionic, and so $\tilde{V}$ is a representation space of $GL(2|2)$. In this case the multi-strand Ramond ground states in definite representation spaces $\Vt_{\lam}$ can again be constructed using the above method of Young symmetrizers, however, with the addition of signs coming from the fermions. These signs enter upon the rearrangement of strands in the second line of \eqref{eq:Sn_Vn_action_strand}.  As an example, for the cases of the states $\ket*{\dot{-}}^{[1]}\ket*{\Ad}^{[2]}$ and $\ket*{\Ad}^{[1]}\ket*{\Bd}^{[2]}$, the Young symmetrizer $s_{T_{\ydiagram{2}}}$ yields the following states in $\Vt_{\ydiagram{2}}$ (\textit{c.f.} \eqref{eq:N=2_states_Rep_2})
\begin{subequations}
    \begin{align}
    \tilde{\cR}(s_{T_{\YTscriptscriptsize\ydiagram{2}}}) \big(\ket*{\dot{-}}^{[1]}\ket*{\Ad}^{[2]}\big)
    &= \ket*{\dot{-}}^{[1]}\ket*{\Ad}^{[2]} + \ket*{\dot{-}}^{[2]}\ket*{\Ad}^{[1]} \notag\\
    &= \ket*{\dot{-}}^{[1]}\ket*{\Ad}^{[2]} + \ket*{\Ad}^{[1]}\ket*{\dot{-}}^{[2]} \ .\\
    \tilde{\cR}(s_{T_{\YTscriptscriptsize\ydiagram{2}}}) \big(\ket*{\Ad}^{[1]}\ket*{\Bd}^{[2]}\big)
    &= \ket*{\Ad}^{[1]}\ket*{\Bd}^{[2]} + \ket*{\Ad}^{[2]}\ket*{\Bd}^{[1]} \notag\\
    &= \ket*{\Ad}^{[1]}\ket*{\Bd}^{[2]} - \ket*{\Bd}^{[1]}\ket*{\Ad}^{[2]} \ .
\end{align}
\end{subequations}
In this case of $GL(2|2)$ the dimensions of the representation spaces $\tilde{V}_{\lam}$ can be found from the formula \cite{Moens2004Dimension,469926}
\begin{equation} \label{eq:V_lam_dim_super}
    \dim(\tilde{V}_{\lam}) = \det \left[\,\sum_{k=0}^{\lam_i +j-i} \frac{2(\lam_i + j -i - k + 1)}{k!(2-k)!}\,\right]_{1\leq i,j\leq\rho_{\lam}} \ ,
\end{equation}
where $\rho_{\lam}$ is the number of rows in the Young diagram $\lam$ and the determinant is of the square matrix whose rows and columns are labelled by $i$ and $j$ respectively. The result of the formula \eqref{eq:V_lam_dim_super} is equivalent to setting the bosonic variables to 1 and the fermionic variables to $-1$ in the appropriate super Schur functions \eqref{eq:Super_Schur_power_sum}; \textit{i.e.}, $\dim(\tilde{V}_{\lam}) = S_{\lam}(1|{-1})$ with $b=f=2$.

Finally, let us address the issue mentioned below \eqref{eq:Hn_LxR}.
When we put together the left- and right-moving states, each of which has a Schur-Weyl expansion \eqref{eq:SW_app}, we must take the $S_n$-invariant part of $M_\lambda^L \otimes M_\lambda^R$ to extract the physical state (we put $L$ and $R$ to distinguish $M_\lambda$ that come from left and right-moving sectors).  Let us see how this goes explicitly in the case of $S_3$.
Assume that we have same $\ket{\pm}$ states in \eqref{eq.BosonicToyModel} in the left and right-moving sectors.  Focusing on the non-trivial case of $\lambda=\YTnormalsize\ydiagram{2,1}$, we can take the relevant $S_3$-representation space to be spanned by the first lines of 
\eqref{eq:V21(1)} and \eqref{eq:V21(2)}:
\begin{align}
\YTscriptsize
M_{\ydiagram{2,1}}^L=M_{\ydiagram{2,1}}^R= {\rm span}\Bigl\{
    2\ket{{+}{+}{-}} - \ket{{+}{-}{+}} - \ket{{-}{+}{+}} ,\
    2\ket{{-}{+}{+}}-\ket{{+}{+}{-}}-\ket{{+}{-}{+}}\Bigr\}.
\end{align}
Acting on these with $GL(2)$ matrices gives states in the second line of \eqref{eq:V21(1)} and \eqref{eq:V21(2)}, but as long as we are interested in $S_3$ we can just take the first lines, which do span the two-dimensional space $M_{\ydiagram{2,1}}$.
To ease notation, let us define
\begin{align}
    \ket{1}=\ket{{-}{+}{+}},\qquad
    \ket{2}=\ket{{+}{-}{+}},\qquad
    \ket{3}=\ket{{+}{+}{-}},
\end{align}
so that
\begin{align}
M_{\ydiagram{2,1}}^L=M_{\ydiagram{2,1}}^R= {\rm span}
\Bigl\{ 
2\ket{3}-\ket{1}-\ket{2},\
2\ket{1}-\ket{2}-\ket{3}
\Bigr\}.
\end{align}
Our task is to find the $S_3$-invariant combination 
in the four-dimensional space $M_{\ydiagram{2,1}}^L\otimes M_{\ydiagram{2,1}}^R$.
One brute force way to do it is to define
\begin{align}
    \ket{a}=2\ket{3}-\ket{1}-\ket{2},\qquad
\ket{b}=2\ket{1}-\ket{2}-\ket{3}
\end{align}
and write the $S_3$ action in this basis.  We find, for the generating elements of $S_3$,
\begin{align}
\begin{split}
    e_{(12)} \ket{a} &= \ket{a},\qquad
    e_{(12)} \ket{b} = -\ket{a}-\ket{b},\\
    e_{(13)} \ket{a} &= \ket{b},\qquad
    e_{(13)} \ket{b} = \ket{a}.
\end{split}
\end{align}
Then we consider the general left-right combination $x\ket{aa}+y\ket{ab}+z\ket{ba}+w\ket{bb}$ and fix the coefficients by requiring $S_3$ invariance. We find that the invariant combination is
\begin{align}
2\ket{aa}+2\ket{bb}+\ket{ab}+\ket{ba},
\end{align}
which can be written in the $\{\ket{1},\ket{2},\ket{3}\}$ basis in a more covariant way as
\begin{align}
\sum_{i,j=1}^3 \left(\delta_{ij}-{1\over 3}\right)\ket{ij},
\end{align}
up to an overall factor.  So, this is the state that spans
$(M_{\ydiagram{2,1}}^L\otimes M_{\ydiagram{2,1}}^R)_{\text{$S_3$-inv}}$.
This is for one of the four components that appear in $V_\lambda\otimes V_\lambda$ for $GL(2)$, but the other components are obtained by considering the second lines of \eqref{eq:V21(1)} and \eqref{eq:V21(2)} as the left- and right-moving states. Finding the $S_n$-invariant state in $M_\lambda^L\otimes M_\lambda^R$ in more general cases should follow in the same way.

\section{Contributions of $GL(2\,|\,2)$ characters to the MEG}
\label{app:GL22MEG}

In this Appendix we give a derivation of the result that the modified elliptic genus only receives non-zero contributions from symmetry sectors labelled by Young diagrams of the single hook type $\lam\in H(1|1)$, \textit{i.e.}\ $\lam=\{\lam_1,\dots,\lam_{\rho_{\lam}}\}$ with $\lam_{2}\leq1$. In the case of the symmetric orbifold of $T^4$ that we discuss in Section \ref{ssec:SW_T4} the contributions of right-moving parts of states to the multi-strand (and grand canonical) partition function come in the form of characters of $GL(2|2)$ irreducible representation subspaces. These are given by the super Schur functions detailed in Appendix \ref{app:symmPoly} constructed from (super) power sum functions made of fundamental $GL(2|2)$ characters
\begin{equation}
    \tilde{z}(\ty,\te) \equiv \tr_{\tilde{V}_{\YTscriptscriptsize\ydiagram{1}}} \big[(-1)^{F_R} \ty^{2\tilde{J}^3_0}\te^{2\tilde{K}^3_0}\big] = \ty + \ty^{-1} - \te - \te^{-1} \ ,
\end{equation}
where we have included fugacities for generators of the Cartan of both $su(2)_R$ and $\widetilde{su}(2)_2$. Simply setting $\te=1$ recovers the character given in \eqref{eq.gl22_char_def1}. Using the definitions in \eqref{eq:Super_Schur_power_sum} and \eqref{eq:def_omega_lambda} these (super) Schur functions are then given in this case by
\begin{align} \label{eq:Schur_ito_power_sum_T4}
    \St_{\lam}(\ty,\te) &= \sum_{\mathbf{i}\,\vdash n_{\lam}} \frac{\omega_\lambda({\bf i})}{z(\bf i)} \prod_{j=1}^{n_{\lam}} \tilde{z}(\ty^j,\te^j)^{i_j} \ ,
\end{align}
where $\mathbf{i}=(i_1,\dots,i_{n_{\lam}})$ and $\sum_{\alpha=1}^{n_{\lam}} \alpha i_{\alpha} = n_{\lam}$.

The contribution of this single right-moving character to the modified elliptic genus \eqref{eq:MEG_T^4} is then obtained by the action of the differential operator ${\mathcal D}[\,\cdot\,] \equiv \frac12 (\tilde{y}\partial_{\tilde y})^2[\,\cdot\,]|_{\ty=\te=1}$ which provides the weights for the various left-moving characters in the Schur-Weyl expression \eqref{eq:Z=sum_SS_T4}. Acting with this differential operator on \eqref{eq:Schur_ito_power_sum_T4} gives
\begin{align} \label{eq:DSchur1}
    \mathcal{D}\big[\St_{\lam}(\ty,\te)\big] &= \frac12 \sum_{\mathbf{i}\,\vdash n_{\lam}} \frac{\omega_\lambda({\bf i})}{z(\bf i)} \Bigg[ \sum_{r=1}^{n_{\lam}}\big(\ty\pd_{\ty}\big)^2 \tilde{z}(\ty^r,\te^r)^{i_r} \prod_{\substack{j=1\\j\neq r}}^{n_{\lam}} \tilde{z}(\ty^j,\te^j)^{i_j} \nonumber\\
    &\quad+ \sum_{\substack{r_1,r_2=1\\r_1\neq r_2}}^{n_{\lam}} \Big(\ty\pd_{\ty}\,\tilde{z}(\ty^{r_1},\te^{r_1})^{i_{r_1}}\Big) \Big(\ty\pd_{\ty}\,\tilde{z}(\ty^{r_2},\te^{r_2})^{i_{r_2}}\Big) \prod_{\substack{j=1\\j\neq r_1,r_2}}^{n_{\lam}} \tilde{z}(\ty^j,\te^j)^{i_j}\Bigg]_{\ty,\te=1} \ ,
\end{align}
where the various derivatives of the $GL(2|2)$ characters can be found to be
\begin{subequations}
\begin{align}
    \ty\pd_{\ty}\,\tilde{z}(\ty^{r},\te^{r})^{i_{r}}\Big|_{\ty,\te=1} &= r i_r \,\tilde{z}(\ty^{r},\te^{r})^{i_{r}-1} \big(\ty^r - \ty^{-r}\big)\Big|_{\ty,\te=1} = 0\quad \forall\ r,i_r \ ,\\
    \big(\ty\pd_{\ty}\big)^2 \tilde{z}(\ty^r,\te^r)^{i_r}\Big|_{\ty,\te=1} &= \bigg[r^2 i_r(i_r-1) \,\tilde{z}(\ty^{r},\te^{r})^{i_{r}-2} \big(\ty^r - \ty^{-r}\big)^2 \nonumber\\
    &\hspace{2cm} + r^2 i_r \,\tilde{z}(\ty^{r},\te^{r})^{i_{r}-1} \big(\ty^r + \ty^{-r}\big) \bigg]_{\ty=\te=1} \nonumber\\
    &= 2r^2 \delta_{i_r,1} \ .
\end{align}
\end{subequations}
Using these results in \eqref{eq:DSchur1} we find
\begin{align} \label{eq:DSchur2}
    \mathcal{D}\big[\St_{\lam}(\ty,\te)\big] &= \frac12\sum_{\mathbf{i}\,\vdash n_{\lam}} \frac{\omega_\lambda({\bf i})}{z(\bf i)} \sum_{r=1}^{n_{\lam}} 2r^2 \delta_{i_r,1} \prod_{\substack{j=1\\j\neq r}}^{n_{\lam}} \tilde{z}(\ty^j,\te^j)^{i_j} \bigg|_{\ty=\te=1} \nonumber\\
    &= \sum_{\mathbf{i}\,\vdash n_{\lam}} \frac{\omega_\lambda({\bf i})}{z(\bf i)}\sum_{r=1}^{n_{\lam}}r^2 \delta_{i_r,1} \prod_{\substack{j=1\\j\neq r}}^{n_{\lam}} \delta_{i_j,0} \nonumber\\
    &= \sum_{\mathbf{i}\,\vdash n_{\lam}} \frac{\omega_\lambda({\bf i})}{z(\bf i)} n_{\lam}^2 \,\delta_{i_1,0}\cdots\delta_{i_{n_{\lam}-1},0}\,\delta_{i_{n_{\lam}},1} \nonumber\\
    &= \frac{\omega_{\lam}\big((0,\dots,0,1)\big)}{z\big({(0,\dots,0,1)}\big)} n_{\lam}^2 \nonumber\\
    &= n_{\lam}\, \omega_{\lam}\big((0,\dots,0,1)\big) \ ,
\end{align}
where in the third line the fact that $\bf i$ is a partition of $n_{\lam}$ localises the sum over $r$ to $r=n_{\lam}$, and in the fifth line the definition \eqref{eq:Schur_ito_power_sum} of the coefficients $z(\bf i)$ was used.

The $\omega_{\lam}({\bf i})$ can be expressed using the Frobenius formula \eqref{eq:def_omega_lambda}, which for the particular conjugacy class $\mathbf{i}=(0,\dots,0,1)$ is
\begin{align} \label{eq.charexample}
    \omega_{\lam}\big((0,\dots,0,1)\big) = \Delta_{\lam}(x) P_{n_{\lam}}(x)\Big|_{x^{\ell}} = \prod_{\substack{i,j=1\\i<j}}^{\rho_{\lam}}(x_i-x_j)\sum_{r=1}^{\rho_{\lam}}x_r^{n_{\lam}}\bigg|_{x^{\ell}} \ ,
\end{align}
where the notation of Appendix \ref{app:symmPoly} is used, including $\ell=(\ell_1,\dots,\ell_{b})$ and $\ell_i=\lam_i+b-i$. We emphasise that here $()|_{x^{\ell}}$ means to project onto the coefficient of $x_1^{\ell_1}\cdots x_{b}^{\ell_{b}}$. Some useful inequalities satisfied by the various quantities involved are:
\begin{subequations}
\begin{align}
    \lam_1\geq\lam_2\geq&\cdots\geq\lam_{\rho_{\lam}}\ \Lto\  \ell_1 > \ell_2 > \cdots > \ell_{\rho_{\lam}}\ ,\label{eq.lamelllimits}\\[0.7ex]
    1\leq\lam_1\leq n_{\lam} \quad&,\quad 1\leq \rho_{\lam}\leq n_{\lam} \quad,\quad \floor{\sqrt{4n_{\lam}-3}} \leq \ \ell_1 \leq n_{\lam}\ .\label{eq.lamklimits}
\end{align}
\end{subequations}
Focussing on the $x_1$ dependence in \eqref{eq.charexample}
\begin{align} \label{eq.charexmaple2}
    \omega_{\lam}\big((0,\dots,0,1)\big) &= (x_1-x_2)\cdots(x_1-x_{\rho_{\lam}}) \prod_{\substack{i,j=2\\i<j}}^{\rho_{\lam}}(x_i-x_j)\bigg(x_1^{n_{\lam}} + \sum_{r=2}^{\rho_{\lam}}x_r^{n_{\lam}}\bigg)\bigg|_{x^{\ell}} \nonumber\\
    &= \sum_{t=0}^{\rho_{\lam}-1} c_{t}(x_2,\dots,x_{\rho_{\lam}}) x_1^{t} \prod_{\substack{i,j=2\\i<j}}^{\rho_{\lam}}(x_i-x_j)\bigg(x_1^{n_{\lam}} + \sum_{r=2}^{\rho_{\lam}}x_r^{n_{\lam}}\bigg)\bigg|_{x^{\ell}} \nonumber\\
    &= c_0(1,\dots,1)\prod_{r=2}^{\rho_{\lam}}x_r \prod_{\substack{i,j=2\\i<j}}^{\rho_{\lam}}(x_i-x_j) x_1^{n_{\lam}}\bigg|_{x^{\ell}} \nonumber\\
    &\quad + \sum_{t=0}^{\rho_{\lam}-1} c_{t}(x_2,\dots,x_{\rho_{\lam}}) x_1^{t} \prod_{\substack{i,j=2\\i<j}}^{\rho_{\lam}}(x_i-x_j)\sum_{r=2}^{\rho_{\lam}}x_r^{n_{\lam}}\bigg|_{x^{\ell}} \nonumber\\
    &= (-1)^{\rho_{\lam}-1} \prod_{r=2}^{\rho_{\lam}}x_r \prod_{\substack{i,j=2\\i<j}}^{\rho_{\lam}}(x_i-x_j)\bigg|_{x_2^{\ell_2}\cdots x_{\rho_{\lam}}^{\ell_{\rho_{\lam}}}} \delta_{\ell_1,n_{\lam}} \ ,
\end{align}
where in the second line we expand in a power series in $x_1$ with expansion coefficients $c_t$. In the third line the second term vanishes due to each $x_r$ for $r\geq2$ having a power of at least $n_{\lam}$ and from \eqref{eq.lamelllimits} and \eqref{eq.lamklimits} the $\ell_r<n_{\lam}$ for $r\geq2$. The resulting condition $\ell_1=n_{\lam}$ implies that the number of rows after the first $\rho_{\lam}-1=n_{\lam}-\lam_1$ which in turn implies that $\lam_2\leq1$. \textit{i.e.}\ $\lam$ is of the single hook type, $\lam\in H(1|1)$. 

This is a necessary condition for a non-zero contribution to the MEG. For it to be a sufficient condition one also needs to show that when $\ell_1=n_{\lam}$ there are no values of the other $\ell_i$ such that \eqref{eq.charexmaple2} vanishes. We can see this by noticing that if $\ell_1=n_{\lam}$ then (if they exist) $\ell_i=\rho_{\lam}-i+1$ for $i\geq2$. Then expanding \eqref{eq.charexmaple2} further yields
\begin{align} \label{eq.charexmaple3}
    \omega_{\lam}\big((0,\dots,0,1)\big) &= (-1)^{\rho_{\lam}-1} \prod_{r=2}^{\rho_{\lam}}x_r \prod_{\substack{i,j=2\\i<j}}^{\rho_{\lam}}(x_i-x_j)\bigg|_{x_2^{\rho_{\lam}\!-1}x_3^{\rho_{\lam}\!-2}\cdots x_{\rho_{\lam}}^{1}} \delta_{\ell_1,n_{\lam}} \nonumber\\
    &= (-1)^{\rho_{\lam}-1} \sum_{R=0}^{\rho_{\lam}-2}x_2^{\rho_{\lam}-1-R}(-1)^R\sum_{\substack{r_1,\dots,r_R=3\\r_1\neq\cdots\neq r_R}}^{\rho_{\lam}}x_{r_1}\cdots x_{r_R} \nonumber\\
    &\hspace{3.5cm} \times\prod_{r=3}^{\rho_{\lam}}x_r \prod_{\substack{i,j=3\\i<j}}^{\rho_{\lam}}(x_i-x_j) \bigg|_{x_2^{\rho_{\lam}\!-1}x_3^{\rho_{\lam}\!-2}\cdots x_{\rho_{\lam}}^{1}} \delta_{\ell_1,n_{\lam}} \nonumber\\
    &= (-1)^{\rho_{\lam}-1} \prod_{r=3}^{\rho_{\lam}}x_r \prod_{\substack{i,j=3\\i<j}}^{\rho_{\lam}}(x_i-x_j) \bigg|_{x_3^{\rho_{\lam}\!-2}\cdots x_{\rho_{\lam}}^{1}} \delta_{\ell_1,n_{\lam}} \nonumber\\
    &= (-1)^{\rho_{\lam}-1} \delta_{\ell_1,n_{\lam}}\ ,
\end{align}
where we iteratively expand as a power series in $x_i$ for $i=2,3,\dots$ and each time only one term ($R=0$) in the expansion survives the projection. Thus in total, by combining \eqref{eq:DSchur2} and \eqref{eq.charexmaple3}, we get the contribution from one right-moving Schur polynomial to the MEG as
\begin{align} \label{eq.MEGcontrib}
    \mathcal{D}\big[\St_{\lam}(\ty,\te)\big] = (-1)^{\rho_{\lam}-1} n_{\lam} \,\delta_{\ell_1,n_{\lam}}\ ,
\end{align}
where the Kronecker delta implements the condition $\lam_2\leq 1$, \textit{i.e.}\ that $\lam\in H(1|1)$.

\section{Derivation of the generating function for the REG}
\label{app:genFuncREG}

In this Appendix, we derive the formula for the generating function for the REG, \eqref{genFuncREG}.

Let us start by defining the signed partition function,
\begin{align}
Z_N(q,y,\yt,\etat)=\Tr_N\!\big[(-1)^{2(J^3_0-\Jt^3_0)}q^{L_0-{c\over 24}}y^{2J^3_0}\yt^{2\Jt^3_0}\etat^{2\Kt^3}\big] \ ,
\end{align}
and expanding it in the $\scrA$-algebra character as
\begin{align}
Z_N(q,y,\yt,\etat)=\sum_{\tj,\tj_2}c_{\tj,\tj_2}(q,y)\,\chi^{\scrA}_{\tj,\tj_2}(\yt,\etat) \ ,
\label{jwzl16Jan26}
\end{align}
where $\chi^{\scrA}_{\tj,\tj_2}(\yt,\etat)$ was defined in \eqref{eq:A_alg_chars} and the range of summation is $\tj,\tj_2=0,\half,1,\dots$.
As discussed in Section \ref{ssec:T4_MEG}, the MEG is obtained by acting with $\cD=\half (\yt \p_{\yt})^2|_{\yt=\etat=1}$ on $Z_N(q,y,\yt,\etat)$:
\begin{align}
\cE_N(q,y)=\cD Z_N(q,y,\yt,\etat)
=\sum_{\tj,\tj_2} c_{\tj,\tj_2}(q,y)\,
(-1)^{2\tj_2} (2\tj+1)(2\tj_2+1) \ ,
\label{jrqs16Jan26}
\end{align}
where we used
\begin{align}
\cD \chi^{\scrA}_{\tj,\tj_2}=(-1)^{2\tj_2}(2\tj+1)(2\tj_2+1).\label{ewfq28Nov25}
\end{align}
The REG, $\cE_{N,\tj_2}$, is defined as the partial sum of
\eqref{jrqs16Jan26} over $\tj$ with fixed $\tj_2$.  

To define the generating function for the REG,
we want to introduce a bookkeeping parameter $\etatb$ to grade terms in the expansion in \eqref{jrqs16Jan26} as
\begin{align}
\cE_N(q,y,\etatb)
&\equiv \sum_{\tj,\tj_2} c_{\tj,\tj_2}(q,y)\,
(-1)^{2\tj_2} (2\tj+1)(2\tj_2+1)
{\etatb^{2\tj_2+1}+\etatb^{-2\tj_2-1}\over 2}
\label{kjcn16Jan26}
\\
&= \sum_{\tj_2} \cE_{N,\tj_2}(q,y)\,
{\etatb^{2\tj_2+1}+\etatb^{-2\tj_2-1}\over 2} \ .
\end{align}
We denote the bookkeeping parameter by boldface $\etatb$ to distinguish it from the fugacity $\etat$ inserted in the partition function's trace.
Obviously, we can transform \eqref{jwzl16Jan26} into $\cE_N(q,y,\etatb)$ if we can find a linear map that transforms
$\chi^{\scrA}_{\tj,\tj_2}(\yt,\etat)$ into $(-1)^{2\tj_2} (2\tj+1)(2\tj_2+1){\etatb^{2\tj_2+1}+\etatb^{-2\tj_2-1}\over 2}$.  Or, if we rewrite
\eqref{jwzl16Jan26} with $\yt=1$ as
\begin{align}
{Z_N(q,y,1,\etat)
\over \chi_{\half}(1)-\chi_{\half}(\etat)}
=\sum_{\tj,\tj_2}
    c_{\tj,\tj_2}(q,y)\,
    (-1)^{2\tj_2} (2\tj+1)\chi_{\tj_2}(\etat) \ ,
    \label{khiy16Jan26}
\end{align}
what we want is a linear map that sends
\begin{align}
\chi_{\tj_2}(\etat) 
~\to~
(2\tj_2+1){\etatb^{2\tj_2+1}+\etatb^{-2\tj_2-1}\over 2} \ .
\end{align}
A map that does the job is
\begin{align}
 (\cT f)(\etatb)
 &\equiv
 {\etatb-\etatb^{-1}\over 4}
 \left(\oint_{\etatb}-\oint_{\etatb^{-1}}\right) {d\etat\over 2\pi i \etat}
 {(\etat^2-1)^2\over (\etat-\etatb)^2(\etat-\etatb^{-1})^2}
 f(\etat)\ ,
 \label{echq28Nov25}
\end{align}
where $f(\etat)$ is an arbitrary function.
We can see that this indeed has the desired property:
\begin{align}
 (\cT \chi_j)(\etatb)
 &\equiv
 {\etatb-\etatb^{-1}\over 4}
 \left(\oint_{\etatb}-\oint_{\etatb^{-1}}\right) {d\etat\over 2\pi i}
 {\etat^2-1\over (\etat-\etatb)^2(\etat-\etatb^{-1})^2}
 \big(\etat^{2j+1}-\etat^{-2j-1}\big)\notag\\
 &=
 {\etatb-\etatb^{-1}\over 4}
 \left(\Res_{\etat=\etatb}- \Res_{\etat=\etatb^{-1}}\right)
 \biggl[
 {\etat^2-1\over (\etat-\etatb)^2(\etat-\etatb^{-1})^2} \big(\etat^{2j+1}-\etat^{-2j-1}\big)
 \biggr]
 \notag\\
&=
{1\over 4}\Big[
 \big((2j+1)\etatb^{2j+1} + (2j+1)\etatb^{-2j-1}\big)
 -
 \big(-(2j+1)\etatb^{-2j-1}-(2j+1)\etatb^{2j+1}\big)
\Big]
\notag\\
&=(2j+1){\etatb^{2j+1}+\etatb^{-2j-1}\over 2}\ .
\label{mbje16Jan26}
\end{align}
We will discuss the construction of this operator later toward the end of the current Appendix.

With the operator $\cT $, computing the generating function for the REG is a simple matter of applying it to \eqref{khiy16Jan26}:
\begin{align}
\cE_{N}(q,y,\etatb)
&=\cT \biggl({Z(q,y,1,\etat)\over \chi_\half(1)-\chi_\half(\etat)}\biggr)(\etatb)
\notag\\
&=-{\etatb-\etatb^{-1}\over 4}\left(\oint_{\etatb}-\oint_{\etatb^{-1}}\right)
 {d\etat\over 2\pi i}
 {(\etat+1)^2\over (\etat-\etatb)^2(\etat-\etatb^{-1})^2}
Z(q,y,1,\etat)
\notag\\
 &={1\over (\etatb^{\half}-\etatb^{-\half})^2}
 \left(1-{\etatb-\etatb^{-1}\over 2}\etatb\p_{\etatb}\right)Z(q,y,1,\etatb) \ ,
 \label{cE(q,y,etatb)_app}
\end{align}
where we used that fact that the spectrum is invariant under $\Kt^3\to
-\Kt^3$ and therefore $Z(q,y,1,\etatb)=Z(q,y,1,\etatb^{-1})$.  
The same relation holds between generating functions
\begin{align}
\cZ(p,q,y,\yt,\etat)
\equiv\sum_N Z_N(q,y,\yt,\etat)\,p^N
=\sum_{\jt,\jt_2}c_{\jt,\jt_2}(p,q,y)\,\chi^{\scrA}_{\jt,\jt_2}(\yt,\etat)
\end{align}
and 
\begin{align}
    \cE(p,q,y,\etatb)
    &\equiv \sum_N \cE_N(q,y,\etatb)\,p^N\notag\\
    &=\sum_{\jt,\jt_2}c_{\jt,\jt_2}(p,q,y)\,(-1)^{2\jt_2}(2\jt+1)(2\jt_2+1)
    {\etatb^{2\tj_2+1}+\etatb^{-2\tj_2-1}\over 2} \ ;
    \label{cE(p,q,y,etabtb)_def_app}
\end{align}
namely,
\begin{align}
\cE(p,q,y,\etatb)
 &={1\over (\etatb^{\half}-\etatb^{-\half})^2}
 \left(1-{\etatb-\etatb^{-1}\over 2}\etatb\p_{\etatb}\right)\cZ(p,q,y,1,\etatb)\ ,
 \label{cE(p,q,y,etatb)_app}
\end{align}
where
\begin{align}
 \cZ(p,q,y,\yt,\etatb)
= \prod_{k,r,\ell}\left[{(1-p^k q^r y^{\ell} \etatb)(1-p^k q^r y^{\ell} \etatb^{-1})
\over (1-p^k q^r y^{\ell}\yt)(1-p^k q^r y^{\ell}\yt^{-1})}\right]^{c(k,r,\ell)} \ .
    \label{cZ(p,q,y,yt,etabtb)_def_app}
\end{align}
This  is what we 
presented in \eqref{genFuncREG}.

We can check that this reduces to the usual MEG in the
$\etatb\to 1$ limit.  If we write $h(\etatb)\equiv \log \cZ(p,q,y,1,\etatb)$ for simplicity, the behavior as $\etatb\to 1 $ is
\begin{align}
\cE(p,q,y,1+\epsilon)
&=e^{h(1)}\left[{1\over \epsilon^2}+{1\over \epsilon}
 -{h'(1)+h'(1)^2+h''(1)\over 2}+\cO(\epsilon)\right]\ .
\end{align}
Noting that
\begin{gather}
 h(1)=0,\qquad
 h'(1)=0,\qquad
 h''(1)=-\sum_{k,r,\ell} {2c(kr,\ell) p^k q^r y^{\ell}\over (1-p^k q^r y^{\ell})^2 }\ ,
\end{gather}
we obtain
\begin{align}
\cE(p,q,y,1+\epsilon)
&={1\over \epsilon^2}+{1\over \epsilon}
 +\sum_{k,r,\ell}{c(kr,\ell)\, p^k q^r y^{\ell}
 \over (1-p^k q^r y^{\ell})^2}+\cO(\epsilon)\ .
\end{align}
Up to a divergent constant which is $\cO(p^0)$ and irrelevant, this correctly reduces to the MEG \eqref{eq.MMS_generating}.

Let us discuss the construction of the operator $\cT $ in \eqref{echq28Nov25}.
Let's say we want a map that sends
\begin{align}
\chi_{j}(\etat) 
~\to~
(2j+1)\etatb^{2j+1}~.
\label{kmnc16Jan26}
\end{align}
This can be done by introducing a kernel
\begin{align}
 K(\theta,\etatb)=\sum_j(2j+1)\etatb^{2j+1}\chi_j(\theta)\ , \qquad
 \etat=e^{i\theta/2} \ ,
 \label{mruj27Nov25}
\end{align}
and defining a map $\cT '$ acting on a function $f(\theta)$ as follows:
\begin{align}
 (\cT 'f)(\etatb)
 &\equiv {1\over 2\pi}\int_0^{4\pi}d\theta\,
 \sin^2\!{\theta\over 2}\, K(\theta,\etatb) f(\theta)\ .
 \label{mrim27Nov25}
\end{align}
Indeed, by orthogonality \eqref{eq:su(2)_char_orth_theta},
\begin{align}
 (\cT ' \chi_j)(\etatb)
 &\equiv {1\over 2\pi}\int_0^{4\pi}d\theta\,
 \sin^2\!{\theta\over 2}\, 
 K(\theta,\etatb) \chi_j(\theta)
\notag\\
 &=\sum_{j'}(2j'+1)\etatb^{2j'+1}\,{1\over 2\pi}\int_0^{4\pi}d\theta\,\sin^2{\theta\over 2}\, 
 \chi_j(\theta) \,\chi_{j'}(\theta)
 \notag\\
&=(2j+1)\etatb^{2j+1}\ .
\end{align}
Carrying out summation over $j$ in
\eqref{mruj27Nov25}, we find an explicit expression for the kernel:
\begin{align}
 K(\theta,\etatb)
 ={\etatb(1-\etatb^2)\over (1+\etatb^2-2\etatb\cos{\theta\over 2})^2}
 =-{\etat^2 (\etatb-\etatb^{-1})\over (\etat-\etatb)^2(\etat-\etatb^{-1})^2}\ .
\end{align}
For convergence, we assumed that $|\etatb|<1$.
The action on general $f$, \eqref{mrim27Nov25}, can then be written as
\begin{align}
 (\cT 'f)(\etatb)
 &\equiv
 {\etatb-\etatb^{-1}\over 2}
 \oint {d\etat\over 2\pi i \etat}
 {(\etat^2-1)^2\over (\etat-\etatb)^2(\etat-\etatb^{-1})^2}
 f(\etat)\ ,
\end{align}
where the contour integral is counterclockwise along the unit circle $|\etat|=1$.  Depending on  $f(\etat)$, we must close the contour appropriately to pick up poles.  Or, instead, we can define the map
\begin{align}
 (\cT f)(\etatb)
 &\equiv
 {\etatb-\etatb^{-1}\over 4}
 \left(
 \oint_{\etatb}-\oint_{\etatb^{-1}}
 \right){d\etat\over 2\pi i \etat}
 {(\etat^2-1)^2\over (\etat-\etatb)^2(\etat-\etatb^{-1})^2}
 f(\etat)\ ,
\end{align}
which picks up poles at $\etat=\etatb,\etatb^{-1}$ independent of $f(\etat)$.  In particular, this gives
\begin{align}
 (\cT \chi_j)(\etatb)
 &=
 (2j+1){\etatb^{2j+1}+\etatb^{-2j-1}\over 2}\ ,
\end{align}
as we saw in
\eqref{mbje16Jan26}.

\subsection{Generating function for the ``$(\jt,\jt_2)$-resolved REG''}
\label{ss:gen_fun_(jt,jt2)-REG}

Here we derive the generating function for the  ``$(\jt,\jt_2)$-resolved REG'' $\cE_{\jt,\jt_2}(p,q,y)$ introduced in Section \ref{sec:summary}.

The generating function for the original, $\jt_2$-resolved REG $\cE_{\jt_2}(p,q,y)$, defined in \eqref{cE(p,q,y,etabtb)_def_app}, sums over $\scrA$-representations $\cR^{\scrA}_{\jt,\jt_2}$ with the bookkeeping parameter $\etatb$ that records the $\widetilde{su}(2)_2$ label $\jt_2$.  We can straightforwardly generalize this to a generating function for $\cE_{N,\jt,\jt_2}(p,q,y)$ by introducing another parameter $\ytb$ that records the $su(2)_R$ label $\jt$ as well:
\begin{align}
    &\cE(p,q,y,\ytb,\etatb)
    \equiv 
    \sum_{\jt,\jt_2}\cE_{N,\jt,\jt_2}(p,q,y)\,
        {\ytb^{2\jt+1}+\ytb^{-2\jt-1}\over 2}
        {\etatb^{2\jt_2+1}+\etatb^{-2\jt_2-1}\over 2}\notag\\
    &=\sum_{\jt,\jt_2}c_{\jt,\jt_2}(p,q,y)(-1)^{2\jt_2}(2\jt+1)(2\jt_2+1)
        {\ytb^{2\jt+1}+\ytb^{-2\jt-1}\over 2}
        {\etatb^{2\jt_2+1}+\etatb^{-2\jt_2-1}\over 2}.
\end{align}
We can derive a closed expression for $\cE(p,q,y,\ytb,\etatb)$ just as we derived \eqref{cE(q,y,etatb)_app}, now by using the $\cT$-operator for both $\etatb$ and $\ytb$.  The final result is:
\begin{align}
&\cE(p,q,y,\ytb,\etatb)
=\frac{\etatb \ytb}{4 (\ytb-\etatb)^3 (\etatb \ytb-1)^3}
\notag\\
&\quad\times
 \Biggl\{
 2 \Bigl[-12 \etatb^2 \ytb^2+4 \etatb (\etatb^2+1) (\ytb^2+1) \ytb-(\etatb^4+1) (\ytb^4+1)\Bigr]
\notag\\
&\qquad\quad
+ (\ytb-\etatb) (\etatb \ytb-1)\notag\\
&\qquad\qquad \times \biggl[
 \Bigl((\etatb^2+1)(\ytb^2+1)-4 \etatb \ytb\Bigr) 
 \Bigl((\ytb^2-1) \p_{\ytb}-(\etatb^2-1) \p_{\etatb}\Bigr)
\notag\\
 &\qquad\qquad\qquad
 +(\etatb^2-1) (\ytb-\etatb) (\etatb \ytb-1) (\ytb^2-1) \p_{\ytb}\p_{\etatb}
 \biggr]
 \Biggr\} \cZ(p,q,y,\ytb,\etatb),
\end{align}
where $\cZ(p,q,y,\ytb,\etatb)$ is defined in \eqref{cZ(p,q,y,yt,etabtb)_def_app}.

\bibliographystyle{utphys}
\bibliography{REGPaper.bib}

\end{document}